\documentclass {article}

\usepackage {psfig}

\newcommand {\be}{\begin {equation}}
\newcommand {\ee}{\end {equation}}
\newcommand{\etal}{et al.}

\newcommand{\mph}{\,h^{-1}_{100}\,{\rm \mbox{Mpc}}}
\newcommand{\bef}{\begin {figure}}
\newcommand{\enf}{\end {figure}}

\begin {document}

\hoffset-3cm
\voffset-2cm
%\title{\large\bf 
\begin {center}
{\bf\Large Fractal Approach to Large--Scale Galaxy Distribution}
\end {center}
\begin {center}
%Galaxy Distribution. Part I.
%\end{center}
% \author{
{\bf\large Yurij Baryshev and Pekka Teerikorpi}
\\
 Institute of Astronomy, St.-Petersburg University,\\
 Staryj Peterhoff, 198904 St.-Petersburg, Russia,\\
 e-mail: yuba@astro.spbu.ru
\\
 Tuorla Observatory, University of Turku,\\
21500 Piikkio, Finland, \\
 e-mail: pekkatee@utu.fi
\end {center}
%\date{~}

%\title{
%\begin{center}
%Fractal Approach to Large--Scale Galaxy Distribution
%\end{center}
%Fractality of Galaxy Clustering: Historical Milestones and
%Challenges of Modern Research
%}
%\author{Yu.Baryshev\inst{a}\thanks{Email: yuba@astro.spbu.ru}
%	  \and P. Teerikorpi\inst{b}
%}
%\institute{
%Institute of Astronomy, St-Petersburg State University,
%Staryi Peterhoff 198504,
%Russia
%\and
%Tuorla Observatory, University of Turku,
%21500 Piikki\"{o}, Finland
%}

%\maketitle

\begin{abstract}
\noindent
We present a review of the history and the present state of the
fractal approach to
the large-scale distribution of galaxies.
The roots of the modern idea of cosmic fractality
go to the beginning of the 20th century,
when hierarchical world models were proposed by Fournier and Charlier.
Fundamental aspects
  of hierarchical matter distribution 
was discussed by Einstein and Selety in the correspondence
concerning inhomogeneous cosmological  models.

~~The Great Debate on the nature of spiral nebulae went over to
a struggle on the structure and  
extension of the galaxy clustering. 
Already during the epoch of galaxy angular catalogues
astronomers detected  superclusters of galaxies on the sky.
However, early counts of bright galaxies were close to
the $0.6m$-law, pointing to homogeneity.
Also the debated variable extinction seemed to be a reasonable
explanation of apparent galaxy clustering.
Angular correlation function analysis gave small
a homogeneity scale $R_{\mathrm{hom}}\approx 10$ Mpc
within which the correlation exponent corresponds to  
a fractal dimension $D \approx 1.2$ and outside which
the galaxy distribution is homogeneous.

~~It was realized later that a normalization condition
for the reduced correlation function estimator results in
distorted values for both  $R_{\mathrm{hom}}$  and $D$.
Moreover, according to a theorem on projections of fractals,
galaxy angular catalogues can not be used for detecting a structure with
the fractal dimension $D \geq 2$. For this 3-d maps are required, and indeed
modern extensive redshift-based
3-d maps have revealed the ``hidden'' fractal dimension of about 2, and
have confirmed superclustering at scales even up to 500 Mpc
(the Sloan Great Wall).
On scales, where the fractal analysis is possible in completely embedded
spheres, a power--law  density field has been found.
 
~~Two new fundamental cosmic numbers have appeared,
the fractal dimension $D$ and the crossover
scale to homogeneity $R_{\mathrm{hom}}$. Their values have been  debated, and
$D = 1.2$ indirectly deduced from angular catalogues has
been replaced by $D =2.2 \pm  0.2$ directly obtained from 3-d maps and
$R_{\mathrm{hom}}$ has expanded from 10 Mpc to scales approaching 100 Mpc.
In concordance with the 3-d map results, modern all sky galaxy counts in the
interval $10^m \div 15^m$ give a $0.44m$-law which corresponds to $D=2.2$
within a radius of 100$h^{-1}_{100}$  Mpc.
The narrow cones of the existing deep galaxy surveys
and poorly known peculiar velocities at small scales
are still the main limiting factors hampering precise estimates of $D$ 
and $R_{\mathrm{hom}}$.
We emphasize that the fractal mass--radius law of galaxy clustering has become
a key phenomenon in observational cosmology. It creates novel challenges
for theoretical understanding of the origin and evolution of the galaxy
distribution, including the role of dark matter and dark energy.
\end{abstract}

\newpage

{\LARGE CONTENTS}
\vskip 0.5 cm
%{\Large \bf PART I}
%\vskip 0.5 cm

{\Large \bf 1. Introduction}

\vskip 0.5 cm

{\Large \bf 2. The fractal view of large-scale structure of the Universe}

\vskip 0.2 cm

{\large 2.1.The idea of a self-similar universe }

{\it Protofractals }

{\it Fournier d'Albe and Carl Charlier }

{\large 2.2.Genuine fractal structures}

{\it Self-similarity and power law }

{\it Fractal dimension }

{\large 2.3. Concepts of density field}

{\it Ordinary fluid-like density fields}

{\it Ordinary stochastic discrete processes}

{\it Fractal density fields}

{\large 2.4.Exclusive properties of fractals}

{\it Power-law density-radius relation}

{\it Massive, zero-density
universes }

{\it Lower and upper cutoffs}

{\it Lacunarity}

{\it Projection and intersection}

{\it Multifractal structures}

{\large 2.5.Modern redshift and photometric distance
surveys}

{\it Redshift surveys}

{\it Limits of a survey}

{\it Galaxy catalogues based on photometric distances}

{\it How to discover fractal structure?}

\vskip 0.5 cm

{\Large \bf 3. Statistical methods to detect
fractal structures}

\vskip 0.2 cm

{\large 3.1.Definitions for correlation functions}

{\it Complete and reduced correlation functions}

{\it Mass variance in spheres and characteristic scales}

{\large 3.2.The method of $\xi$ correlation function}

{\it Peebles' $\xi$ correlation function}

{\it $\xi$-function estimators}

{\it The normalization condition for $\xi$ estimators }

{\it A systematic distortion of the true power-law }

{\it Redshift space and peculiar velocity field}

{\large 3.3.The method of conditional density $\Gamma$}

{\it Definitions}

{\it $\Gamma$-function estimator }

{\it $\Gamma (s)$ and peculiar velocities}

{\it $\Gamma$-function for intersections}

{\large 3.4.Comparision of $\xi$ and $\Gamma$ analyses}

{\it The relation between $\Gamma $ and $\xi $ }

{\it Power-law correlation
and fractal density field }

{\it Dependence of $r_0$
on the sample parameters}

{\it Geometry and characteristic scales of a survey}

{\large 3.5.Nearest neighbours distribution}

{\it Poisson distribution }

{\it Fractal distribution }

{\large 3.6.Two-point conditional column density}

{\it Definitions}

{\it Estimation}

{\large 3.7.Fourier analysis of the structures}

{\it Ordinary density field}

{\it Fractal density field}

{\it Geometry of surveys}

{\large 3.8.Multifractals and luminosity function}

{ \it Spectrum of fractal dimensions}

{ \it Schechter's luminosity function}

{ \it Space-luminosity correlation in multifractal
model}

\vskip 0.5 cm

{\Large \bf 4. The epoch of galaxy angular
position catalogues }

\vskip 0.2 cm

{\large 4.1.The birth of the debate}

{\it Einstein--Selety
correspondence }

{\it Retrospective view on the posed
questions }

{\large 4.2.Early arguments for galaxy clustering }

{\it Observations disclose clusters
of galaxies }

{\it Santa Barbara 1961
conference}

{\it The cosmological de Vaucouleurs
law}

{\large 4.3.Early arguments for  homogeneity }

{\it Hubble's counts of bright
galaxies }

{\it Hubble's deep galaxy counts }

{\it Variable dust extinction }

{\it The classical ``linearity'' argument}

{\it Isotropy in a homogeneous
universe }

{\large 4.4.Results from angular catalogues}

{\it Main catalogues of
galaxies and clusters}

{\it The angular correlation
function analysis}

{\it Angular and spatial correlation
 functions }

{\it Hierarchical $D = 1.2$ models}

{\it Tallinn 1977 conference }

{\it First evidence for $D = 2$
galaxy distribution }

{\it Why did angular catalogues lose information
about the $D=2$ structures?  }

%\vskip 0.5 cm
%{\Large \bf PART II}
\vskip 0.5 cm

{\Large \bf 5. Debate on fractality:
the epoch of spatial maps }

\vskip 0.2 cm

{\large 5.1.The fractal breakthrough in the 1980s.}

{\it Davis \& Peebles $\xi$ correlation function analysis
of the CfA sample of galaxies }

{\it The puzzling behaviour of the $\xi$-function }

{\it Pietronero's solution
the mystery of $r_0$ }

{\it Balatonfurd 1987 conference }

{\it Multifractal controversy
for $D_2 = 1.2$ estimation }

{\it The size of a fractal cell in the Universe}

{\large 5.2.Further steps in the debate}

{\it  Princeton ``Dialogues'96'': 
Davis's evidence for
homogeneity }

{\it  Princeton ``Dialogues'96'':
Pietronero's arguments for
fractality  }

{\it The problem of sky projection of fractals}

{\it Modern research topics on fractality }

{\large 5.3.Recent results from the $\xi$ and $\Gamma$ 
functions analyses}

{\it The redshift space $\xi$-  and $\Gamma$- functions }

{\it The problem of peculiar
velocity field }

{\it Dependence
of $r_0$ on sample parameters }

{\it Power spectrum and intersection of fractals}

{\large 5.4.Other results of the fractal 
approach}

{\it The two-point conditional
column density }

{\it Number counts of all-sky
bright LEDA galaxies }

{\it Radial counts of KLUN galaxy sample}

\vskip 0.5 cm

{\Large \bf 6. Why fractality is important for
cosmology}

\vskip 0.2 cm

{\large 6.1.Basic elements of cosmological models}

{\it Three major empirical laws in 
cosmology}

{\it Theoretical basis of modern cosmology}

{\it The standard cosmological model}

{\it Fractal sources for gravity field}

{\large 6.2.The origin and evolution of large scale
fractals: challenges for theoretical models}

{\it Hubble law within fractal galaxy distribution}

{\it Problem of the origin of the fractal structure}

{\large 6.3.The Cosmological Principle}

{\it Einstein's cosmological principle}

{\it Derivation of homogeneity from isotropy}

{\it Mandelbrot's cosmological principle}

{\it Towards Einstein-Mandelbrot concordance}

\vskip 0.5 cm

{\Large \bf 7. Concluding remarks}

\vskip 0.5 cm

{\Large \bf References}

%\vskip 0.5 cm
%{\Large \bf Appendix}

\vskip 1.5 cm

\section {Introduction}

In this review we discuss the historical
roots, methodological problems and  modern results
that branch  of observational
cosmology, which has given
rise to the
``Great Debate on Fractality''. This sharp and sometimes
dramatic debate
has been in the limelight almost the whole
20th century and still is. It has involved such persons as Einstein,
Hubble, Sandage, Peebles,
Charlier, Selety, Lundmark, de Vaucouleurs,
Mandelbrot, Pietronero and many
others who  appear in the story.

The debate on the fractality of the large scale galaxy
distribution has been going on around
two new fundamental empirical cosmic numbers, --- the fractal
dimension $D$ and the
bordering scale where fractality transforms
into homogeneity $R_{\mathrm{hom}}$.
The discussion of galaxy clustering started from
scales $1 \div 10$ Mpc, then
observations of the large scale structure (LSS) have
shifted to the scales of $10 \div 100$ Mpc, and now
we are entering gigantic scales of $100 \div 1000$ Mpc.

There are several recent papers and books devoted to the LSS
analysis, which are close
to the subject of our review
(Sylos Labini et al. 1998;
 Martinez \& Saar 2002;  Gabrielli et al. 2004;
Jones et al. 2004).
However, what has been lacking is
a picture of historical steps in the studies of
the large scale structure with a focus on
methodological problems and challenges
of current research.
Our review addresses these questions.

A fundamental task of practical cosmology is to study
how matter is distributed in space and how it has evolved in cosmic time.
The discovery of
the strongly inhomogeneous spatial distribution of
galaxies,
at scales from galaxies to superclusters, i.e. over
four
orders of magnitude in scale, was of profound cosmological
significance. Only faintly anticipated
from photographic
surveys, the surprisingly rich texture of galaxies
became visible
thanks to a large
progress in measuring distances by redshifts for
thousands of galaxies.
The observed clustering is not just random clumping, but
obeys a universal
law.
The two major theoretical tools, the
correlation function and the
conditional density analyses, both have
revealed a power law dependence of the galaxy number density  on the
length scale, on scales $0.1 \div 10$ Mpc.
This signature of the scale invariance or
fractality of the galaxy distribution has opened
a new application of fractal geometry, widely used in
modern statistical physics.

In this review we present
the growing evidence for fractality of the large scale
galaxy clustering.
In sect.2 we give a brief introduction to the concept of fractal
density field.
In sect.3 basic statistical methods for 
analysis of galaxy distribution are described.
Sect.4 is devoted to the main arguments for and against galaxy clustering,
debated during the epoch of angular galaxy catalogues.
Such 2-d data do not contain information on distance, and what is
even more important, they are not sensitive to a distribution with
the fractal dimension $D \geq 2$.
In sect.5 we summarize results obtained in the modern 3-d epoch from extensive
redshift surveys which  have revealed the ``hidden'' fractal dimension
of about two, and have detected structures at scales up to 500 Mpc.
One has seen
how the value of $D = 1.2$ derived from the angular catalogues has been
replaced by $D =2.2 \pm  0.2$ in 3-d maps and the maximum observed scale of
fractality $R_{hom}$ has increased
from 10 Mpc to scales approaching 100 Mpc.
In sect.6 we consider cosmological importance of the observed large scale
fractality.

\section { \Large \bf The fractal view of the large scale
structure of the Universe}

The last two decades have seen the first extensive
surveys of galaxy
redshifts, which have permitted us to move from the study of
the angular distribution of galaxies on the celestial
sphere to
the analysis of their three-dimensional distribution in
space.

Already the first redshift surveys revealed a rich
variety of structures in
the galaxy universe. These have been characterized by
astronomers using
terms such as binaries, triplets, groups, rich, regular,
and irregular
clusters, walls, superclusters, voids, filaments, cells,
soap bubbles, sponges,
great attractors, clumps, concentrations, associations
\ldots Of course
each form of structure deserves separate studies, but
they can be also viewed as natural appearances of one
global
master entity which is called a fractal. 
By using this self-similar structure
one can describe an inhomogeneous galaxy
distribution by means of 
one basic parameter, the fractal dimension $D$,
which determines the global mass--radius behaviour
of the Universe.

However, it took a lot of time to recognize in the clumpy
galaxy distribution on the celestial vault the signature
of fractal structures. It is interesting and useful to see
how the idea of self-similarity emerged in ancient times and
developed into first protofractal models.

\subsection {The idea of a self-similar universe}

\subsubsection { Protofractals }
The ancestors of fractals appeared in old cosmological
thinking.
In the presocratic Greece the philosopher Anaxagoras of
Clazomenae
(about 500 - 428 B.C.), who spent a large part of his
life teaching
at Athens,
put forward his theory of seeds. Only fragments of his
text have been
preserved, but these give the impression that the theory
was based on
ideas reminiscent of self-similarity (e.g. Gruji\'{c} 2001,
2002).
Contrary to atomists, he regarded the matter divisible
without a limit, and
his famous words are: ``In all things there is a portion
of everything.''

What has been called by Mandelbrot a protofractal,
appeared as
simple hierarchies in the 18th century works by
Swedenborg, Kant and Lambert.
Emanuel Swedenborg (1688--1772)
was a Swedish scientist and visionary
who  was appointed, at the
 age of 28, assessor extraordinar in the Swedish College
of Mines.
A very productive thinker and writer,
he discussed practically all fields of science of his
time.
In 1734, in his \emph{Principia},
 Swedenborg put forward the remarkable view of self-
similarity and cosmic
 hierarchy,
which  was related to his general opinion that
everything in the
 world is constructed according to a common plan.

The next steps were made by Immanuel Kant (1724-1804)
and Johann Lambert
(1728-1777) with their hierarchic systems.
For example, Kant wrote in his book available in
translation
as Kant (1755):
{\small \it We see the first
 members of a progressive relationship of worlds and
systems; and
 the first part of this infinite progression enables us
already
 to recognize what must be conjectured of the whole.
There is no
 end but an abyss $\cdots$ without bound.}

But there was an essential difference between these
world
 models.  Kant, and perhaps Swedenborg, imagined that
the hierarchy
 continues without end towards larger levels of
 celestial systems: it was an infinite hierarchy.
Lambert
 thought that after a large (he cites the figure 1000 as
an
 example), but finite number of steps, the hierarchy
ends.  He
 thought that the stellar systems are kept together by
the
 gravity of a dark mass: {\small \it ... in the end you
arrive at the middle
 point of the whole world structure and there I find my
ultimate
 mass which governs the whole creation}.

 John Herschel (1792--1871) was, among
 many others, intrigued by the dark sky or Olbers's
Paradox
 (Harrison 1987).  And he outlined a solution in a
 private letter:
{\small \it $\ldots$ it is easy to imagine a
constitution of a universe
 literally infinite which would allow of any amount of
such
 directions of penetration as not to encounter a star.
Granting
 that it consists of systems subdivided according to the
law that
 every higher order of bodies in it should be immensely
more
 distant from the centre than those of the next inferior
order -
 this would happen.}

Herschel had in mind some kind of a hierarchical system.
In another text he gives as examples the satellites of
the planets
of the solar system and the large distances between
stars and asserts that
{\small \emph { ``the principle of subordinate grouping''} }
assumes
{\small \emph {
``the character and
importance of a cosmical law''.}}

\subsubsection { Fournier d'Albe and Carl Charlier }
\begin {figure}[th]
\centerline {\psfig {file=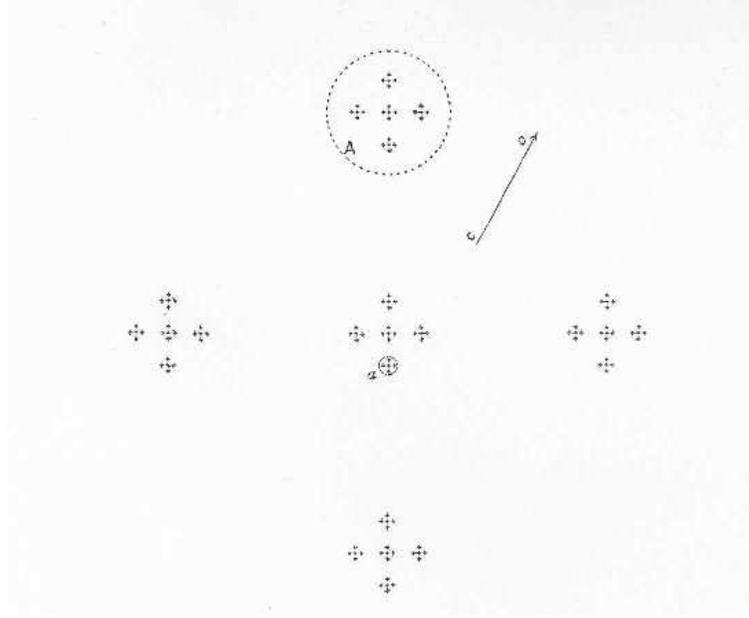,width=10cm}}
%\resizebox{\hsize}{!}{\includegraphics{fig1.ps}}
\caption {\footnotesize
%\centerline {\small Fig.1. 
Fournier's (1907) regular hierarchic world model. 
The same elements are repeated on different levels in a 
self--similar way.
 The average density $\rho (R)$ within the scale $R$
decreases with increasing scale $R$, hence $\rho (R) \rightarrow  0$ for
$R \rightarrow \infty $.}
\label {fournier}
\end {figure}

The idea
of a hierarchic structure of the universe
 was blown into life again by Edmund Fournier d'Albe
(1868--1933).
In 1907 he published the
 remarkable book \emph{Two new worlds} where one finds a
 mathematical description of a possible hierarchical
distribution
 of stars.  In Fournier's world (Fig.\ref{fournier}) the stars are
distributed in a hierarchy of spherical clusters in
 an infinite space, so that the mass inside each sphere
 increases directly proportionally to its radius:

\be
M(R) \propto R\, .
\label{fourniermass}
\ee

Of course, this makes it dramatically
different from the mass--radius
behaviour in a homogeneous universe where $M(R) \propto R^3$. This
was Fournier's idea of how to avoid cosmological paradoxes
appearing in Newton's universe.

The cosmic
 hierarchy was further studied by the Swedish astronomer
 Carl Charlier (1908)
 in his article \emph {How an infinite world
 may be built up}.  After
 enthusiastic inspection of Fournier's book, he
developed more
 general models of stellar distributions which also
solve the
 Olbers's paradox and the riddle of infinite
 gravitational potential.
 
Charlier found a criterion which the hierarchy must
fulfil so
 that it will solve the mentioned paradoxes.  The
decisive factor
 is how fast the density decreases from one level (i) to
the next
 one (i+1), and this depends on the ratio of the sizes
of the
 successive elements and on the number $N_{i+1}$ of the
lower elements
 forming the upper element.  Denoting the sizes (radia)
with
 $R_i$ and $R_{i+1}$, Charlier's first criterion may be
written as
\be
R_{i+1} / R_i \geq N_{i+1}
 \label{charlier1}
\ee
or the size of upper level element divided by size of
lower level
 element is larger or equal to the number of lower
elements
 forming the upper elements.  For example, Fournier's
 illustration fulfils Charlier's first criterion:  there $N_i
= N_{i+1} = 5$, and the size ratio is always about 7.

In Charlier (1922), after a note by Selety (1922), a
second criterion
was derived:
\be
R_{i+1} / R_i \geq \sqrt{N_{i+1}}
 \label{charlier2}
\ee
In terms of a continuous mass--radius behaviour for identical
particles with mass $m$ the first
criterion corresponds
to $M(R) = m N(R) \propto R^1$. The second criterion, $M(R)
\propto R^2$, and it
is sufficient to cope with Olbers' paradox and the
infinite gravity force.
The original Fournier's (and the first Charlier's)
criterion is a stronger condition, and allows also
a finite gravitational potential and finite stellar
velocities.

\subsection {Genuine fractal structures}

\begin {figure}[th]
\centerline {\psfig {file=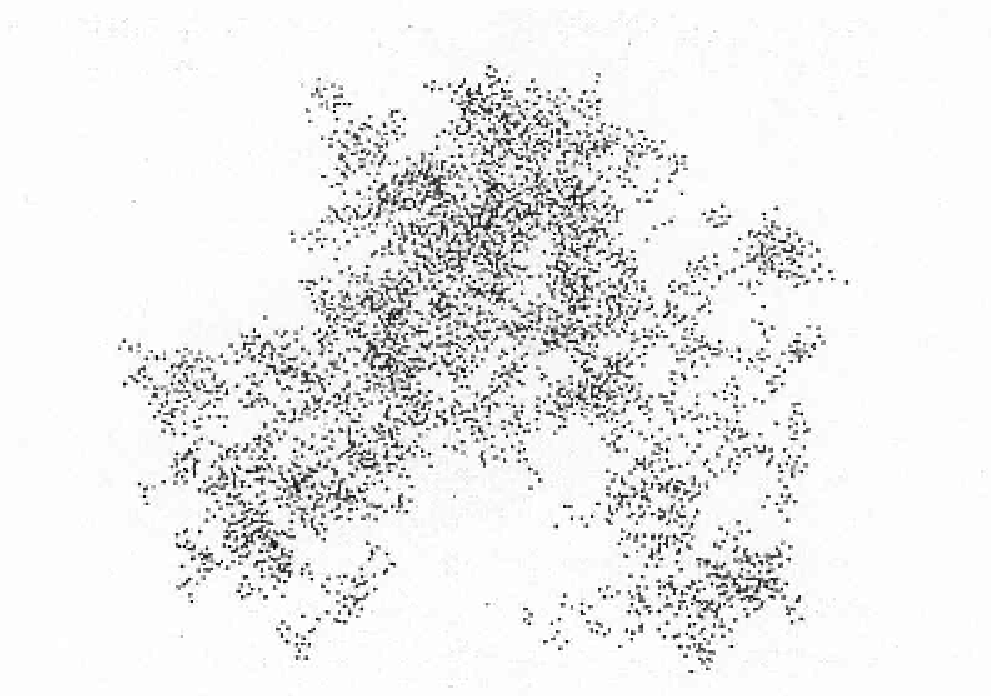,width=10cm}}
\vspace {1cm}
\caption{\footnotesize An example of a
genuine stochastic
fractal structure
with the fractal dimension $D= 1.26$
in two-dimensional space (B. Mandelbrot 1989).}
\label {mandelbrot}
\end {figure}

Even though Fournier's and Charlier's hierarchies were
overly simple
for the real world, as we now know, they contained the
seeds of
the modern concept of genuine fractal.
Regular hierarchical models (like in Fig.\ref{fournier})
have a number
of preferred scales, corresponding to the sizes of
clusters at the
level ``$i$''. A great advantage of stochastic fractals
is that they can be used to model scale-invariant galaxy clustering without
preferred scales (Fig.\ref{mandelbrot}).

Though fractal geometry emerged just a few decades ago,
some
elements of it can be found already in the works
 of  Poincar\'e  and Hausdorff  about a
century ago.
Mandelbrot (1975) introduced the name "{\it Fractals}"
and gave the following definition:
{\it \small A fractal is a set  for which the Hausdorff dimension
strictly exceeds the topological dimension.}

Mandelbrot realized that fractal geometry is a
powerful tool to characterize intrinsically
irregular system.
Nature is full of strongly irregular structures;
trees, clouds, mountains and lightnings are
familiar objects,
%but are very different from the
% structures of euclidean geometry.
which have in common the  property
that if one magnifies a small portion
of them, a complexity comparable
to that of the entire structure is revealed. This is geometric
\emph {self-similarity}.

Fractals are simple but subtle, as Luciano Pietronero
likes to say,
and becoming familiar with them requires
some training of intuition.
Here we briefly describe the essential properties of fractals,
which are used in studies of the galaxy distribution.
We recommend Mandelbrot
(1982) for an original presentation by the father of fractals
and Falconer (1990) for a special mathematical treatment.
Here we emphasize
observational consequences of
self-similarity so that if this important property is
actually present in galaxy data, one will
 be able to detect it correctly.

\subsubsection {Self-similarity and power law. }

The difference between a self-similar distribution and a distribution
with an intrinsic characteristic scale was clearly discussed by
Coleman \& Pietronero (1992).
%From Fig.\ref {fournier} the geometrical self-similarity
%is evident in the construction.
From a mathematical point of view self-similarity
implies that the rescaling of the length $r$
by a factor $\:b$
\be
\label {l7}
r \rightarrow r ' = br
\ee
leaves the considered property, presented by an arbitrary function
$f(r)$, unchanged
apart from a renormalization that depends on
$\:b$ but not on the variable $\:r$.
 This leads to the functional relation
\be
\label {l8}
f ( r') = f (b\cdot r) =
A(b)\cdot f (r)
\ee
This is satisfied by a power law
with any exponent.
In fact for
\be
\label {l9}
f (r) = f_{0} r ^{\alpha}
\ee
we have
\be
\label {l10}
f (r') = f_{0}(br)^{\alpha}
= (b)^{\alpha}f (r).
\ee
Here the exponent $\alpha$ defines the behaviour of the
function everywhere.
There is no preferred scale.
It is true that the condition on the amplitude
$\:f(r_{0})=1$
implies a certain length $r_0$
\be
\label {l12}
r_{0}=f_{0}^{-1/\alpha},
\ee
and one might be tempted to call $r_0$ a characteristic
length, but this is quite misleading for self-similar
structures!
The power law refers to a fractal
structure that was at the beginning constructed as self-similar
and therefore cannot posses a characteristic
length.

In eq.\ref{l12} the value of $\:r_{0}$
is just related to the amplitude of the
power law, and the amplitude has nothing to
do with the scaling property.
Instead of the value
1 for the amplitude,
one could have used any other number in the relation
$\:f(r_{0})=1$
to obtain other lengths. This is a subtle point of self-similarity;
there is no reference value (like
the average density) with respect
to which one can define what is big or small.

This behaviour of the power-law is in contrast with the case of
the exponential decay function where there is an
intrinsic
characteristic parameter $r_0$
\be
\label {l11}
g ( r ) = g _{0} e ^{-r/r_{0}}
\ee
which determines a preferred length scale for the
behaviour of the function.
In the power-law the constant, dimensionless exponent $\alpha$
is not related to length scales at all.

\subsubsection { Fractal dimension }
The basic characteristics of a fractal
structure is its dimension.
The fractal dimension is a measure of the ``strength of
singularity''
around the structure points.
If there is a zero-level of structure elements, as in physical
fractals where there is no mathematical singularity, the rate of
growth of density with a decreasing spatial scale still defines
the fractal dimension.

Let us consider the simplest example of a regular
fractal structure
to illustrate how the fractal dimension can be
determined.
 Starting from a
 point occupied by an object we count how
many objects are present within a sphere
of radius $r$ in
 order to establish a number-radius
relation from which one can define the fractal
dimension $D$.
Suppose that in the structure of Fig.\ref {fournier}  we
 can find $N_0$ objects with mass $m_0$ in a volume of
 size $\:r_{0}$. If we consider a larger volume
of size $\:r_{1} = k_r \cdot r_{0}$ we will find
$\:N_{1} =k_N \cdot N_{0}$ objects.
In a self-similar structure the parameters $\:k_r$
 and $\:k_N$ will be the same also for
other changes of scale. So, in general in a
structure of size $\:r_n = k^{n}_r \cdot r_{0}$
we will have $\:N_{n} =k^{n}_N \cdot N_{0}$
objects. We can then write the relation
 between the number $\:N$ (or mass $M=m_0N$) 
and length $\:r$  in the form
\be
\label{l2}
N(r) = B\cdot r^{D}\,,
\,\,\,\, \mathrm{or} \,\,\,\, M(r)=m_0B\cdot r^D
\ee
where the fractal dimension is the exponent $D$ of the power
law, i.e.
\be
\label{dimknkr}
D = \frac{\log{k_N}}{\log{k_r}}
\ee
depends on the rescaling factors $\:k_r$
and $\:k_N$. The prefactor $\:B$
is related to the zero-level parameters $\:
N_{0}$ and $\:r_{0}$,
\be
\label{l4}
B =\frac{N_{0}}{r_{0}^{D}} \, .
\ee
We note that Eq.\ref{l2} corresponds
to a smooth continuous presentation of a strongly
fluctuating function as is evident
from Fig.\ref {fournier}.
The smooth average power-law for a fractal
structure is always accompanied by large
 fluctuations and clustering at all scales.

Simple algorithms for constructing stochastic fractal structures
with given fractal dimension one may find e.g. in Gabrielli et al.
(2004). With these algorithms one may build artificial
galaxy catalogues useful for testing different methods of structure
analysis.

\subsection {Concepts of density field}

The definition of density field is the basis of the large scale
structure analysis.
There is an essential conceptual difference between ordinary
and fractal density fields. The former kind of model is
usually utilized for the description of gas or fluid dynamics
having short range of correlations,
while the latter one emerges in physical systems with strong
long range scale-invariant fluctuations.

\subsubsection {Ordinary fluid-like density fields}

The concept of the density of a continuous medium (approximating
fluid or gas), as it is
normally used in hydrodynamics, contains the assumption that
there exists
a value of density independently of the size of the
volume element
$dV$. Then one may define the density $\rho(\vec{x})$
at a point $\vec{x}$ and regard it as a usual continuous
function of
position in space:
\be
\rho_{_{fluid}} =
\rho(\vec{x}) = \lim_{V \rightarrow 0} \frac{M(\vec{x}, V)}{V}
\label{limMV}
\ee
where $M$ is the mass of the fluid inside the volume
$V$ around the point $\vec{x}$. For ordinary continuous
media the limit exists and does not depend on the volume $V$,
because at sufficiently small  scales homogeneity is reached,
 i.e. $M = \rho V$.

 When one studies fluctuations of an ordinary fluid, one
can regard
$\varrho(\vec{x})$ as one realization of a stochastic
process for
which the usual statistical moments are defined -- average,
dispersion etc.

\begin {figure}[th]
\centerline {\psfig {file=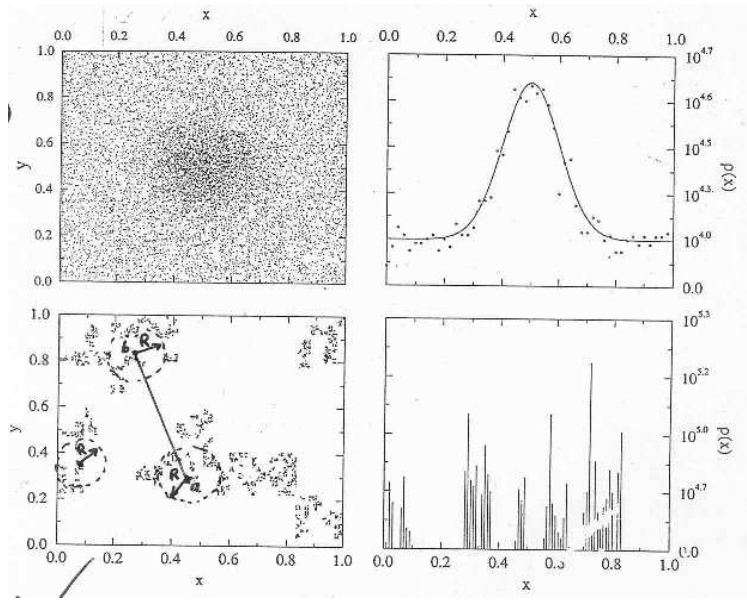,width=10cm}}
%\resizebox{\hsize}{!}{\includegraphics{fig4.ps}}
\caption{\footnotesize Top: an ordinary
(fluid-like) discrete density field: a fluctuation
on a Poisson background; 
bottom: a stochastic fractal density field
(from Sylos Labini et al. 1998).}
\label {example}
\end {figure}

Models for ordinary density fields go
under the common name \emph{fluid-like correlated distributions}.
These have a uniform average background with superimposed fluctuations
that are correlated. An example of such a fluctuation may be seen
in Fig.\ref{example}, top. Main properties
of such distributions, in comparison with fractal distributions,
are discussed by Gabrielli et al. (2004).

An ordinary \emph{stationary} stochastic density field $\rho
(\vec{x})$ may be represented as a sum of density fluctuations
$\delta \rho (\vec{x})$ and the mean constant density
$\rho_0 = \langle \rho (\vec{x}) \rangle$, so that
\begin {equation}
\rho(\vec{x}) = \rho_0 + \delta \rho \, ,
\end {equation}
or in terms of the dimensionless relative density
fluctuation:
\begin {equation}
\label {relfluct}
\delta(\vec{x}) = \frac{\rho - \rho_0}{\rho_0} =
\frac{\delta \rho (\vec{x})}
{\rho_0}\, .
\end {equation}

Note that the relative fluctuation
$\delta (x)$
in Eq.\ref{relfluct}
has positive and negative values in the range $-1 \leq
\delta (x) < \infty $,
while the density field is always positive $\rho (x) > 0$
for positive masses of particles.

One usually considers $\delta \rho (\vec{x})$ as a realization
of a Gaussian
stochastic process, for which the phases of fluctuations
are
uncorrelated. Here a fundamental role is played by the
average density
$\bar{\rho} = $ const $> 0$ which should exist and be
well defined
and positive for each
realization of the random process $\rho (x)$. ``Well
defined'' means that
the average density does not depend on the size
and location of a {\it fair} test volume.

\subsubsection {Ordinary stochastic discrete processes}

Ordinary  density field
may be also presented by a discrete stochastic process,
called a stochastic point process or a point-particle
distribution.
Here discreteness introduces some new aspects
related to point-like singularity of particles.

An important example of a \emph{homogeneous}
stochastic discrete density field is the Poisson process,
giving rise to a number density of particles $n(\vec{x})
=\Sigma_{i=1}^N \delta (\vec{x} - \vec{x}_i)$.
According to the Poisson law
the probability $P$ to
find $N$ particles in a volume $V$ is
\be
\label{poisson}
P(N,V) = \frac{\langle N \rangle ^N \exp (-\langle N
\rangle)}{N\mathrm{!}}\, ,
\ee
where $\langle N \rangle = n_0 V$ is the average number
of particles
in the volume $V(r)$. The only parameter of the Poisson
distribution is
the constant number density $n_0 =\langle n(\vec{x}) \rangle $,
or intensity of the Poisson process.
It determines a characteristic scale $\lambda_0$ for this process
\be
\label{lambda0}
\lambda_0 \approx n_0^{-1/3} \approx R_{sep}\,\,,
\ee
which is about the average distance between the particles
also denoted as $R_{sep}$.

The normalized number (mass) variance $\sigma^2(r)$ in a
sphere with radius $r$ is defined as
\be
\label{sigmapoisson}
\sigma^2(r) = \frac{\langle N(r)^2 \rangle - \langle N(r) \rangle^2}
{\langle N(r) \rangle^2}\,\,,
\ee
which is an important quantity to characterize a stochastic process
at both small and large spatial scales $r$.

{\it Discreteness noise. }
Finite distances between point-particles of a discrete
stochastic process make an essential {\it noise of discreteness} or
{\it shot noise} appear when considered scales are less that
average distance between particles $r< R_{\bar{n}}$.
This noise increases with decreasing $r$ as
\be \label{sigma-r}
\sigma(r)\approx \frac{1}{\sqrt{N}} \approx 
\left(\frac{r}{\lambda_0}\right)^{-3/2}
\ee
and becomes infinitely large at very small scales.

{\it Large spatial scales. }
On large spatial scales $r$ the distribution becomes homogeneous
with its normalized dispersion (eq.\ref{sigma-r}) approaching zero
as $1/\sqrt{N}$ when the number of particles increases indefinitely.
The scale $R_{\mathrm{hom}} \approx \lambda_0$ is the homogeneity scale
of the Poisson process, it is
defined from the condition that the normalized number variance
$\sigma^2 (R_{\mathrm{hom}}) = 1$.  

The Poisson distribution is a homogeneous stochastic discrete
density field \emph{without correlations}, so that its
correlation function is zero (neglecting the so-called `` diagonal''
term corresponding to the point singularity at $r=0$).
This means that the positions of points are independent of each other,
so that there are no genuine structures. Though the eye may see
apparent structures, these are just random fluctuations in
an outcome (realization) of the process.

Gabrielli et al. (2004) considered another important example,
the \emph{superhomogeneous discrete process} like particles in
a lattice with small correlated shifts of the particles around the
regular lattice knots.
Such a process is used for generating initial conditions for cosmological
N-body simulations.

\subsubsection {Fractal density fields}

In the large scale structure analysis one applies a concept
of discrete fractal point processes for which one may
consider the density field in the form of a spatial distribution
of $N$ particles in positions $\vec{x}_a$
with masses $m_a$ in a volume $V$, then
\begin{equation}\label{rho-r}
  \rho(\vec{x}) = \sum_{a=1}^N m_a \delta (\vec{x}-\vec{x}_a)\,.
\end{equation}
In the case of identical particles with $m_a = m$ one may simply use
the concept of the number density field
\begin{equation}\label{n-r}
  n(\vec{x}) = \sum_{a=1}^N  \delta (\vec{x}-\vec{x}_a)\,.
\end{equation}

In order to describe the continuous hierarchy of clustering, which
is a new characteristic of a fractal stochastic process,
we first take into use
a new independent variable, the radius $r$ of the spherical volume $V(r)$
in which the particles are counted. Then the {\it fractal mass
density} may be defined as
a function of two variables, the position $\vec{x}_a$ of a particle
and the radius $r$ of the volume.
\be
\rho_{_{\mathrm{fractal}}} = \rho (\vec{x}_a, r) = \frac{M(\vec{x}_a,V)}{V} \, ,
\label{rho(x,r)}
\ee
where $M(\vec{x}_a,V)$ is the mass inside the volume $V(r)$ located around the
particle position $\vec{x}_a$.

For a mathematical infinite ideal fractal the number of
points of the structure in a finite volume is infinite, so the mass
is also infinite.
In physics this problem may be avoided
by some natural lower limit in sizes of elements, making the zero-level
of the hierarchy. In this case one may speak about basic structure
elements --- point mass particles. Then
the number of particles and the mass within the volume $V(r)$ is finite,
i.e. in this case Eq.\ref{rho(x,r)} defines a physically measurable density.
Though the quantity $\rho(\vec{x}_a,r)$ is highly fluctuating from
one particle position to another, 
as we shall see below
it is possible
to consider statistical average which is more stable characteristics
of a fractal structure.

In Fig.\ref{example} an ordinary (fluid-like)
density fluctuation on a
Poisson background is compared
with a stochastic fractal density fluctuations.
In case of fractal structures the ordinary concept of
the mass density of a continuous medium is not applicable.
This is because the mass density can be defined only if
both the position $\vec{x}$ and the volume $V$ are considered.
In every volume $V$
containing a part of the structure there is a hierarchy of clusters
and the value of
the mass density strongly depends on the size of the volume.
This is totally different from the usual calculus.
Now there is no limit for the mass-volume ratio (Eq.\ref{limMV}) and
the density increases
indefinitely when the volume tends towards zero:
\be
\{\rho_{_{fractal}}\,\, \rightarrow \infty \,\,, \,\,\,\,\,
\mathrm{for} \,\,\,\,\,
V \rightarrow 0 \} \,\,\,\,.
\label{limzero}
\ee
If the basic zero level elements exist then it determines
the maximum value of the fractal density for the structure.

\subsection {Exclusive properties of fractal density fields}

\subsubsection  { Power-law density-radius relation}

A characteristic feature of the fractal density field
is a power-law behaviour of the density with increasing
of the radius of the spherical volume $V(r)$ centered at a structure
point.
We can illustrate this property for the case of regular
fractal structure (Fig.1), where the number of  subelements within
an element of the higher level is given by Eq.\ref{l2}.
So in continuous representation,
using Eq.\ref{rho(x,r)}, one can define the
fractal number density, which related to the elements
of radius $r$,  as
\be
\label{volfracdens}
n_{V(r)} =\frac{ N(r)}{V(r)} = \frac{3}{4\pi } B r^{-(3-D)}
\ee
where $D$ is the fractal dimension of the structure.

Hence, in order to calculate the fractal density one can start
from a basic element which belongs to the structure and count the
number of objects within the sphere $V(r)$ and
divide it by the volume of the sphere $V(r)$ centered at
the point. Apart from  some fluctuation, depending on the actual
position of the point within the structure the power-law
Eq.\ref{volfracdens} will be obtained.

Note that the result is unexpected for our
usual intuition --- the density decreases from each
point of the fractal structure.
It seems like each point were the centre of the
structure from which the density decreases outwards
following the
law in Eq.\ref{volfracdens}.
This property is essential for cosmological implications of the
fractal structures (sec.6.5).

\subsubsection { Massive, zero-density universes }

An important consequence from Eq.\ref{volfracdens} is
that in infinite space 
the fractal density field differs from an
ordinary fluid-like density field
at the limit of large volumes $V$ where
\be
\{\rho_{_{fractal}}\,\, \rightarrow 0 \,\, , \,\,\,\,\,
\mathrm{for} \,\,\,\,\,
V \rightarrow \infty \}
\label{liminf}
\ee
This property is due to a growing dilution of the hierarchy
with increasing scales, so that
a fractal structure is asymptotically
dominated by voids.

Hence an infinite fractal universe
can contain
an infinitely large number of objects (hence an infinite mass)
simultaneously with the zero density of the whole Universe.
This unusual property of a hierarchical structure
was exploited in old cosmological models to avoid
gravitational
and photometric paradoxes of the Newtonian infinite
universe. This follows from the relations
$\varphi \propto M/R $ and $F \propto M/R^2$.

\subsubsection {The role of lower and upper cutoffs}

In the realm of physics real structures usually have a lower scale
$R_{\mathrm{min}}$
and an upper scale $R_{\mathrm{max}}$ between which the physical system follows
fractal self-similar behaviour. These scales are called
{\it lower } and {\it upper cutoffs}.

In studies of large-scale galaxy distribution
the lower cutoff $R_{\mathrm{min}}$ is assumed
to be equal to the size of a galaxy, while galaxies play a role of
point-like particles. For different cosmological problems
there could be different choices of the lower cutoff: dark matter
clumps of $10^{6\div 8}M_{\odot}$, stars, comet-size objects,
atoms, elementary particles. So the lower cutoff is usually
a well-defined quantity.

The upper cutoff presents a much more complicated problem
in studies of the galaxy distribution.
In principle, one should apply such methods of the large
scale structure analysis which allow one to determine 
directly from a galaxy survey
the scale $R_{\mathrm{max}} = R_{\mathrm{hom}}$ where the galaxy distribution
becomes homogeneous.
However, such methods need a large survey volume whose size
should correspond to several times the scale.

Up to now the largest galaxy redshift surveys cover a small part
of the sky which hinders a firm estimation of the size $R_{\mathrm{hom}}$.
Is there an upper cutoff for the large-scale galaxy distribution
and what is its value? These are the primary questions
around which the most acute discussion is going on.

\subsubsection { Lacunarity }
Two fractal structures with the same fractal dimension may
look very different. In particular, when one makes fractal models
of the galaxy distribution, it is quite essential how
large a relative
volume is occupied by voids, on a given scale.

This property was termed
\emph{lacunarity} by Mandelbrot (1982), from the word ``lacuna'' meaning hole
or gap in Latin.
Quantitatively lacunarity may be characterized by the constant of
proportionality $F$ in the relation

\be
N_v(\lambda > \Lambda) = F \Lambda^{-D}
\label{lacunar1}
\ee
where $N_v$ is the number of voids 
with size $ \lambda > \Lambda $ within a fixed volume inside the structure. 

Another definition was introduced by Blumenfeld \&
Mandelbrot (1997).
They used the variation of the prefactor $B(\vec{x}_a)$ in the density law
computed for each structure point $\vec{x}_a$ within a ball with
a fixed radius $R$,
$N(r, \vec{x}_a, R) = B(\vec{x}_a, R)\,r^D$ (Eq.\ref{l2}). Then lacunarity is
defined as a normalized dispersion of the distribution of the prefactor
\be
\label{lacunar2}
\Phi = \frac{\langle (B - \bar{B})^2 \rangle _{\vec{x}_a}}{\bar{B}^2}
\ee
Concrete examples of structures inside fixed sample volumes, having different
lacunarities according to this definition may be found in Martinez \& Saar
(2002).

We note that the high lacunarity of the Rayleigh--
L\'{e}vy flight fractal
was the reason why it was rejected as a model for the
real distribution of galaxies (Peebles 1980). However, later
Mandelbrot (1998)
demonstrated that fractal
structures with
a small lacunarity resemble more closely the
arrangement of galaxies (see Fig.\ref{mandelbrot}).

\subsubsection {Projection and intersection}

The properties of orthogonal projections and intersections of a fractal
structure play an important role in the analysis of galaxy samples
with different geometries, both from angular 2-d and spatial 3-d catalogues.

{\it Orthogonal projection. }
Let an object (structure) with a fractal dimension $D$, embedded in
an Euclidean space of dimension $d = 3$, be orthogonally projected
onto an Euclidean plane with $d' = 2$. Then according to a general
theorem of fractal projections (see Mandelbrot 1982; Falconer 1990),
the projection as a fractal object receives the fractal dimension
$D_{pr}$ so that

\be
D_{pr} = D \,\,\,\, \mathrm{if} \,\,\, D < 2
\label{Dproj1}
\ee
and 
\be
D_{pr} = 2 \,\,\,\, \mathrm{if} \,\,\, D \geq 2 \, .
\label{Dproj2}
\ee
This means that in 3-d space a cloud having the fractal dimension
$D \approx 2.5$ satisfies eq.\ref{Dproj2} and hence gives rise to a homogeneous
shadow ($D_{pr} = 2$) on the ground. Consequently, the orthogonal
projection hides from view fractal structures with $D > 2$. This has an
important implication for the apparent distribution of galaxies on the sky
(sec. 4.4).

{\it Intersection of a fractal. }
If an object with a fractal dimension $D$, embedded in a $d =3$ Euclidean
space, intersects an object with the dimension $D'$, then according to
the law of co-dimension additivity (see Mandelbrot 1982; Falconer 1990),
the dimension of the intersection $D_{\mathrm{int}}$ becomes
\be
 D_{\mathrm{int}} = D + D' - d \, .
\label{Dint}
\ee
For example, if a fractal structure with $D = 2$ in 3-d space is intersected by
a plane with $D' = 2$, then we obtain for the fractal dimension of the
thin intersection $D_{\mathrm{int}} = 2 + 2 - 3 = 1$. This property of 
intersections
explains why a fractal structure with $D \approx 2$ may look as a fractal
with $D \approx 1$ when inspected on large scales from a sample
coming from a thin slice-like galaxy survey.
 
\subsubsection{Multifractal structures }

In fractal models of the galaxy distribution one usually utilizes
only spatial positions for a sample of galaxies.
This allows one to describe the distribution
by means of only one parameter --- the exponent in the power-law
or the fractal dimension of the structure.

Real observational data contain also other important astrophysical
information for each galaxy, such as luminosity, morphology,
spectral properties, stellar contents e.t.c.
In this case the scaling properties
can be different for different types of galaxies.
To take into account the dependence of the distribution on
these parameters
one has to introduce a more general model, called
\emph{multifractal} structures, which is characterized by a continuous
set of fractal dimensions.
Such approach was firstly suggested for galaxy distributions
by Pietronero (1987). For recent discussions of this subject see
Gabrielli et al. (2004),  Martinez \& Saar (2002) and
Jones et al. (2004).

We note that multifractality may be viewed differently thanks
to the complexity of the problem (even for fractals there is
no unique definition).
Multifractals are in contrast with homogeneity exactly
like fractals are.
In fact, the multifractal picture is a refinement and
generalization of fractal properties
 (Paladin \& Vulpiani 1987; Benzi et al. 1984).  
%\cite{paladin87}
%\cite{benzi84}).

\subsection {Modern redshift and photometric distance
surveys}

For many years, astronomers could make only indirect
conclusions
about the distribution of galaxies on the basis of their
two-dimensional
projected locations on the celestial sphere.
Such studies of projections are well reviewed in
Peebles (1980). 

In recent years the situation was completely changed
when it became
possible to measure the 3-dimensional 
distribution of galaxies using data from massive surveys
of galaxy redshifts.
At the present time there are several approaches for
investigating the space
distribution of matter (luminous and dark):
photometric distance measurements,  extensive redshift
surveys of galaxies
and quasars, the analysis of counts of galaxies, and the
study of
image distortion effects produced by weak gravitational
lensing.
All these observational studies have shown that clustering
is a common phenomenon in the realm of nebulae.

%When considering structures around the observer
%in  the Local Group (LG), it
%is convenient to divide
%the space into three parts: the Local Volume (LV) within
%the distance about
%10 Mpc, the Local Hypervolume (LH) within about 100
%Mpc and Local Hubble Volume (LHV) for distances up
%to 1000 Mpc.

\subsubsection { Redshift surveys}

Nature has given the astronomer, in the form of the linear redshift--
distance law,
an accurate way to measure extragalactic distances, which
is generally more precise than photometric methods. For
example,
for a velocity dispersion of  50 km/s, typical for field
galaxies,
one can measure distances with an accuracy of about 1
Mpc.
In order to get in this way a deep, 3-dimensional map of
the
 surrounding galaxy universe, it is necessary to make
deep
 surveys of redshifts, complete up to sufficiently faint
 magnitude limits.

Over 2700 galaxies had their redshift listed in the
\emph{Second Reference
Catalogue} by de Vaucouleurs, de Vaucouleurs \& Corwin
(1976).
This was the breakthrough which made it possible to use redshifts for
mapping
the structures made by galaxies.

Giovanelli \& Haynes (1991)
 emphasized that "In the last fifteen years, advances in
detector
 and spectrometer technology at both optical and radio
 wavelengths have spurred a tremendous explosion in the
galaxy
 redshift tally."  Indeed, this explosion has continued
with
an exponential rate up to present. Currently more than
one million redshifts are known, almost all from optical
spectra.
This ``redshift industry'' continuously produces new points
for
the 3-d maps of spatial galaxy distribution. 
Special telescopes are  dedicated to measurements of redshifts.

\begin{table}[h!]
\label{surveys}
\caption {\footnotesize
Some recent galaxy
surveys.
The columns give: name of survey,
solid angle $\Omega$ covered by survey, apparent
magnitude limit,
total number of galaxies $N$, the method of distance
determination.}
\vskip 0.3cm
\begin {tabular}{|c|c|c|c|c|c|}
\hline
     &           &          &   &   &       \\
Catalogue & $\Omega$ ($sr$) &
$m_{lim}$ & $N$ &  distance & reference \\
 &       &    &  & indicator  &  \\
\hline
CfA1               & 1.83          & 14.5           &
1845 & z & Huchra
\etal 1983  \\
CfA2 (North)       & 1.23          & 15.5           &
6478 & z & de Lapparent et al. 1988  \\
PP                 & 0.9           & 15.5           &
3301 & z &   Haynes \& Giovanelli 1988 \\
SSRS1              & 1.75          & 14.5           &
1773 & z &   Da Costa et al. 1990\\
SSRS2              & 1.13          & 15.5           &
3600 & z &   Da Costa et al. 1994 \\
Stromlo-APM        & 1.3           & 17.15          &
1797 & z &   Loveday et al. 1992 \\
LEDA               & $4 \pi$       & 16.0           &
25156 & z &    web site      \\
LCRS               & 0.12          & 17.5           &
26000 & z &  Shectman et al. 1996  \\
IRAS $2 Jy$        & $4 \pi$       & 2. Jy          &
2652  & z & Strauss et al. 1992\\
IRAS $1.2 Jy$      & $4 \pi$       & 1.2 Jy         &
5313  & z &  Fischer et al. 1996\\
ESP                & 0.006         & 19.4           &
4000  & z &  Vettolani et al. 1997\\
KLUN               & $4\pi$        & 15             &
6500  & TF &  Theureau et al. 1997b\\
KLUN+              & $4\pi$        & 16             &
20000 & TF & Theureau et al. 2004\\
Local Volume       & $4\pi$        & $ < 500$ km/s  &
300 & RGS & Karachentsev et al. 2003 \\
2dF                &   0.27            &  19.5      &
250 000 & z &  web site\\
SDSS               &  $ \pi $       &     19         &
$10^6$   & z & web site \\
 &           &                   &               &
\\
\hline
\end {tabular}
\end {table}

Many extensive redshift surveys have already been
completed, among them what are known by the
abbreviations
 CfA, SSRS, LCRS, ESP.
Their relevant parameters are presented in
Tab.\ref{surveys}.
 For a more detailed review the reader may consult
Sylos Labini et al. (1998).
Based on these surveys several 3-d maps of
galaxy
distribution have become available: both wide angle such
as
CfA1, CfA2,
SSRS1,
SSRS2,
Perseus-Pisces,
LEDA,
and narrow angle such as LCRS, ESP.

\begin {figure}[th]
\centerline {\psfig {file=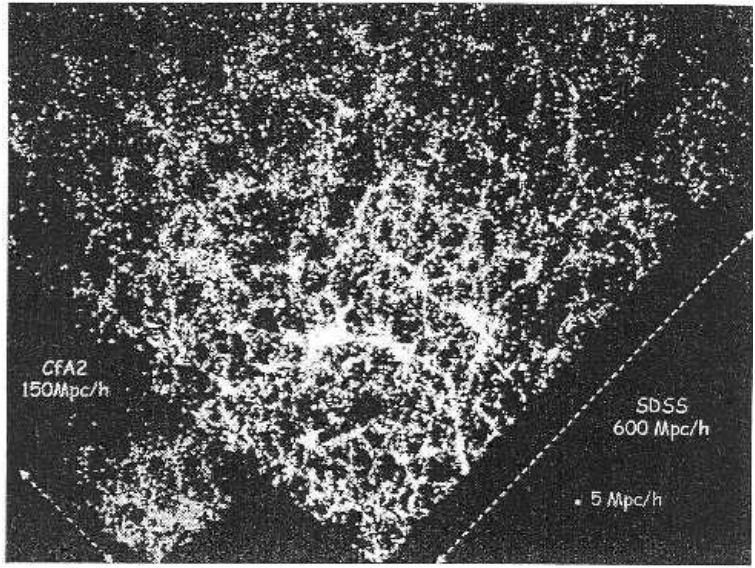,width=10cm}}
%\resizebox{\hsize}{!}{\includegraphics{fig5.ps}}
\caption{\footnotesize
The large scale structure from
the CfA and the first
release of SDSS
galaxy redshift
survey (from Courtois et al. 2004a).
One may easily recognize different kinds of
structure on scales of about 100 Mpc. }
\label{sdss}
\end{figure}

The last decade saw the appearance of essentially larger
surveys with
hundreds of thousand of
redshifts: the two-degree field galaxy redshift survey
2dF
and the Sloan Digital Sky Survey SDSS.
The depth of these galaxy catalogues allows one
to detect and analyze structures with sizes up to 100
Mpc (see Fig.\ref{sdss}).

{\small
\emph{ Limits of a survey.}
A redshift survey is basically restricted by two limits:
the apparent magnitude limit of the survey and the
distance modulus limit
which is different for different absolute magnitudes. In
addition, it is important
for the structure analysis to have a sufficiently large
sky coverage.

The number of galaxies per square degree increases
steeply with
 magnitude.  There is roughly one galaxy/deg$^2$ up to
$m \approx 15.5$
 and about 85 up to 19 mag.  On the other hand, this may
be coped
 with the modern MOS (multiple object spectrograph)
techniques
 which allow a simultaneous measurement of several tens
of spectra
 of galaxies in the field of the telescope.
Even if the sample is complete up to a certain
apparent
 magnitude limit, its completeness in volume depends on
the
 absolute magnitude: for galaxies with absolute
magnitude $M$,
 the spatial completeness limit of the survey is at the
distance modulus
\begin{equation}
 \mu_{\mathrm{lim}} = m_{\mathrm{lim}} - M
\end{equation}
For example, with $m_{\mathrm{lim}} = 15.5$ and $M = -
20$, $\mu = 35.5$
which corresponds to the distance $r_{\mathrm{l}} = 125$
 Mpc (corresponding to $9300h_{_{75}}$ km/s).  On the other
hand,
 galaxies with $M = - 18$ will be completely sampled
only up to
$r_{\mathrm{lim}}  = 50$ Mpc ($3700h_{_{75}}$ km/s).
 This shows that redshift surveys probe the
 distribution of only the most luminous galaxies at
large
 distances, which may cause a biased picture.  In order
to cope
 with this problem, the only way is to push the surveys
to
 fainter and fainter magnitude limits, because one
cannot tell
 beforehand whether a galaxy taken is intrinsically
bright or faint
(or whether it is distant or relatively near).
}

\subsubsection{ Galaxy catalogues
with photometric distances.  }
For the study of  large-scale structure, redshift
catalogues are generally
superior to those based on photometric distances: 1)
Redshift catalogues are
much less time-consuming to make, 2) generally
photometric distances are
less accurate than redshift distances, and 3) redshifts
may be measured for
all Hubble types of galaxies.

However, there are questions which necessarily require
large samples of
galaxies for which both redshifts and independent
photometric distances
have been measured (e.g. the Hubble constant, deviations
from the Hubble
law, peculiar velocities and large scale streams). On
the other hand,
even photometric distances as such are valuable for
investigating structures.
They can be used for mapping the environment of the
Local Group and
for deriving the average number density law around us.

As radial velocities of galaxies, especially of those
within groups and
clusters, give only an approximate estimate of
distances, Igor Karachentsev
started a
vast program of distance measurements to
nearby galaxies, using the luminosity of their brightest
blue and red
stars  (Karachentsev \& Tikhonov, 1994; Karachentsev et
al., 1997)
and the luminosity of the tip of the red giant branch stars
(Karachentsev et al. 2003). Over the
last 10--15 years many nearby galaxies have been
resolved into stars
for the first time. This
labour-consuming program requiring a lot of observing
time with the
largest ground-based telescopes, as well as the Hubble
Space Telescope,
is not yet complete. So far, the distances have been
measured for about 150
galaxies situated within 9$h_{50}^{-1}$ Mpc
 (Karachentsev et al. 2003).

During the last years,
a special effort has
been undertaken to increase the Local Volume sample.
``Blind'' surveys
of the sky in the 21 cm line (Kilborn et al. 2002),
infrared and radio
surveys of the Zone of Avoidance (Kraan-Korteweg \&
Lahav, 2000) and searches
for new dwarf galaxies of very low surface brightness
based on the POSS-II
and ESO/SERC plates (Karachentseva \& Karachentsev,
1998, 2000) have made the
total number of the Local Volume galaxies increase
more than two times.

A formidable example of a galaxy sample with photometric
distances up to $\approx 100$ Mpc is
the KLUN galaxy sample and its growing version KLUN+
(see KLUN+ home page at http://klun.obs-nancay.fr/KLUN+).
These contain
thousands of spiral galaxies for which both photometric
magnitudes
and the width of the neutral hydrogen
21cm line have been measured.
Using the Tully--Fisher relation these two quantities (properly corrected and
reduced to homogenized systems) allow one to make an estimate of
the distance of a galaxy. Originally planned for measuring the value of
the Hubble constant, the  KLUN with its 5500 galaxies was
used to study the radial number density distribution of galaxies
around us (Teerikorpi et al. 1998). The new KLUN+ project is based
on a Cosmological Key Project at the refurbished Nancay radio telescope
in France. The HI survey plans to build a uniquely large magnitude-limited
sample
of 20~000 spiral galaxies distributed on the 80 percent of the sky
visible from Nancay ($\delta > -40^o$). Theureau et al. (2004) give
information and first results on the new HI measurements.
The photometry comes from the
DENIS (Near Infrared Survey) and 2MASS (2 Micron All Sky Survey).
The aim is to build a sample complete to well defined magnitude limits
in five photometric bands B, I, J, H and K (the earlier KLUN had
only B magnitudes and diameters).

All these data give us the possibility to study the large scale
galaxy distribution by means of different statistical methods
which we describe below.

\subsubsection {How to discover fractal structures}

From the 3-d map shown in fig.\ref{sdss} (SDSS) one may recognize by
eye structures with different sizes up to 100 Mpc. However, a quantitative
analysis of the observed inhomogeneities in the spatial distribution of
galaxies is a hard problem around which there is going on the
debate on fractality of observed structures.

In order to understand the meaning of conclusions from different
analyses of the observational data, 
it is highly advisable to investigate idiosyncracies
and limitations of the used methods.
The next section addresses this issue and discusses main mathematical tools
used to discover fractal structures. We restrict our consideration to
the methodological questions arising in practical applications
of the methods to analysis of galaxy samples.

\section{ Statistical methods to detect
fractality in galaxy distribution}

Several statistical
methods have been used
for the analysis of both 2-d
(angular distributions)
and 3-d (spatial distributions) galaxy catalogues.
A comprehensive  review of  mathematical
approaches for
the description of the large scale
galaxy distribution may be
found in the
book by Martinez \& Saar (2002).

Here we confine our discussion to the fractal approach.
Detailed descriptions of such methods for the
analysis of the
distribution of galaxies are given by
Sylos Labini et al. (1998) and
Gabrielli et al. (2004).

One of our goals is to compare
the standard method of the $\xi$ correlation
function (Peebles, 1980; 1993) with
the fractal approach
%originated in statistical physics
(Pietronero 1987;
%\cite{pietronero87};
Gabrielli et al. 2004).
% \cite{gabrielli04}).
We demonstrate that
it is essential for the study of the distribution of
galaxies to utilize
mathematical instruments which are adequate for the
existing structures. It is especially important to use
undistorted estimators for measuring the fractal dimension
and the range of fractality.

\subsection {Definitions for correlation functions}

The theory of stochastic processes
introduces and studies different functions intended for
 the correlation analysis
(see e.g. sect.2 of Gabrielli at al. 2004).

\subsubsection{Complete and reduced correlation functions}

The {\it complete two-point correlation function} $R_{\mu\mu}$ 
(we call it simply the complete correlation function)
of a stationary isotropic process 
$\mu(\vec{r})$ is defined as
\begin{equation}\label{R-xx}
 R_{\mu\mu} (r) = \langle \mu(\vec{r}_1) \mu(\vec{r}_2)\rangle
\end{equation}
where $r = \mid \vec{r}\mid = \mid \vec{r_1} - \vec{r_2}\mid $ is
the mutual distance between considered points, and 
$\langle \cdot \rangle$ is the ensemble average over all
realizations of the stochastic process. 

Taking into account the truly constant mean value $\mu_0$ of the process
\begin{equation}\label{x_0}
\mu_0 = \langle \mu(\vec{r})\rangle = \mathrm{const}
\end{equation}
one may define the {\it reduced two-point correlation function} $C_2$
for the fluctuations around $\mu_0$
(we call it simply the reduced correlation function) as
\begin{equation}\label{K-xx}
 C_2 (r) = \langle (\mu(\vec{r}_1)-\mu_0)
( \mu(\vec{r}_2)-\mu_0)\rangle = R_{\mu\mu}(r) - \mu_0^2\,.
\end{equation}
For $r=0$ it expresses the squared dispersion of the process as
$\sigma_{\mu}^2=C_2(0)$.

We emphasize an important difference between the 
complete and reduced correlation
functions $R_{\mu\mu}(r)$ and $C_2(r)$.
For a stochastic process with long range power-law correlations
the complete correlation function $R_{\mu\mu}(r)$ has a power-law form,
while the reduced correlation function $C_2(r)$, according to its
definition (eq.\ref{K-xx}), cannot be a power law in this case.

\subsubsection { Mass variation in spheres and characteristic scales }

In the applications that we discuss the stochastic process $\mu(\vec{r})$ will
describe the density field $\rho(\vec{r})$.

{\it Conditional and unconditional functions. }
It is important to distinct between {\it conditional} and
{\it unconditional} functions (or statistics).
For instance, when one considers such statistics which are defined
with the condition that there is a fixed point-particle relative to which
other particles of a process are considered, then one speaks about a
conditional function. We will see below that the two major tools
of the LSS analysis, the $\Gamma$ and $\xi$ functions, are both
conditional correlation functions.

As an example of unconditional statistics we consider mass (number)
fluctuations inside a sphere of radius $R$. Let us define in addition
to $\rho(\vec{r})$ a new stochastic variable $M(R)$ as
\be
\label{M1}
M(R) = \int_{V(R)} \rho(\vec{r})d^3 r
\ee
For a given radius $R$ fluctuations of this mass calculated in different
positions in space can be characterized by the {\it normalized mass variance}
$\sigma^2_M (R)$:
\be
\label{sigmamass}
\sigma^2_M (R) = \frac{\langle M(R)^2 \rangle - \langle M(R)\rangle^2}
{\langle M(R) \rangle^2}\,\,,
\ee
where
\be
\label{Mave}
\langle M(R) \rangle = \frac{4 \pi}{3} \rho_0 R^3
\ee
and for volume $V=V(R)$
\be
\label{Msquareave}
\langle M(R)^2 \rangle = \int_{V}d^3r_1 \int_{V} d^3 r_2  \langle
 \rho(\vec{r}_1) \rho(\vec{r}_2)\rangle
\ee
Here there is no condition on the locations of the centre of the sphere,
which may be put anywhere in the space within the sample regardless
of the positions of the particles, also ``between'' them.

{\it The scale of homogeneity. }
Our intuitive vision of uniformity may be formalized by means of
the variable $M(R)$. E.g. {\it the homogeneity scale} $R_{\mathrm{hom}}$ may be 
defined
as the scale at which $\sigma^2_M(R_{\mathrm{hom}}) = 1$ (or some other
threshold value). This means that it is possible to regard a distribution
of particles as approaching homogeneity if the average mass
fluctuation within spheres of radius $R_{\mathrm{hom}}$ is about the average 
mass
$\langle M(R_{\mathrm{hom}})\rangle$.
This scale is well defined when $\sigma^2_M(R)
\rightarrow 0$ for $R > R_{\mathrm{hom}}$.

{\it The correlation length. }
The second scale is the \emph{correlation length} $R_{\mathrm{cor}}$,
which does not depend
on the amplitude of the correlation function and just characterizes the rate of
decrease of the correlation function. The correlation length may be infinite,
as it is for a power law correlation
$R_{\mu\mu}(r) \propto r^{-\gamma}$, or finite, as for an exponential
correlation function $R_{\mu \mu}(r) \propto e^{-r/R_{\mathrm{cor}}}$.

\subsection {The method of the $\xi$-correlation function}

A widely used classical approach to the analysis of the
large scale structure
is the method of correlation function. 
It was first introduced to the galaxy analysis by Totsuji \&
Kihara (1969) who
adopted this method from the statistical
physics of ordinary gas density
fluctuations (e.g. Landau \& Lifshitz 1958).
It was further developed and extensively applied to
galaxy data
by Peebles (1980, 1993) and others (for recent reviews see
Martinez \& Saar 2002; Jones et al. 2004).

\subsubsection {Peebles' $\xi$-correlation function}
According to Peebles (1980) the {\it two-point correlation function}
$\xi (r)$
is defined as the dimensionless {\it reduced} correlation function 
of the density fluctuations $\delta \rho (\vec{r}) = \rho (\vec{r}) -
\rho_0$ around
the average density $\rho_0$
\begin{equation}\label{ordinary_ksi}
  \xi (r) = \frac{\langle \delta \rho(\vec{r}_1)  \delta \rho(\vec{r}_2)
\rangle}{\rho_0^2} = 
 \frac{\langle \rho(\vec{r}_1) \rho(\vec{r}_2)
\rangle - \rho_0^2}{\rho_0^2}
\end{equation}
In fact the $\xi$-function is simply the reduced correlation function
(eq.\ref{K-xx}) divided by the squared mean value of the process
(eq.\ref{x_0}),
i.e. 
\begin{equation}\label{ksiC}
\xi(r)=C_2(r)/\rho_0
\end{equation}
In the case of a distribution of identical particles 
(with masses $m=m_0$) one uses
a number density  $n(\vec{r})=\rho(\vec{r})/m_0$,
whose average is $\langle n(r) \rangle = n_0$. Then
\begin{equation}\label{ksi2}
  \xi(r)=\frac{\langle n(\vec{r}_1) n(\vec{r}_1 +
\vec{r})\rangle}{n_0^2}
  - 1\,.
\end{equation}
This dimensionless function measures correlations of fluctuations
relative to a constant average number density $n_0$.

In the theory of stochastic processes one usually considers 
a {\it normalized correlation function} which is defined as
$K_{\mu\mu}(r)=C_2(r)/\sigma_x^2 =  (R_{\mu\mu}(r) - x_0^2)/\sigma_x^2 $. 
Then there is
the normalization condition $K_{\mu\mu}(0)=1$. The definition
for the $\xi$-function (eq.\ref{ordinary_ksi}) 
implies the condition $\xi(0) = \sigma_{\rho}^2/\rho_0^2 $. 

{\it Definition via Poisson process. }
The correlation function may also be defined as a measure of the
deflection of a distribution of particles from
a Poisson (uniform) distribution (Peebles 1980).
In this case one considers
two infinitesimal
spheres at the points  $\vec{r}_1$ and $\vec{r}_2$
with volumes $dV_1$ and $dV_2$ and
with the mutual distance $\vec{r}_{12}$. Then
the joint probability to find one particle in the volume
$dV_1$ and
another particle in the volume $dV_2$ is proportional to the
number of pairs $dN_{12}$
\begin{equation}\label{ksi1}
  dN_{12} = n_0^2 dV_1 dV_2 [1+\xi(\vec{r}_{12})]\,,
\label{ksidef2}
\end{equation}
where $n_0$ is the ensemble average number density of particles and
$\xi(\vec{r}_{12})$ measures the deflection from the Poisson distribution.
This definition implies that $\xi(\vec{r}_{12}) = 0$ automatically 
for a Poisson process. 

For a statistically isotropic distribution the function
$\xi(\vec{r}_{12})
= \xi(r)$ depends on the separation $r$ only.
For the case when an object is chosen at random from the
sample, the
probability of finding that it has a neighbour at a
distance $r$ in
$dV$ (e.g. $4 \pi r^2 dr$) is proportional to the expected number
of neighbours $dN$
\be
dN = n_0 dV [1+\xi(r)]\,.
\label{ksidef}
\ee
Here $\xi(r)$ is considered as the same two-point correlation function
as defined by eq.\ref{ksi2} (see sect.33 of Peebles 1980).
It is a measure of finding an excess number of particles relative
to the Poisson distribution, at the distance $r > 0$
provided that there is a particle at $r=0$.

\subsubsection {$\xi$-function estimators}

In the theory of stochastic processes it is important to make
a distinction between functions (e.g. $\xi (r)$) defined by ensemble averages
and
their estimators ($\hat{\xi} (r)$), e.g. applied to a finite galaxy sample.

To estimate the two-point correlation function
from an available data sample of $N_s$ objects within a volume $V_s$, 
one generally uses a method based on
artificially generated Poisson process, which
fills the same volume $V_s$ of the sample.
Then the $\xi$-function for a given scale $r$ is estimated as
the ratio of the number of pairs with such mutual distance in the sample
to the number of such pairs in the artificial Poisson distribution.
There are several different pairwise estimators
(Kerscher et al. 2000; Martinez \& Saar 2002; Gabrielli et al. 2004),
and the difference between them
lies mainly in their method of edge correction.
% i.e. the way how to calculate the number of pairs when the
%points are so close to the border of the sample that in some pairs one
%point will lie outside the sample volume if we could repeat
%the counts in spheres outside the observable volume. 

For instance, the Davis--Peebles estimator weights the points according to
the part of the spherical shell volume which is contained in the volume of
the sample. It has the form

\begin{equation}\label{ksi3}
  \hat{\xi}(r)=\left(\frac{N_{rd}}{N_s-1}\right)
\frac{N_p(r)}{N_{p,rd}(r)}
  - 1\,,
\end{equation}
Here $N_p(r)$ is the number of data-data pairs of observed objects in the
catalogue having their
mutual distance in the interval
$(r,\,r+dr)$.  $N_{p,dr}(r)$ comes from the joint catalogue of
data and artificial random distributions in the same volume $V_s$.
It is the number of data-random pairs with
the distance $r$ in the joint catalogue.
$N_s$ and $N_{rd}$ are the total numbers of objects in the
real sample and the
random distribution, respectively.

An essential assumption 
of the correlation function method is the hypothesis of
homogeneity according to which 
the true average of objects $n_0$
is estimated from the observed sample as
\be \label{n0}
\hat{n}_0 = \bar{n} = \frac{N_s}{V_s}  
\ee
with a high formal accuracy $\sigma_{n_0} \approx 1/\sqrt{N_s}$,
where $N_s$ is the total number of objects in the volume $V_s$
of the ``fair'' sample, which is assumed to be representative of
the homogeneous
distribution of galaxies in the whole Universe.

\subsubsection {The normalization condition for $\xi$ estimators}

A significant point related to $\xi$-function estimators
 was emphasized by Pietronero (1987) and
Calzetti et al. (1988), and in more detail by Gabrielli et al. (2004). 
Namely, the definition of the correlation function
as a deflection from the Poisson distribution
(eq.\ref{ksidef}) implies an integral condition for
the $\xi$ function estimated from a finite sample of galaxies.

This comes from the fact that for any sample with a
finite
number of galaxies $N_s$ in a volume $V_s$ the estimation of the
average number density is $\bar{n} = N_s/V_s$.
Integrating the left side of eq.\ref{ksidef} over the sample volume we get

\be
\int_{V_s}dN = N_s -1
\ee
where $N_s -1$ is the number of neighbours, i.e. the total number of particles
in the volume $V_s$ without the one whose neighbours are counted. Then
the integration of the right side of eq.\ref{ksidef} over the sample volume
gives
\be
N_s - 1
% = \int_{V_S} \bar{n} (1+ \hat{\xi} (r))dV
 = \int_{V_S} \bar{n}dV +
       \bar{n} \int_{V_S} \hat{\xi} (r) dV.
\label{normcond}
\ee
The first term on the right side is $\int_{V_s} \bar{n} dV = N_s$, the total
number of the particles in the sample.
Hence the second term will satisfy the condition
\be
\int_{V_S} \hat{\xi} (r) dV = - \frac{1}{\bar{n}}
\label{int_ksi1}
\ee
or in dimensionless form:
\be
\int_{V_S} \hat{\xi} (r) \frac{dV}{V_s} = - \frac{1}{N_s}
\label{int_ksi2}
\ee
In the case of fluid-like correlated distributions the effective
number density of particles may be arbitrarily large and hence
the condition of eq.\ref{int_ksi1} becomes
\be
\int_{V_S} \hat{\xi} (r) dV = 0
\label{int_ksi3}
\ee
These restrictions lie behind some 
controversial results obtained by the $\xi$ function method
of the large scale structure analysis.
In particular, we have in mind
the inevitable non-power law behaviour of the $\xi$ estimator.
From eqs.\ref{int_ksi1} and \ref{int_ksi3} follows that there is a distance
$r_z$ where
$\hat{\xi} (r_z) = 0$. Here
the correlation function estimator changes its sign from positive
to negative values, which is impossible for a power-law function.

\subsubsection{ A systematic distortion of the true power-law correlation due
to the $\hat{\xi}$-estimator}

As was shown above, if the complete correlation function is a
power-law then neither $\xi(r)$ nor $\hat{\xi}(r)$ can be a power-law
function. Nevertheless, in practice $\hat{\xi}(r)$ is usually presented
in the form
\be \label{ksi-gamma}
\hat{\xi} (r) = \left( \frac{r}{r_0} \right)^{-\gamma} \;,\;\;\;
r_1 < r < r_2 \,,
\ee
valid for some range of scales $r_1<r<r_2$. Here the parameter $r_0$
defines the amplitude.

From such a power-law presentation
one usually derives two numbers: the {\it unit scale} $r_0$
and the \emph{correlation exponent} $\gamma$.
We emphasize that due to
the normalization condition (eq.\ref{int_ksi2})
 both numbers give systematically distorted values for 
the homogeneity scale and the power-law exponent of the true
complete correlation function $R_{\mu\mu (r)}$ describing
the density field.

The unit scale $r_0$
(which is often called, somewhat misleadingly, as correlation length)
is defined from the relation 
\be \label{ksi-r0}
\hat{\xi} (r_0) = 1\, ,
\ee
which characterizes the amplitude of density fluctuations
at the scale $r_0$.
In fact, it is a distorted value of a true homogeneity scale
of the process if the true value is equal to or larger than
the size of a maximum sphere embedded completely in the sample
volume $V_s$.

The correlation exponent $\gamma$
in the power-law representation of $\hat{\xi} (r)$ (eq.\ref{ksi-gamma})
describes correctly only a restricted interval of
scales $r_1<r<r_2$ .
On scales $r>r_2$ this does not represent the true value of the exponent,
because there the estimated value
 is distorted as 
the normalization condition (eq.\ref{int_ksi2})
makes  $\hat{\xi} (r)$ deflect from the 
inherent power-law and to cross zero level.
For example, below it will be shown that for the exponent $\gamma$
is derived twice the true correlation exponent at the unit scale $r_0$
(sect. 3.4.1).  

On scales $r<r_1$ the true value of the exponent is distorted
due to the noise of discreteness, 
which behaves as $\sigma \propto r^{-3/2}$ (eq.\ref{sigma-r}).
The error will essentially
increase for scales smaller than the average distance
between particles in a sample (e.g. $r<\lambda_0$).
 We will see later examples of 
how this has happened in data analysis.

\subsubsection {Redshift-space and the peculiar velocity field}

From a galaxy redshift survey one obtains a redshift-space map,
i.e. ($\alpha$ , $\delta$, $z_{\mathrm{obs}}$) coordinates, where
 ($\alpha$ , $\delta$) give the position on the sky and
$z_{\mathrm{obs}}$ gives the observed redshift of a galaxy in the sample.
The value of $z_{\mathrm{obs}}$ in principle contains all possible
contributions from different physical causes according to 
the relation
\be
1 + z_{\mathrm{obs}} = (1+z_{\mathrm{cos}})(1+z_v)
(1+ z_{\phi})(1 + z_{\mathrm{new}})
\label{redshiftadd}
\ee
Here $z_{\mathrm{cos}}$ is the cosmological redshift which
determines the true distance to the galaxy $r_{gal}(z)$ calculated
from the empirical distance--redshift relation (i.e. from the Hubble
law, which gives $r_{\mathrm{gal}}=cz_{\mathrm{cos}}/H_0$)
or from an adopted cosmological
model for large redshifts.
The $z_v$ is the redshift component caused by the peculiar velocity of 
the galaxy, $ z_{\phi}$ is the gravitational part of the redshift
caused by the local gravitational potential of the galaxy
(e.g. in a cluster), and  $z_{\mathrm{new}}$ is a possible component
due to unknown cosmological physics.

{\it Distance error due to peculiar velocity. }
Let us consider the influence of the peculiar velocity field
on the distance estimation. For peculiar velocities
$v << c$ the Doppler part of the observed redshift is determined
by the radial component $v_r$ of the velocity as
\be
\label{z-dop-rad}
z_v \approx  v_r/c\,\,. 
\ee
So for small radial peculiar velocities $v_r$
we obtain
\be
 z_{\mathrm{obs}} =  z_{\mathrm{cos}} +
\frac{v_r}{c}    
(1 + z_{\mathrm{cos}})
\label{zobs}
\ee
Hence the true spatial galaxy distribution will be distorted by a peculiar
velocity field in the line of sight direction
\be
\label{r-par}
 r_{\parallel \mathrm{obs}} = r_{\parallel} + \Delta r_v 
\ee
by the value
\be
\Delta r_v = \frac{v_r (1+z_{\mathrm{cos}})}{H_0} =
5 \mathrm{Mpc}\, h^{-1}_{100}
  \frac{v_r (1+ z_{\mathrm{cos}})}{500\mathrm{km/s}} 
\label{zdisterror}
\ee
Note that the factor $(1+z_{\mathrm{cos}})$ leads to an increasing
influence of $v_{\mathrm{pec}}$ on the distance distortion for deep
redshift surveys.

{\it Real-space and redshift-space $\xi$ functions. }
A directly observed $\xi$-function
is called  the {\it redshift-space correlation function}
$\xi_z (s)$. In order to obtain the {\it real-space correlation
function} $\xi_{\mathrm{real}}=\xi(r)$ one should extract and delete
all non-cosmological contributions to $z_{\mathrm{obs}}$.
This is a hard problem because it requires a priori knowledge
of the peculiar velocity field, the total mass around the galaxy,
and restrictions on new physics.

The shape of the observed  $\xi_z(s)$ is determined by the character of
the peculiar velocity field. In virialized clusters the
velocity dispersion leads to the so-called ``fingers-of-God'' effect,
i.e. an elongated shape along the line of sight direction $r_{\parallel}$. 
The mean tendency of galaxies at larger scales to approach each other
due to the gravity of large-scale structures will appear as a compression
of  $\xi_z(s)$  in the direction $r_{\parallel}$.
As these two effects are related to different spatial scales, they do
not compensate each other.

Peebles (1980; sec.76) and Davis \& Peebles (1983) suggested a procedure
for the restoration of both the true $\xi (r)$ 
and the relative peculiar velocity distribution $f(v)$
from the observed correlation
function $\xi_z (r_{\perp}, r_{\parallel \mathrm{obs}})$
where $r_{\perp}$ an $r_{\| \mathrm{obs}}$
are the observed perpendicular
and parallel to the line of sight components of the separation
$s = \sqrt{r_{\perp}^2 + r_{\parallel \mathrm{obs}}^2}$.

{\it Derivation of the real-space $\xi$-function. }
The method is based on the calculation of the projected correlation function
$w(r_{\perp})$ which does not depend on the peculiar velocity field, if
the distribution of the radial peculiar velocities is symmetrical around
each galaxy of the sample.
Then integrating along the line of sight we obtain:
\be
w(r_{\perp}) = 2 \int_0^{\infty} 
\xi_z 
(r_{\perp}, r_{\parallel obs}) dr_{\parallel obs} \, ,
\label{corproject}
\ee  
where in practice the interval of integration is restricted by
chosen radial velocity limits. Then the wanted inverse is the Abel integral
\be
\xi (r) = - \frac{1}{\pi} \int_r^{\infty} \frac{w' dr_{\perp}}
{(r^2 - r_{\perp}^2)^{1/2}} \, ,
\label{cortrue}
\ee 
where $w' = dw(r_{\perp})/dr_{\perp}$. Eq.\ref{cortrue} gives
the solution for the problem of restoration of the real-space
$\xi(r)$ correlation function.

For the power law $\xi = (r_0/r)^{\gamma}$ the integral (\ref{corproject})
gives
\be \label{ksi-proj}
w(r_{\perp}) = A r_{\perp}^{1-\gamma} \, ,
\ee
where
\be \label{A}
A = r_0^{\gamma} \frac{\Gamma_e(1/2) \Gamma_e((\gamma -1)/2)}
{\Gamma_e(\gamma /2)}
\ee
and $\Gamma_e(x)$ is the Euler gamma function.

{\it Limitations of the projection method. }
It is clear from eq.\ref{ksi-proj} 
that such a solution for the real-space $\xi$ correlation
function is valid only if the exponent
$\gamma \geq 1$. For example, $\gamma = 1$ gives
$w(r_{\perp}) =$ constant, which demonstrates that 
a uniform background galaxy distribution
may be confused with the projection of real-space non-uniform
distribution. 
%This happens because
%the considered method contains a forbidden (for structures
%with $\gamma \leq 1$) logical  chain of arguments:
%{\it 3-d redshift-space distribution $ \, \, \Longrightarrow \, \,$ 
%projection on the sky $ \, \, \Longrightarrow \, \,$ 
%deprojection to the 3-d true distribution}.

According to the theorem on fractal projections (sec.2.4.5) such a method
inevitably leads to elimination of information on structures with
the fractal dimension $D \geq 2$.
Therefore to take into account the peculiar velocity field
within fractal structures with $D \geq 2$ (such a structure is
actually observed,
see sec.5), it is necessary to use another method of restoration for the
correlation function, which is free from the above limitation.
Also  a more careful study of distance errors in
both $r_{\parallel}$ and $r_{\perp}$ components is required.

{\it Estimation of the relative velocity dispersion. }
In the case, when density and velocity fields are weakly coupled,
the observed correlation function $\xi_z(s)$ can be modelled as
a convolution of the real space correlation function $\xi(r)$
with the galaxy pairwise velocity distribution $f(v)$.
According to Peebles (1980, sec.76) and Davis \& Peebles (1983)
this equation may be presented in the form
\be \label{ksi-vel}
1+\xi_z (r_{\perp}, r_{\parallel \mathrm{obs}}) =
H_0 \int \left[ 1 + \xi (\sqrt{r_{\perp}^2 + r_{\parallel}^2}) \right]
f(v) dr_{\parallel}
\ee
where
\be \label{v-arg}
v = H_0r_{\parallel \mathrm{obs}} - H_0r_{\parallel} + \bar{v}_{12}(r)
\ee
and  $\bar{v}_{12}(r)$ is the mean radial pairwise velocity
of galaxies at separation $r$, which is represented by a 
model. Davis \& Peebles (1983) adopted the model
\be \label{v-12-r}
 \bar{v}_{12}(r) = \frac{H_0r_{\parallel}}{1+(r/r_0)^2}
\ee
and an exponential form for $f(v)$:
\be \label{f-v}
f(v)=B~\exp \left(-2^{1/2} \mid v \mid /\sigma_{12}\right)
\ee
As a result of this approach one obtains the 
{\it pairwise velocity dispersion} $\sigma_{12}$.

\subsection {The method of conditional density}

\subsubsection { Definitions}

The method of conditional density has been successfully used 
for analysis of fractal structures in modern statistical physics.
This method was proposed for extragalactic astronomy 
by Pietronero (1987) and has been applied to
3-d galaxy catalogues  by many authors
(see reviews in  Coleman, Pietronero 1992; Sylos Labini, Montuori,
Pietronero 1998; Gabrielli et al. 2004).
Conditional density method has the advantage that it gives
an undistorted estimation of the true power law correlation
and the true fractal dimension. It also may be used for finding
an undistorted value of the homogeneity scale of a galaxy sample.

{\it Continuous stochastic processes. }
The {\it conditional density} $\Gamma (r)$ may be defined by means
of the complete correlation function (eq.\ref{R-xx})
in the following form:

\begin{equation}\label{eta_autocor}
  \Gamma (r)=\frac{R_{\mu\mu}(r)}{\mu_0} =
\frac{\langle \rho(\vec{r}_1) \rho(\vec{r}_1 +
\vec{r})\rangle}{\rho_0}
\end{equation}

Here $\rho(\vec{r})$ is the stochastic density field  
and $\rho_0$ is the ensemble average density.
The $\Gamma$-function has the physical dimension of density [g/cm$^3$],
and it is a measure of correlation in the total density field
without subtraction of the average density.
The physical dimension
of the $\Gamma$ function  agrees with
the common interpretation
of $\Gamma (r)$ as an average density law around each
point of the structure. This makes its estimator a natural
detector of fractality.

As we shall see below this definition allows one to construct
such an estimator which has no additional restrictions like
the normalization in eq.\ref{int_ksi1}, and hence is able to give an 
undistorted value of the exponent for true power-law 
correlations.

{\it Discrete stochastic fractal processes. }
Let us consider a discrete stochastic process, one
realization of which is
a set of identical particles at randomly selected positions
$\{\vec{x}_a\}, ~a=1,...,N$, so that the number
density $n(\vec{x})$ is given by the expression

\begin{equation}\label{n-x}
  n(\vec{x}) = \sum_{a=1}^N\delta (\vec{x}-\vec{x}_a)\,.
\end{equation}

If the stochastic process generates a fractal,
then it is natural to define 
the number density  as a function of
two variables: $n=n(\vec{x},r)$. The first variable describes the
position $\vec{x}_a$
of a structure particle, and the second  variable 
gives the radius $r$ of a ball inside which
one calculates the number of particles of the structure.
The variable $r$ serves for constructing
a statistics which can measure
the \emph{strength of the singularity} around a particle
of the fractal structure where the number of particles
grows as a power-law $N(r) \propto r^D$.

Denote by $N_V(\vec{x}_a, r)$ the number of particles in
a sphere of
radius $r$, centered at the particle $a$ with the
coordinates $\vec{x}_a$,
belonging to the structure:

\begin{equation}\label{N-x-r}
  N_V(\vec{x}_a, r)= \int_0^r n(\vec{x_a} +
\vec{x})4\pi x^2 dx\,,
\end{equation}
and $N_S(\vec{x}_a, r)$ is the number of particles in
the spherical shell
 $r, r+\triangle r$, with the centre at $\vec{x}_a$:

\begin{equation}\label{S-x-r}
  N_S(\vec{x}_a, r)= \int_r^{r+\triangle r} n(\vec{x_a}
+ \vec{x})4 \pi x^2dx\,.
\end{equation}
From one realization to another these quantities
fluctuate, but
after averaging over many realizations the stable
power-law dependence
on the scale $r$ emerges. In the case of ergodic processes
averaging over
many realizations may be replaced by many points in one
realization.
Following the work by Pietronero (1987) we define the
\emph{conditional (number)
density} of a stochastic fractal process in the form:

\begin{equation}\label{gamma-r}
  \Gamma (r) =\left\langle \frac{N_S(\vec{x}_a, r)}{4\pi
  r^2\triangle r}\right\rangle _{\vec{x}_a} =
  \frac{DB}{4\pi}~r^{-(3-D)}\,,
\end{equation}
and the \emph{conditional volume density} as

\begin{equation}\label{gamma-V-r}
  \Gamma^*(r) = \left\langle \frac{N_V(\vec{x}_a,
r)}{(4\pi/3)r^3}
  \right\rangle _{\vec{x}_a}= \frac{3B}{4\pi}~r^{-(3-
D)}
  \,,
\end{equation}
where $\langle \cdot \rangle_{\vec{x}_a}$ means
averaging over all points $\vec{x}_a$ in one realization
with the condition that the centres of the spheres lie
at the  particles of
a realization (this explains the word
``conditional'').
The last equalities in (\ref{gamma-r}) and (\ref{gamma-V-r})
relate to
ideal fractal structures, for which
\begin{equation}
\Gamma^*(r)=\frac{3}{D}\Gamma (r)
\end{equation}

\subsubsection { $\Gamma$-function estimator.}

Consider a stochastic fractal process where
the number of particles $N_V(\vec{x}_a, r)$
in a sphere of radius $r$, centered at the point
$\vec{x}_a$ and the number $N_S(\vec{x}_a, r)$ of
particles in the shell  ($r, r+\triangle r$)
are given by eqs.\ref{N-x-r} and \ref{S-x-r}.
Taking into account definitions of conditional densities
(eqs. \ref{gamma-r} and \ref{gamma-V-r}) one can use following two
statistics for their estimation from one realization (a finite
galaxy sample):

\begin{equation}\label{S-eta-est}
  \hat{\Gamma} (r) =
  \frac{1}{N}\sum_{a=1}^N \frac{1}{4\pi r^2 \triangle r}
  \int_r^{r+\triangle r} n(\vec{x_a} + \vec{x})4 \pi x^2 dx
%  \frac{\mathcal{D}B}{4\pi}~r^{-(3-\mathcal{D})}
\end{equation}
for the shell conditional density $\Gamma$, and

\begin {equation}\label{V-eta-est}
  \hat{\Gamma}^*(r) =
   \frac{1}{N}\sum_{a=1}^N \frac{3}{4\pi r^3}
  \int_0^{r} n(\vec{x_a} + \vec{x})4 \pi r^2dx
%= \frac{3B}{4\pi}~r^{-(3-\mathcal{D})}
\end{equation}
for the volume conditional density $\Gamma^*$.
So the use of the conditional density method is in principle
quite simple, just counting the number of particles
inside the spherical
volume $V(r)$ or inside the shell $S(r) \Delta r$. This
is done
for each structure point and then the average is
calculated. For the $\Gamma$-function estimation
one need not generate artificial Poisson distributions,
which was necessary for the $\xi$-function method.

{\it Fractal dimension and co-dimension. }
For a fractal structure both the $\Gamma$ function (eq.\ref{gamma-r})
and the estimator
(eq.\ref{S-eta-est}) have a power-law form
\be
\label{est-gam} 
\hat{\Gamma} (r) = \Gamma_0 r^{-\gamma}
\ee
This very important property of the $\Gamma$-estimator allows one to
obtain an undistorted value of the fractal dimension in a galaxy sample.

The exponent that defines the decay
of the conditional density 
\be
\label{gamma}
\gamma = D - 3
\ee
is called the \emph{co-dimension}, where $D$ is the fractal
dimension (or the correlation dimension $D_2$).
The amplitude $\Gamma_0$ of the estimator $\hat{\Gamma} (r)$ does not 
change when the sample volume $V_s$ is increased,
only the range of available scales $r$ increases. This corresponds to
the meaning of $\Gamma (r)$ as characterizing the number density behaviour.

{\it Homogeneity scale. }
For a fractal structure which has an upper cutoff at a
homogeneity scale $R_{\mathrm{hom}}$,
after which the distribution becomes uniform,
the fractal dimension $D=3$, and the estimator of the $\Gamma$-function is 
\be \label{Rhom}
\hat{\Gamma}(r) =  \mathrm{constant}, \,\,\,\,\mathrm{for}\,\,\,r>
R_{\mathrm{hom}}\,\,.
\ee
Thus the method of
conditional density is a powerful instrument when one
searches for the crossover from the regime of fractal clustering
to the realm of homogeneity.

So, for processes with a finite upper cutoff scale  of fractality,
beyond which the distribution of particles turns into
homogeneity,
the statistics
(\ref{S-eta-est}) and  (\ref{V-eta-est}) give constant
values corresponding
to the fractal dimension  $D =$ 3 as it indeed should
be for uniform structures.

\subsubsection {Redshift-space $\Gamma_z(s)$ and  $v_{pec}$}

As we discussed in sec.3.2.5 an estimation of the spatial
distribution of galaxies from redshift catalogues is based on
the redshift-space ($\alpha$ , $\delta$, $z_{obs}$).
Using the same notations as in sec.3.2.5 for the $\xi$ function ($s$, 
$r_{\perp}$, $r_{\parallel}$)
we can right for the redshift-space conditional density
\be \label{gam-s} 
\Gamma_z(s) = \Gamma_z(r_{\perp}, r_{\parallel \mathrm{obs}} )\,\,.
\ee

The relation between the real-space $\Gamma(r)$ and redshift-space
$\Gamma_z(s)$ conditional density is
\be \label{gam-s-gam-r}
\Gamma_z(r_{\perp}, r_{\parallel \mathrm{obs}}) =
\int g(\vec{r},\,\vec{w}) \Gamma (r) d^3 w \,\,,
\ee
where $\vec{w}= \vec{v}_1 - \vec{v}_2$ is the relative peculiar
velocity of a galaxy pair at separation $\vec{r}$,
and $g(\vec{r},\,\vec{w})$ is the relative peculiar velocity distribution.
Here the components of the relative distance $\vec{r}$ are given by
the following formulae:
$r_1 = r_{\perp}$, $r_3=r_{\parallel \mathrm{obs}} - w_3/H_0$, and
$r^2 = r_{\perp}^2 + (r_{\parallel \mathrm{obs}} - w_3/H_0)^2$.

In order to restore the real-space conditional density from the directly
observed redshift-space conditional density it is necessary
to make computer simulations of artificial fractal structures
with known  peculiar velocity fields and then compare
the modelled redshift-space $\Gamma_{\mathrm{mod}}(s)$
 with the observed $\Gamma_z(s)$.

\subsubsection {$\Gamma$-function for 2-d intersections }

If in 3-d space a fractal structure is intersected by 
a plane then the expected value of the fractal dimension
for the intersection is given by eq.\ref{Dint}:
\be
D_{\mathrm{int}} = D - 1
\ee
To make the $\Gamma$-function analysis for the sample which
presented the 2-d intersection we shall use the 2-d coordinate system
$\vec{y}=(y_1,\, y_2)$ for which we can calculate
the $\Gamma$-function for the intersection  $\Gamma_{\mathrm{int}}(y)$. 

Such a situation may occur in a slice-like galaxy
survey for scales $r$ larger than the thickness of the survey.
E.g. for the true fractal dimension $D=2$ the  
fractal dimension of the intersection will be $ D_{\mathrm{int}} = 1$.
Hence we expect to obtain a power-law
behaviour of the corresponding $\Gamma_{\mathrm{int}}(y) \propto y^{-1}$.
We will see below that the intersection theorem helps one to understand
the behaviour of the power-spectrum derived from
slice-like galaxy surveys.

\subsection {Comparison of correlation function and
conditional density}

\subsubsection { The relation between $\Gamma$ and
$\xi$.  }
From the definitions of the conditional density $\Gamma$
(eq.\ref{eta_autocor})
and correlation function (eq.\ref{ksi2}) we have the
relation:
\be
\xi (r) =\frac{\Gamma(r)}{n_0} - 1
\label{ksi-eta}
\ee
if the average number density $n_0$ of a considered
stochastic process exists.
Both $\Gamma$ and $\xi$ functions are
conditional characteristics of a stochastic process, i.e. they are
defined on the condition that the
centres of counting spheres are set to structure particles.
However, there is still
a deep difference between them. The $\Gamma(r)$ represents
the \emph{complete} correlation function,
while the $\xi(r)$ represents the \emph{reduced} correlation function
of the stochastic process. This fact makes the properties of the
corresponding $\xi$ and $\Gamma$ estimators very different. Finally,
it results in
conflicting values for the estimated correlation exponent and homogeneity
scale of a galaxy sample.

From eq.\ref{ksi-eta} follows a similar relation between the estimators
for a finite sample of galaxies:
\be
\hat{\xi}_{FS} (r) =\frac{\hat{\Gamma}_{FS}(r)}{\bar{n}} - 1
\label{ksi-eta2}
\ee
Here $\hat{\xi}_{FS}$ is called the ``full shell'' estimator because
it is defined through the $\Gamma$ estimator which is calculated using
full shells completely embedded in the sample volume.

The estimator $\hat{\Gamma}$ (eq.\ref{S-eta-est}) is always a positive
function and has a power-law form for fractal
structures. On the contrary, the estimator $\hat{\xi}$ (eq.\ref{ksi3})
inevitably changes its sign and hence cannot
be presented as
a power law even for scale invariant structures.
All estimators of the $\xi$-function, which are based on
counting of pairs relative to an artificial Poisson distribution,
have a common drawback. They
give essentially distorted values for the true correlation
exponent of the complete correlation function
of long range power-law correlated processes
and for fractal distributions.
But the $\Gamma$-function estimator is specially constructed in order to
give undistorted values of the correlation exponent and
fractal dimension.

The $\Gamma$-function estimator relates to
\emph{intrinsic
properties} of the sample, while the $\xi$-function estimator
depends on both intrinsic
and external properties of the sample. In fact,
$\hat{\Gamma}$  measures
the behaviour of the total density inside spheres within
a sample,
while $\hat{\xi}$  measures density fluctuations relative to
the average density, which is assumed to be valid for
all space outside a finite sample. 
This can be illustrated also by the following reasoning.
Let us consider counts around a fixed point. The
expected number of points in
a shell with radius $r$ and volume $dV$ is $n(r)dV$,
where
$n(r)$ is the conditional density $\Gamma (r)$ describing
the density--radius
law. On the other hand, the same expected number may be
calculated
with the correlation function $\xi (r)$ as $n_0 (1 +
\xi (r))dV$.
So
\be
\Gamma(r) =(1 + \xi (r)) n_0
\label{eta-ksi}
\ee
It is important to note that the right-hand side of
eq.\ref{eta-ksi}
becomes defined only after the mean density $n_0$ is
calculated for
the whole sample, while $\Gamma(r)$ always exists
locally. Remember that $\xi(r)$ was defined for fluctuations
around the mean $n_0$.

{\it Fractal density field.  }
For the case of a scale-invariant stochastic
fractal density field, 
the complete correlation function $\Gamma(r)$ has the power-law form
\be
\Gamma(r) = \frac{BD}{4\pi} \;  r^{-\gamma} \,\, ,
\label{eta_power}
\ee
while for the same fractal structure the reduced correlation
function $\xi (r)$ will be
\be
\xi(r) = \frac{BD}{4\pi
 n_0}
     \; r^{-\gamma} -1
\label{ksi_power}
\ee
which is not a power-law.
This difference between complete and reduced correlation functions was
pointed out by Pietronero \& Kuper (1986).
 
Thus the $\xi$-function may be approximated by a power-law only for such $r$ 
when $\xi (r) \gg 1$, which corresponds to small scales
$r<<r_0$. However, on small scales the noise of discreteness
is essential, which also leads to distortion of a true
power-law. Hence a $\xi$-function estimation gives a distorted
values of the correlation exponent not only on large scales
(normalization), but also on small scales (discreteness).

It is instructive to calculate the exponent
$\gamma_{\xi}$ of
 the correlation function
(eq.\ref{ksi_power}) on scales close to the unit
scale $r_0$. Taking the logarithmic derivative of eq.\ref{ksi_power}
one obtains
(Joyce, Montuori, Sylos Labini 1999):

\be
\gamma_{\xi}(r) = - \frac{d[\log \xi (r)]}{d \log r} =
                \frac{2\gamma (r/ r_0)^{-\gamma}} {2
(r/r_0)^{-\gamma} -1}
\label{gamma_ksi}
\ee
Therefore at the unit scale $r =
r_0$ we get the
remarkable result:
\be
\gamma_{\xi} (r_0) = 2 \gamma
\label{gamma_ksi_r0}
\ee
For example, for a true density power-law with $\gamma =
1$, one would
infer an apparent slope $\gamma_{\xi} = 2$ for the
correlation $\xi$-function,
if measured at scales close to the ``correlation length''
$r_0$!

\subsubsection { The dependence of $r_0$ on sample depth.   }

The sample depth $R_s$
%the mean separation $\bar{d}$, and the luminosity
%$L$ are major 
is an important global parameter of an observed galaxy
distribution. 
For a fractal structure sampled inside a
spherical volume
$V_s = \frac{4\pi}{3} R_s^3$ with 
$\bar{n} = N_s/V_s = 3BR_s^{D-3}/4\pi $,
the eq.\ref{ksi_power} yields
\be
\hat{\xi} (r) = \frac{D}{3}
\left(\frac{r}{R_s}\right)^{-\gamma} -1\,\,,
\label{ksi_Rs}
\ee
and the unit scale $r_0$ comes to depend on the sample parameters.

%{\it The sample depth $R_s$ and $r_0$. }
%Let us consider the ideal spherical case.
Inserting $r = r_0$ into eq.\ref{ksi-eta} and taking
into account
that $\xi (r = r_0) = 1$ one gets
\be
r_0 = \left(\frac{DB}{8\pi \bar{n}}\right)^{1/\gamma}
\label{r0_n}
\ee
where $\bar{n}$ is the average number density of objects
in the sample.
Hence one obtains:
\be
\Gamma (r_0) = \frac{DB}{4\pi} \;  r_0^{D-3} =  2\bar{n}
=
  \frac{6B}{4\pi} R_s^{D-3}
\ee
From this follows a simple
relation between the correlation length $r_0$ and the
depth of the sample
$R_s$:
\be
r_0 =\left(\frac{3-\gamma}{6}\right)^{1/\gamma} R_s
\label{r_0-R_s}
\ee
We note again that this is true for a spherical sample 
(i.e. $R_s=R_{max}^{sph}$)
and a fixed luminosity (i.e. $L=const$)
of sample galaxies.

\subsubsection {Geometry of a survey and characteristic scales. }
A strong restriction for the practical
application of the $\Gamma$-function  method is the requirement that
there
should be room for the whole sphere in the volume of the
considered sample.
For example, for galaxy surveys with slice-like
geometry,
this makes it impossible to measure the conditional
density
for scales larger than the thickness of the slice, i.e.
the diameter of the maximum sphere fully contained in
the survey volume.

 The above dependencies between
$r_0$ and
various sample parameters were derived for the ideal case of a
simple fractal in
a  sufficiently large
spherical volume inside which the correlation function
is estimated.
For a non-spherical survey geometry (like a slice
or cone) these relations are
valid only for such scales $r$ for which the survey galaxies
are contained completely within
the sphere with the radius $r$. For non-spherical
geometries these dependencies
are expected to differ from the above predictions.

For a galaxy sample under consideration one should always
control the following characteristic distance scales:
\be \label{R-scales}
R_{sep},\,\,R_{max}^{sph},\,\, R_s\,.
\ee

The separation distance $R_{sep}$ between galaxies in a sample
may be roughly estimated as $\bar{n}^{-1/3}$ or calculated
from the nearest neighbours distribution.
We may define $R_{sep}=\alpha R_{sep-av}$, where
$\alpha \approx 0.1$ and $R_{sep-av}$ is the average distance
between nearest neighbours in a sample.
For $r<R_{sep}$ the discreteness noise is important.

The radius of the maximum sphere $R_{max}^{sph}$ is related
to the complete embedded sphere in the sample and play a crucial
role in estimation of correlation properties of the galaxies
from the sample.
The depth $R_s$ of a survey is related to the largest distances
of galaxies in the sample, it  essentially differs from the
radius of the maximal sphere in the case of a slice-like surveys.

As a general rule for a correlation analysis of slice-like samples
one should  consider separately two regions of scales. First,
\be
R_{sep} < r < R_{max}^{sph} \,\, ,
\ee
where it is possible to use the $\Gamma$ function method and
estimate the value of the true fractal dimension, and secondly,
\be
R_{max}^{sph} < r < R_s \,\, ,
\ee
where the $\Gamma$ function cannot be applied. In this region other
methods should be developed such as the $\Gamma$ function for fractal
intersections and the two-point conditional column density.

\begin {figure}[th]
\centerline {\psfig {file=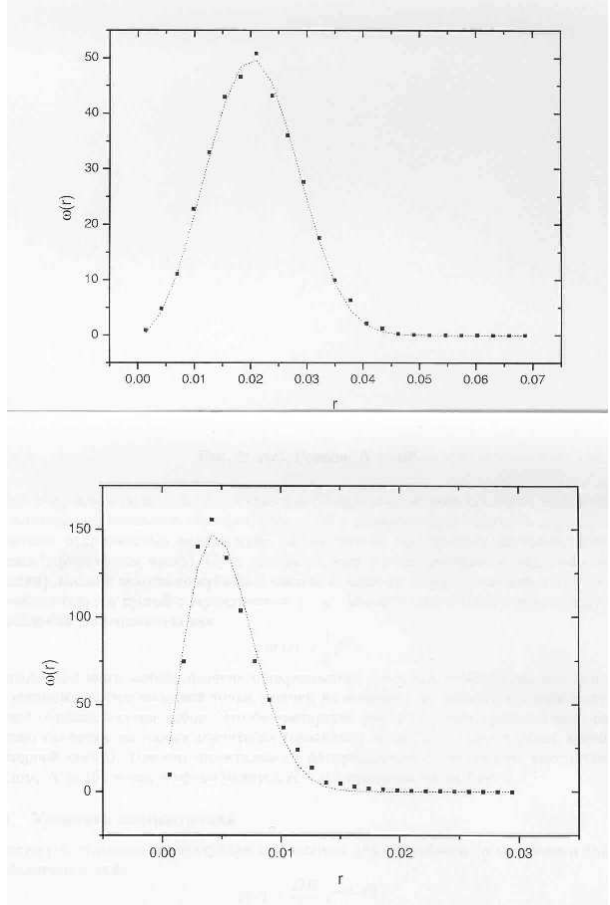,width=8cm}}
%\resizebox{\hsize}{!}{\includegraphics{fig3.ps}}
\caption{ \footnotesize
Probability distributions for finding the
nearest neighbour at
a distance $r$ for two cases:(top) the homogeneous Poisson
distribution
($D=3$) and (bottom) a
fractal distribution with $D=2$
(a random Cantor set). The distributions are in 3-d space
inside a sphere of
diameter = 1, the total number of points in each case is
$N = 2.5 \cdot 10^3$. The dotted curves
correspond to theoretical predictions according to eqs. 99, 100
 (from Vasiliev 2004).
}
\label {neighbours}
\end {figure}

\subsection{ Nearest neighbours 
distribution. }

The distribution of distances to the nearest neighbour
point is a useful property of fractals, which
may be used to make a distinction between fractal and ordinary
distributions.
It is discussed in detail by Gabrielli et al. (2004).

{\it Poisson distribution. }
For example, for a Poisson process in 3-d space
the probability density distribution $\omega (r)$
for finding the nearest
neighbour at
a distance $r$ is
\be
\label{nn_Poisson}
\omega (r)  = 4 \pi n_0 r^2
\exp \left(
-\frac{4\pi n_0 r^3}{3}
\right) \;.
\ee
For Poisson process the average distance between
point-particles is $R_{sep} \approx n_o^{-1/3}$

{\it Fractal distribution. }
There is no exact formulae for the general case 
of a fractal structure. However,
Gabrielli et al. (2004) derived a useful approximation for the
probability density distribution for the nearest neighbour in 
a fractal structure:
\be
\label{nn_D}
\omega (r)  = 4 \pi C r^{D-1}
\exp \left(
-\frac{4\pi C r^{D}}{D}
\right) \;,
\ee
where $C=DB/4\pi $.

Figure \ref{neighbours}
presents $\omega (r)$  for two cases: a homogeneous Poisson
distribution
($D=3$) and a fractal distribution with $D=2$.
The dotted curves correspond to theoretical
predictions from eq.\ref{nn_Poisson} and eq.\ref{nn_D}.
It is seen that the lower is the value of the fractal dimension $D$,
the closer are the neighbours within the fractal structure.
This example shows that the nearest neighbour probability
function may be used as an additional method for
estimating the fractal dimension of a galaxy sample. 

\subsection {Two-point conditional column density}

\subsubsection {Definitions}

The  conditional densities of stochastic fractal
processes above
discussed were one-point, as the centre of the sphere in
which
particles are counted lies at one fixed point $\{a\}$ with the
coordinates
$\{\vec{x}_a\}$ and the counts are made around each point $\{a\}$
of the sample. However, in some cosmological studies, e.g.
related
to gravitational lensing (Baryshev \& Ezova 1997), it
becomes necessary to use two-point
conditional densities, whereby one simultaneously fixes two particles
 $\{a,~b\}$ with the coordinates
$\{\vec{x}_a,~\vec{x}_b\}$ and counts galaxies within a
thin cylinder
between these points.

In order to define the distribution of particles along
such a cylinder,
whose axis connects two structure points
$\{a,~b\}\subset
\{\vec{x}_i\,,\,i=1,...,N\}$, Baryshev \& Bukhmastova (2004)
introduced
the concept of 2-point conditional
column density $\eta_{ab}(r)$ for a stochastic fractal
process.
In applications to galaxies,
according to the cosmological principle of Mandelbrot
the particles
$a$ and $b$ are statistically equivalent, hence for each
of them
the 1-point conditional density is given by the
expression
(\ref{gamma-r}), which is proportional to the probability
of finding particles at the distance $r$ from fixed
structure points.

Let us now take two independent points of the structure
at the distance $r_{ab} =|\vec{x}_a - \vec{x}_b|$ from each
other. Denote by $C$ the event, when particles appear
at the distance $r_a$ from $a$ and independently at the
distance
$r_b$  from $b$. Then $C$ is given by the union
$C=A\bigcup B$ of the two events related to $a$ and
$b$. So the 2-point conditional column density, which is
proportional to
the probability of finding particles around the
particles $a$ and $b$,
may be presented
as a sum of 1-point conditional densities.
The assumption of independent events is a first
approximation.
Numerical simulations have shown that
this approximation is sufficient for the analysis of
usual fractal structures
(Baryshev \& Bukhmastova 2004).
Then the 2-point conditional
column density can be expressed as

$$ \eta_{ab}(r) =
\frac{1}{2} \left[
 \Gamma_a(r) + \Gamma_b(r_{ab} - r) \right] $$

\begin{equation}\label{eta-ab-r}
 = \frac{DB}{8\pi}\cdot~r_{ab}^{-\gamma}\cdot
 \left[\left(\frac{r}{r_{ab}}\right)^{-\gamma} +
 \left(1-\frac{r}{r_{ab}}\right)^{-\gamma}\right]\,,
\end{equation}
where $\gamma = \mathrm{D} -3$.  The distance $r$ is
measured
along the segment of the line
connecting the particles $a$ and $b$, and at the same
time it defines
the radius $r$ of the sphere having its centre at the
first point and
the radius $r_{ab}-r$ for another sphere having its centre
at the second point.
The constant $B$ defines the normalization.
We note that in this formula the volume elements
are taken along the line connecting the two points -- in
this sense the
coordinate $r$ is not a spherical radial coordinate, but
a Cartesian coordinate
labelling volume elements (``tablets'')  with thickness $dr$ along this
cylinder.

\begin {figure}[th]
\centerline {\psfig {file=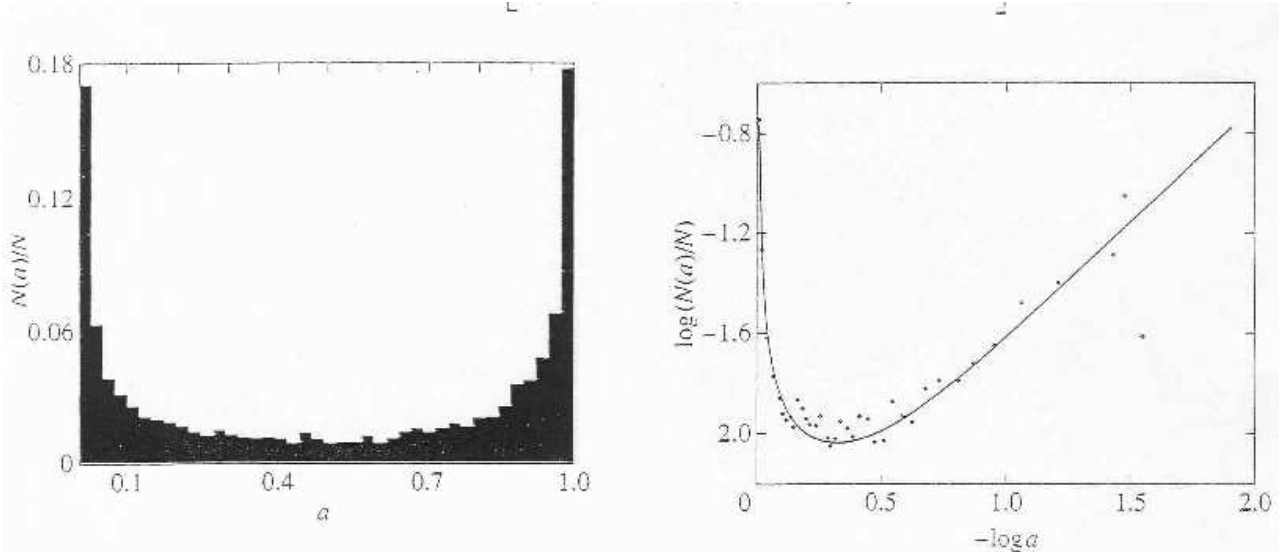,width=17cm}}
%\resizebox{\hsize}{!}{\includegraphics{fig6.ps}}
\caption{\footnotesize
(left) The observed distribution of the galaxy
two-point conditional column density
$\hat{\eta}_{ab}(r)$ of the
LEDA sample along cylinders with $l < 100$ Mpc.
(right) The same distribution in
logarithmic scale and the fitted theoretical curve
$\eta_{ab}(r)$, with
the fractal dimension $D = 2.02$
(from Baryshev and Bukhmastova 2004).}
\label{2pointlineleda}
\end {figure}

\subsubsection {Estimation}
In order to estimate $\eta_{ab}(r)$ one may use the
statistic

$$  \hat{\eta}_{ab}(r) =
\left\langle \frac{N_c(\vec{x}_a,
\vec{x}_b,~r,~h,~\triangle r)}
  {\pi h^2\triangle r}\right\rangle _{\{a,b\}}$$
\begin{equation}\label{eta-ab-r-stat}
 =
  \frac{1}{N_{ab}}\sum_{\{a,b\}}^{N_{ab}} \frac{1}{\pi
h^2 \triangle r}
  \int_r^{r+\triangle r}\int_0^h n(\vec{x})2\pi h
~dh~dr\,,
\end{equation}
where $N_c$ is the number of particles in a volume
element of
the cylinder with a diameter
$h$ and height $\triangle r$ and whose axis connects the
structure
particles $a$ and $b$. The volume element lies at the
distance
$r$ from $a$, which corresponds to the distance $r_{ab}
-r$ from $b$.
Averaging is performed for every pair of particles with
connecting
cylinders having the length  in the interval $(l, l+
\Delta l)$.
The parameters $h$ and $\Delta r$ are chosen to be a
fraction of
the mean separation of the particles in the sample.
Practical modelling of artificial fractals
show that the estimated fractal dimension is robust to
reasonable
variations of the tablet size
(Baryshev \& Bukhmastova 2004; Vasiliev 2004).

A more general situation, which may be encountered in
practice, is
that both a fractal structure and a homogeneous
background exist
in the sample.Then
the practical calculation
of the fractal dimension $D$ by means of the
estimator $\eta_{ab}(r)$ may be done  by fitting through
the observations
the dimensionless probability distribution
with three free parameters $\gamma,\, R_1,\,R_2$
\begin{equation}
\frac{N(y)}{N} = R_1 \cdot \frac{[y^{-\gamma} +
(1 - y)^{-\gamma}]}{2} + R_2
\end{equation}
where
 $N(y)$ is the observed number of points which
are found in each tablet, i.e. within the small
intervals $(y, y + \Delta y)$
along the cylinder with a length $l$. The variable $y$
is the relative distance measured along the line
connecting the
two points ($y = r/r_{ab} = r/l$). $N$ is the total
number of points within
cylinders of length $l$. In the case of genuine self-
similar structures it is
possible to calculate these numbers for all cylinders of
different
length simultaneously. $R_2$ shows the contribution of
a possible Poisson background.

It is useful to introduce a measure of the relative
strength $\beta$ of the
fractal component as was done by Vasiliev (2004):
\be
\beta = \frac{1-R_2}{R_2} \; .
\label{frac_hom_rel}
\ee
When $\beta =1$, the contribution from a fractal
structure is equal to
the Poisson background contribution ($R_2 = 0.5$).

The main advantage of the cylinder method is that it can
work
for slice-like surveys allowing the study of scales
comparable
to the depth of the survey.

\subsection {Fourier analysis of the galaxy distribution}

\subsubsection { Ordinary density fields }
If the spatial distribution of objects is given
by a stochastic
density field, then fluctuations of this field may be
represented
by the Fourier integral as a superposition of plane
spatial waves with
the wave number $k = |\vec{k}|  = 2\pi/\lambda$:
\begin{equation}\label{fourier}
  \delta(\vec{x}) = \frac{\varrho(\vec{x})-
\bar{\varrho}}{\bar{\varrho}}=
  \frac{1}{(2\pi )^3}\int \tilde{\delta}(\vec{k})
  e^{-i\vec{k}\cdot\vec{x}}d^3k\,
\end{equation}
Here the Fourier transform $\tilde{\delta}(\vec{k})$
of the density fluctuations $\delta(\vec{x})$  is a
complex quantity
and thus may be given in the form

\begin{equation}\label{am-phase}
  \tilde{\delta}(\vec{k}) = |\tilde{\delta}(\vec{k})|
  \exp (i\phi(\vec{k}))\,.
\end{equation}
Here we see that for a complete description of the
spatial distribution of
objects the analysis should include both the amplitude
spectrum
 $|\tilde{\delta}(\vec{k})|$ and the phase spectrum
$\phi(\vec{k})$.

In the case of a Gaussian random process the phases of
plane waves are
distributed uniformly in the interval
$[0,\,2\pi ]$ and to describe the density field it is
sufficient to
consider only the power spectrum

\begin{equation}\label{P-k}
  P(\vec{k})=\langle |\tilde{\delta}(\vec{k})|^2
\rangle\,.
\end{equation}
When the distribution is isotropic, the power spectrum
and the correlation function are connected by the
relation

\begin{equation}\label{P-ksi}
  P(k)= 4\pi \int \xi(r) \frac{\sin(kr)}{kr}r^2dr\,.
\end{equation}
Thus for a power law correlation function
$\xi(r)\propto
r^{-\gamma}$ the power spectrum has also a power law
form  $P(k)\propto
k^{\gamma -3}$ for restricted range of scales $k_1<k<k_2$.
For small $k$ (large scales $\lambda > R_s$) there is a limit
due to the size of a survey $R_s$, which give 
$P(k) \rightarrow 0$ for $\lambda \rightarrow R_s$.

The phase spectrum $\phi (\vec{k})$ carries important
information,
complementary to $P(k)$. Possible non-Gaussianity of
the clustering properties i.e. non-uniformity of the phase
spectrum was considered in
Chiang \& Coles 2000 and Coles \& Chang 2000.

The main problem of the power spectrum analysis
is the same as for the correlation function: the average
density
$\bar{\rho}$ should be well defined within the sample
volume.

However an advantage of the $P(k)$-method is that the
correlation exponent $\gamma$ may be estimated without distortion
from $P(k)$ for the scales $\lambda < R_{max}^{sph}$.

\subsubsection { Fractal density fields }

For stochastic fractal processes one may perform the
Fourier-analysis
of fractal density field ( Sylos-Labini \&
Amendola 1996, Sylos Labini et al. 1998).
They introduced the concept of generalized power
spectrum  $\Pi (k)$ of a fractal process, which is
defined by the Fourier transformation of the
conditional density $\Gamma(r)$:
\begin{equation}\label{Pi-ksi}
  \Pi(k)= 4\pi \int \Gamma(r) \frac{\sin(kr)}{kr}r^2dr\,.
\end{equation}
Because of a power-law form $\Gamma(r)\propto
r^{-(3-D)}$ for
the conditional density of a stochastic fractal
process, we also get for
the generalized power spectrum a power law
$\Pi(k)\propto k^{-D}$. Here $D$ is
the fractal
dimension, which hence may be directly measured from the
power spectrum.
The generalized power spectrum is a suitable tool
for describing 
highly irregular systems with fractal properties.

\subsubsection {Role of geometry of a sample}

In the case of a slice-like survey for scales larger than
maximum sphere completely embedded in the sample,
the geometry of considered structure become effectively
be close to the case of an intersection of a 3-d fractal
structure by a plane. Hence according to the theorem on
intersections of fractals (sec.2.4.5) it is expected that
for a structure with the true fractal dimension $D$
the measured fractal dimension of the sample will be
$D_{int}=D-1$. It means that there are three characteristic 
intervals of scales for behaviour of  power spectra $P(k)$ or $\Pi(k)$:
1) for scales $\lambda < R_{max}^{sph}$ we get $P(k) \propto k^{-D}$
with exponent equal the true fractal dimension;
2) for scales $ R_{max}^{sph} < \lambda < R_s$  
we get $P(k) \propto k^{-(D-1)}$ with exponent equal the fractal
dimension of intersection; and
3) for scales $\lambda > R_s$ we get $P(k) \rightarrow 0$ due to
a finite size of the sample.

\subsection {Multifractals and luminosity function}

\subsubsection {Spectrum of fractal dimensions}

There are several different definitions for the fractal
dimension:
Hausdorf dimension $D_H$, box or capacity dimension $D_b$,
correlation dimension
$D_2$, mass-length dimension $D_m$
(for brief descriptions for astronomers see e.g. Gabrielli et al. 2004;
Martinez \& Saar 2002). For self-similar distributions
all these dimensions are usually
equal (and are $\leq d$, the dimension of the embedding space).
In such simple cases
we shall use the symbol $D$ for this common fractal dimension.
For multifractal analysis 
which operates with a whole spectrum of fractal dimensions
all dimensions should be considered separately.

We have already encountered the mass-length fractal dimension $D_m$
in the Fournier world model which was constructed so that
$M(r) \propto r$ and hence $D_m=1$. It was the first mathematical
model of a regular fractal structure applied to the whole universe.

The box dimension or capacity of a set $S \subset \Re^p$ 
is defined as
\begin{equation}
\label{box-dim}
D_b = \lim_{\epsilon \rightarrow 0} 
\frac{\log N(\epsilon)}
{\log (1/\epsilon)}
\end{equation}
where $N(\epsilon)$ is the minimum number of 
p-dimensional boxes of
size $\epsilon$ needed to cover completely the set $S$.

In the case of spatial galaxy distribution
$p=3$ and one should extract from a sample
the cube (3-d box) within which to perform the calculation
of the box dimension $D_b$.
The slope of the plot of $\log N(\epsilon)$ versus
$\log (1/\epsilon)$ gives an estimation of the box dimension.

The correlation dimension $D_2$ is defined via the correlation
exponent $\gamma$ in the complete correlation function or conditional
density $\Gamma$.
Our discussion in the previous
section
was intended to distinguish between homogeneity and scale
invariant
properties and, for this purpose, it remains perfectly
appropriate even if
the galaxy distribution were multifractal.
In this case the correlation functions we have
considered
would correspond to a single exponent of the
multifractal spectrum,
but the issue of homogeneity versus scale invariance
(fractal or multifractal) is exactly the same.

We emphasize also that
a discussion in the literature
around the multifractal properties of galaxy distribution
had its origin in the difference between
the values of the fractal dimension obtained by two methods:
box-counting and correlation function
(Jones et al. 1988; Martinez \& Jones 1991; Martinez \& Saar 2002).
However, as we have seen above, the difference was caused by
the use of the  $\xi$-function estimator which gave a distorted value
for the correlation dimension $D_2$.
If one uses the appropriate $\Gamma$-function estimator
then the unique fractal dimension $D \approx 2$  well fits
observational data and hence eliminates the claimed need 
for that kind of multifractality.

\subsubsection {Luminosity function}
Above we considered fractal structures made of identical
particles with unit mass.
In connection with galaxies, Sylos Labini et al. (1998)
and Gabrielli et al. (2004)
use the term multifractality in
the case where the so-called \emph{fractal support}
(positions of particles) has a unique dimension.
In this case a stochastic process labels the fractal support particles
by different values of some
random quantity $\mu$ (for example luminosity $L$ or mass $m$).

Real galaxies have different masses and different
luminosities which
may be characterized by a luminosity function $\phi(L)$.
This is usually
represented by Schechter's law

\begin{equation}\label{phi-L}
  \phi (L)~dL=\phi^* \cdot (L/L^*)^{\alpha}\cdot
  exp(-L/L^*)~dL\,,
\end{equation}
which gives the fraction of galaxies in the unit volume
in the luminosity interval $(L,~L+dL)$.
The parameters $\alpha$ and $L^*$ are to be determined
from observations and
$\phi^*$ is the normalizing coefficient from the
condition $\int_{\beta}^{\infty}
\phi (L)~dL=1$, so that

\begin{equation}\label{phi-*}
  \phi^* =\frac{\Gamma (\alpha+1,~\beta)}{L^*}\,.
\end{equation}
Here  $\Gamma(n,~x)$ is the incomplete gamma function
and  $\beta$ is a
parameter which determines the luminosity truncation
in the faint tail.

\subsubsection {Space-luminosity correlation}
Let us consider a realization of
a stochastic process for which one may define
the luminosity (or mass) density $\mu$ in the form

\begin{equation}\label{mu-x}
  \mu(\vec{x}) = \sum_{i=1}^N \mu_i ~ \delta (\vec{x}-
\vec{x}_i)\,.
\end{equation}
where $\mu_i$ is the luminosity (or mass) of $i$-th particle.

Pietronero (1987), Coleman \& Pietronero (1992)
%\cite{coleman92},
and Sylos-Labini \& Pietronero (1996)
% \cite{sylos96} 
regard luminosity or mass density functions
as multifractal measures on the set of
realizations. Multifractals are characterized by the
spectrum of
fractal dimensions
 $D(L)$ which determines the dependence of the
fractal
dimension on luminosity or mass. This means that
luminous and faint
galaxies may have different spatial distributions.
One prediction of the multifractal model is that the
fractal
dimension decreases for increasingly luminous objects.

Let $N_{L,S}(\vec{x}_a,~L,~r)$ be the number of galaxies
having the
 luminosity in the interval $(L,~ L+\triangle L)$ in a
spherical shell
$S(r)=4\pi r^2\triangle r $ with its centre at the point
$\vec{x}_a$ belonging to the structure.
In order to generalize the concept of conditional
density 
to the case of particles of different luminosity or
mass, we define the conditional luminosity (mass)
density as

\begin{equation}
  \nu(L,r) =\left\langle \frac{N_{L,S}(\vec{x}_a, L,
r)}{4\pi
  r^2\triangle r~\triangle L}\right\rangle _{\vec{x}_a}
=
\end{equation}
\begin{equation}\label{nu-L-r}
 = \frac{1}{N}\sum_{i=1}^{N} \frac{1}{4\pi r^2 \triangle
r~\triangle L}
  \int_L^{L+\triangle L}
  \int_r^{r+\triangle r} \mu(\vec{x})d^3x~dL\,.
\end{equation}
As shown by Sylos-Labini \& Pietronero (1996)% \cite{sylos96},
for a wide class of multifractal stochastic processes
the conditional
luminosity density has the general form

\be
  \nu (L,r) =\phi_{_r}(L)\cdot\Gamma_{_L}(r) =
\ee
\be \label{nu-phi-eta}
  \phi^* \cdot \left(\frac{L}{L^*(r)}\right)^{-
\alpha}\cdot
 \exp (L/L^*(r))\cdot
  \frac{D(L)B}{4\pi}\cdot~r^{D(L)- 3}\,.
\ee
This means that the Schechter law for the luminosity
function
is an observable consequence of multifractality and not
just a convenient analytical form!

Special features of the expression
(\ref{nu-phi-eta}) are the dependence of the ``knee''
of
the LF, $L^*(r)$,
on the radius of the volume
and also the dependence of the fractal dimension on
luminosity
$D(L)$. These properties may be used for
testing multifractality.
However, this needs large samples of galaxies, because
fractal analysis
must be made for each luminosity interval separately.
If the distribution is multifractal then
 the brightest luminosity $\:L_{\mathrm{max}}$
in a sample is related to its spherical depth $\:R_{s}$ by the
relation
(Coleman \& Pietronero 1992)% \cite{coleman92}
\be
\label{l-max}
L_{\mathrm{max}} \propto R_s^{\beta}\,\,,
\ee
where the exponent $\beta$ depends on multifractal
spectrum.

\newpage

\section {The epoch of angular-position galaxy catalogues}

The history of main events illuminating
the fractal debate in the 20th
century is presented in the Tab.\ref{debate}
"The debate
on large scale fractality".
This subject has always been at the centre of
cosmological thinking, even
sharp debates,
because of its direct link to cosmological principles.
Of course, inhomogeneities in our neighbourhood are
evident and the border
to the assumed uniformity must lie somewhere at a larger
distance.
During the history the border to uniformity has
gradually shifted outwards
from the crystal sphere of fixed stars to Newton's
evenly scattered stars, and
then to
Hubble's uniform galaxy distribution. At our times
the observed superclusters of galaxies have put the
border of uniformity
to a scale of at least 100 Mpc.

\begin{table}[h!]
\label{debate}
\caption{
The history of the debate on large scale
fractality.}
%\vskip 0.2cm
\begin {tabular}{|c|c|c|}
% \hline
%\multicolumn{3}{|c|}{\emph{}} \\
 \hline
%& & \\
Years & Authors & Subject  \\
%& & \\
 \hline
%& & \\
1900 --   & Fournier d'Albe & regular hierarchical
models  \\
1920s   &   Charlier, Selety, Lundmark & criteria for
infinite world   \\
paradoxes   & Einstein, Selety & Mach, stability, middle
point  \\
\hline
        & Shapley, Zwicky, Abell & strong galaxy
clustering  \\
1930s --        & Carpenter, Kiang, Karachentsev &
superclusters up to 100 Mpc    \\
1970s        & Neyman, Scott & 2-level hierarchical
model   \\
  & de Vaucouleurs & cosmic density-radius law  \\
clusters  &             & $\rho (r) \propto r^{-\gamma}$
with $\gamma = 1.7$  \\
and       & -----------------------------------
& --------------------------------------  \\
uniformity & Hubble & uniformity from galaxy counts \\
           &        & $\log N(m) = 0.6 m + $ constant \\
         & Ambartsumian, Holmberg, Fesenko & variable
extinction in MW  \\
        & Sandage, Tammann, Hardy & linear Hubble law at
$ < 30$ Mpc   \\
        & Webster, Longair & isotropy of radio sources   \\
 \hline
        & Wertz, Bonnor, Wesson, Alfven & physics of
hierarchy  \\
1970s -- & Haggerty, Severne, Prigogine & N-body
dynamics in hierarchies  \\
1980s & Totsuji, Kihara, Peebles & $w \propto \theta^{-
0.8} \Rightarrow
         \xi \propto r^{-1.8}$  \\
      & Mandelbrot  & fractals, multifractals   \\
$\xi$ & Baryshev, Perdang & first evidence for $D=2$: \\
and & Lerner, Schulman, Seiden & $M(r)$, $z(r)$,
stability, percolation \\
$\Gamma$ & ---------------------------------- & --------
----------------------------  \\
function & Davis, Peebles  & $\gamma = 1.8$, $r_0 = 5$
Mpc, 2000 galaxies  \\
      & Einasto, Klypin, Kopylov, Bahcall & $r_0$
depends on depth and type  \\
      & Pietronero & the method of $\Gamma$ function \\
     & Pietronero, Sanders, Coleman & the first fractal
analysis of CfA  \\
     & Pietronero, Ruffini, Calzetti et al. &
explanation of $r_0 (R_s)$  \\
 & Jones, Martinez, Saar, Einasto  & $D_2 = 1.2$, $D_0 =
2$, multifractal?  \\
\hline
   & Sylos Labini, Montuori, Pietronero & $D \approx 2$
from all 3-d catalogs  \\
1990s -- & Wu, Lahav, Rees & fractality at $r < 30$ Mpc  \\
2000s & Teerikorpi, Hanski, Theureau et al. & TF $ <
200$ Mpc
       $\Rightarrow D= 2.2$, KLUN  \\
     & Paturel, Teerikorpi, Courtois  & LEDA counts $<
15^m$: $0.44m$,
        $D=2.2$  \\
fractals    & Baryshev, Bukhmastova & 2-point column
density: $D = 2.1$   \\
  in   & Zehavi (SDSS team)  & $\xi (s) \propto s^{-
1.2}$, 29 300 gal.   \\
3-d maps & Hawkins (2dFGRS team) & $\xi (s) \propto s^{-
0.75}$, 200 000 gal.,

$D = 2.25$   \\
     &         & distortion by peculiar velocities?   \\
     & Gott et al. (SDSS team) & 500 Mpc Sloan Great
Wall  \\
\hline
\end {tabular}

\end{table}

Together with the border to homogeneity, the centre of
the universe has
moved from the earth to the sun, and to the Milky Way.
One of the most famous events in 20th-century
astronomy was the Great Debate between Harlow Shapley
and Heber D. Curtis
in 1920 about the scale and structure of the Universe
(Smith 1983).
In fact, this debate
heralded the change of the Milky Way into an ordinary
galaxy, whereby the
centre of the universe finally disappeared into the
realm of galaxies.

Meanwhile, a new debate around the nature of galaxy clustering
and on the border of the large scale inhomogeneity emerged.
For almost the
whole 20th century this struggle was going on, involving
such figures
as Charlier, Einstein, Selety, Hubble, Lundmark, de
Vaucouleurs, Sandage,
Peebles and others. As always in astronomy and in
particular in this new
extragalactic field, the scarcity of available
observations at any historical
moment leads to uncertain interpretations of the data.
The modern phase
of this fractal debate
concerns observational tests with specially dedicated
galaxy surveys
to find the spatial scale where homogeneity becomes
reality.

New aspects and instruments for tackling the
inhomogeneity problem
entered the scene, when
the concept of fractal was introduced by Benoit
Mandelbrot (1975, 1977, 1982, 1988).
Fractals, self-similar structures with long-range
correlations, had been
disclosed in physics and were then extended to
astronomical scales, from the
solar system and interstellar clouds to clusters of
galaxies.
Fractals on the largest scales became observable
entities thanks to new
astronomical techniques for measuring
redshifts for thousands of galaxies, together with
theoretical recipes
for powerful data analysis. The primary open questions
are: 1) Where is
the border of transition from fractality to
homogeneity?, and 2) what is the
value of the fractal dimension of the galaxy
distribution?

\subsection {The birth of the debate}

\subsubsection { Einstein -- Selety correspondence}
In his first paper on cosmology Einstein (1917) gave
arguments for homogeneity. He notes that
any finite stellar system will evaporate into infinity
due to internal
gravitational interactions. He also emphasized that
stellar velocities are small,
which spoke against large potential differences and
supported uniform large
scale
mass distribution. For Einstein a strong theoretical
argument was that
the Poisson equation modified by a cosmological constant
term $\lambda$

\be
\triangle \phi - \lambda \phi = 4 \pi G \rho
\label{lambda-poisson}
\ee
has a solution for the density $\rho = $ constant:

\be
\phi = - \frac{4 \pi G}{\lambda} \rho ,
\label{phi-lambda-sol}
\ee
so that a static, homogeneous matter distribution is
possible, which also explains
the small stellar velocities. The homogeneity scale was
supposed to be
about the mean distance between stars.

In his paper Einstein then extended this result to
general relativity, obtaining the spherical world model
with a finite
radius of curvature. At the end of his article Einstein's
mentions that
he does not ponder if his model is compatible with
available astronomical
observations.
Later this crucial hypothesis of homogeneity came to be
called
Einstein's cosmological principle of homogeneity.

% \marginpar{\em Franz Josef Selety 1893-1933?}
The Austrian scientist Franz Selety was aware of both
Einstein's
homogeneous and Charlier's hierarchical models.
In an article in \emph{Annalen der Physik},
 Selety (1922) argued that it is possible to construct
hierarchical
 worlds which fulfil simultaneously the following
conditions:

\begin{itemize}
\item[$\bullet$]{\emph{infinite space}}
\item[$\bullet$]{\emph{infinite total mass}}
\item[$\bullet$]{\emph{mass filling space so that
locally there is everywhere
 a finite density}}
\item[$\bullet$]{\emph{zero average density of the mass
in the whole world}}
\item[$\bullet$]{\emph{non-existence of a unique middle
point or middle region
 of the world}}
\end{itemize}
In fact, Selety was one of the first to realize that
the cosmological principle of ``no centre'' may also
be valid for hierarchical systems. He expressed the
essence of hierarchic
models in that the universe appears for an observer
in a ``molecular-hierarchic'' system everywhere
basically similar.
He also raised the question of Mach's principle in such
universes and
argued that it can be fulfilled.
He pointed out that in such models a zero average
density for the whole
universe exists simultaneously with its infinite total
mass.

Einstein (1922) quickly replied to Selety. He expressed
his opinion that
Mach's principle is not fulfilled in a zero-density
universe. Selety (1923) did
not agree with Einstein and once more discussed the
crucial points of his model.
Summarizing the arguments which were raised by Einstein
and Selety, we
see as main objections to hierarchical models in the
1920's:

\begin{itemize}
\item[$\bullet$]{\emph{Mach's principle is not valid for
a hierarchic world
model with zero global density.}}
\item[$\bullet$]{\emph{Large potential differences in a
strongly inhomogeneous universe lead to too high a
velocity dispersion for stars, which is not
observed.}}
\item[$\bullet$]{\emph{A hierarchic stellar system will
evaporate and stars
will fill the voids, leading to a homogeneous
distribution.}}
\item[$\bullet$]{\emph{A hierarchic world contains a
preferred
middle point.}}
\end{itemize}

\subsubsection { A retrospective view on the posed
questions.  }
In retrospect, one may say that intriguingly, all these
arguments are still actual in modern cosmology.

First, Mach's principle links the inertial mass of a
body to the large scale
mass distribution in the universe. Only relative to
those distant masses
one can define the acceleration of a test particle. In
fact, the nature of
inertial mass is still a challenge for modern
theoretical physics, including
general relativity, where
the rest mass of a particle is regarded as a
relativistic invariant,
independent of
the cosmologically distributed mass around the particle.
Thus Mach's principle
cannot any more be considered as a reason to reject the
hierarchic models.

Second, the small velocity dispersion of stars is due
to their motion in our
Galaxy and not related to the universe as a whole. In
hierarchic models
with $D =1$ there is a constant velocity dispersion for
each hierarchic
level ($v^2 \propto M/r \propto$ constant). In fact, the
problem of
velocity dispersion has moved from stars to galaxies. In
recent years,
there has been special concern
why the velocity dispersion around the local Hubble flow
is so small
inside the highly inhomogeneous galaxy distribution
(Sandage, Tammann \&
Hardy 1972; Chernin 2001; Baryshev, Chernin \&
Teerikorpi 2001).
One possibility is that the $\Lambda$-term, which
Einstein introduced in his 1917 paper,
or its modern generalization dark energy, is responsible
for the smooth Hubble flow.

Third, the question of stability of hierarchical
(fractal) structures
of gravitating particles is one of the open modern
topics of
gravithermodynamics.
Perdang (1990) and de Vega, S\'{a}nches \& Combes (1996)
concluded that
a statistical equilibrium may be possible for fractal
structures with
$D \approx 2 $.

Fourth, the question of the middle point is
interestingly related to
the cosmological principle. A stochastic fractal
structure does not contain
a privileged centre. Fractals preserve important
properties of the old hierarchical systems and are more
realistic
models of the real galaxy distribution (Mandelbrot 1989;
Pietronero, Montuori \& Sylos Labini 1997).

The arguments of Einstein continue to inspire
physical questions.
Now they are not reasons for rejecting inhomogeneous
world models, but define important directions to
study fractals.

\subsection {Early arguments for galaxy clustering.}

In the meanwhile, Charlier (1922) prepared a sky map for
the distribution of 11475 nebulae, taken from Dreyer's
New General
Catalogue and his two Index catalogues.
This led him to conclude that: {\small \it A remarkable
property
of the image is that the nebulae seem to be piled up in
clouds.}
In the debate on the nature of nebulae, Charlier
considered: {\small \it
it appropriate to regard the spiral nebulae as foreign
Galaxies similar
to our own.}
Thus he related the observed clustering to the global
matter distribution
in the universe, i.e. he saw in it evidence for his
hierarchical world
models.

Charlier's successor in the professor's chair at Lund
University,
Knut Lundmark, in his doctoral dissertation written in
1919, had expressed
his conviction that: {\small \it We can certainly expect
to find a very complicated
structure in the doubtlessly gigantic universal system,
which is formed
by the spiral nebulae.} Later one of his main occupation
was to build
the Lund General Catalogue of galaxies in order to study
the real distribution of galaxies. However, this work on
thousands of galaxies described on
separate cards was never completed.

\subsubsection{ Observations disclose clusters of
galaxies.  }
After the discovery of the galaxy universe it soon
became clear that
in addition to field galaxies there are pairs, groups
and clusters of
galaxies. For example, in the 1930's clusters of
galaxies were already routinely used
for extending the redshift--distance law to larger
distances by Humason (1931)
and Hubble \& Humason (1931).

Several observations were used as evidence for the large
scale clustering:

\begin{itemize}
\item[$\bullet$]{\emph{Shapley's metagalactic clouds.}}
\item[$\bullet$]{\emph{De Vaucouleurs's Local
Supercluster and density-radius
relation.}}
\item[$\bullet$]{\emph{Abell's rich clusters and their
superclusters.}}
\item[$\bullet$]{\emph{Shane-Wirtanen clouds of galaxies
in the Lick counts.}}
\end{itemize}

Shapley initiated wide photographic surveys of galaxies.
The Shapley-Ames catalogue of 1249 bright  galaxies from
the year 1932
formed the basis for de Vaucouleurs's Reference
Catalogue in the 1960's.
Inspecting the distribution of galaxy clusters, Shapley
came to the
conclusion that there are ``metagalactic clouds''
(today's superclusters),
for example in the constellations of Coma, Centaurus and
Hercules.
The Centaurus cloud is nowadays called Shapley's
supercluster.
It is interesting to mention that Clyde Tombaugh, the
discoverer of
the planet Pluto, noted as a by-product of his extensive
planet searches
the Perseus--Pisces supercluster. He counted 1800
galaxies in this
elongated cloud which is now a much studied
agglomeration
of clusters of galaxies at a distance of about 100 Mpc.

Harlow Shapley's book \emph{The Inner Metagalaxy} (Shapley 1957)
is an interesting and illustrative outcome of the 2-d epoch, where he
summarizes the work on the clustering of galaxies performed at the
Harvard observatory. In a section ``Introduction on depth surveys'' (p.77)
Shapley writes:{\small \it The distribution of galaxies on the surface of the 
sky is
easily examined on any uniform collection of long-exposure photographs.
An effective study, however, of the distribution in the line of sight
requires much greater labour. It is complicated by the difficulties of
nebular photometry as well as by uncertainties introduced through the
considerable dispersion in the intrinsic luminosities of galaxies.
Systems side by side in space can differ by five or more magnitudes
in apparent brightness, as for example the Andromeda nebula and its
companions; and a pair with equal apparent brightness may differ in
distance by a factor of ten. [...] In the study of the radial distribution
of population it is necessary to use photometric methods for estimating
distances, relative or absolute.} A modern reader in the middle of large
redshift catalogues is stricken by the fact that nowhere the redshift is
mentioned as a possible indicator of distances. But at that time the
local universe was not yet a subject of redshift surveys, the precious
telescope time went to extending time demanding redshift measurements
to deeper space. But after the first and second Reference Catalogues
by de Vaucouleurs and collaborators were published in 1964 and 1976,with their
compilations of hundreds and thousands of redshifts, many people
started experimenting
with the redshift in order to see the galaxy distribution in the radial
direction.

De Vaucouleurs (1953,1958) presented evidence, from the
Shapley--Ames
catalogue, for a local supercluster
of galaxies centered at the Virgo cluster, and having an
overall diameter
of 30 Mpc. This
system basically causes the well-known asymmetry in the
number counts in the two
hemispheres.
It should be noted that there were also views regarding
this enhancement of
galaxy number density as a chance fluctuation (Bahcall
\& Joss 1976).

A large increase in the number of known galaxy clusters
came with George Abell's (1958) catalogue of 2712
\emph{rich clusters of galaxies}.
According to Abell' s selection criteria.
a richness class is based on the counted number $N$ of
galaxies that are
not more than 2 mag fainter than the third brightest
galaxy. Abell
introduced six richness classes (0,1,...,5), so that for
class 0
$30 \leq N \leq 49$ and for class 5 $N \geq 300$. This
catalogue
covers the sky north of declination $-27$. The rich
clusters of the
southern sky were catalogued by Abell et al. (1989).
Together these
compilations contain 4704 clusters.
Such
clusters are rare, but can be seen from large distances
in space.

Abell's collection was one outcome of the photographic
survey of the
entire northern sky, made by the large 48 inch Schmidt
telescope
at Palomar Observatory -- an incredibly important
observational programme which gave astronomers huge
amounts of data
about stars and galaxies. The nine hundred $60 \times
60$ cm copies
of the Palomar Sky Atlas photographs were a basic tool
of observational
astronomy at observatories all around the world for
decades.
Now the question was: Do Abell's
rich clusters form superclusters?

\subsubsection { Santa Barbara 1961 conference. }
      The international conference ``On the stability
of systems of galaxies'' held in Santa Barbara,
California in 1961
was devoted to the Ambartsumian hypothesis of
instability of stellar
and galaxy systems. Leading
specialists on extragalactic astronomy presented their
works on galaxy clustering
on different scales from binary galaxies to
superclusters.
We wish to mention that here de Vaucouleurs (1961)
already presented his study
of the density-radius relation for galaxy systems of
different scales (see below).
He emphasized that there was no indication of constant
density level being reached
in the observed range, so a constant mean density might
be reached only for
scales 10 times larger than the Local Supercluster (i.e.
300 $h^{-1}_{100}$ Mpc).
At the same conference Abell (1961) described his work
on the rich clusters of
galaxies and their superclustering. Typical second order
clusters contain
10 clusters and have sizes of 50$h^{-1}_{75}$ Mpc.
In their conference summary Neyman, Page, and Scott
wrote:
{\small \it
Both Abell and de Vaucouleurs feel that superclustering
is established beyond
doubt, and that the dimensions of second-order clusters
(clusters of clusters of
galaxies) are 30 to 60 Mpc. This means that clusters
cannot be treated as
isolated systems embedded in an isotropic, homogeneous
medium of field galaxies...
}

Here the statisticians Jerzey Neyman and Elizabeth Scott
 recognize that their classical 2-level hierarchy model
of the galaxy clustering (Neyman \& Scott 1952) is not
enough to explain the
real galaxy distribution, and some radically new concept
of clustering is needed
for describing the observations.

Further evidence for the large scale clustering among
galaxies were obtained from
the Lick Observatory galaxy survey by Shane \& Wirtanen
(1967). Its results
were reviewed by Shane (1975) in an important paper,
close to the end of the
2-dimensional period of galaxy catalogues. He
summarized:
{\small \it Clustering seems to be a general, if not a
universal, property among
the galaxies. We find larger aggregations comprising
numbers of clusters that
extend over linear distances up to 30 Mpc. There is
suggestive evidence of still
larger assemblages of galaxies on a scale of 100 Mpc or
more.}

\subsubsection{ The cosmological de Vaucouleurs
law.  }
A most interesting discovery in the 1930's was
that the non-uniformities of the galaxy distribution
possess an intriguing regularity.
Edwin Francis Carpenter, an American astronomer
studied clusters of galaxies and found that their galaxy
number density depends on the
cluster size so that the density is smaller in larger
clusters.
He calculated that the number of galaxies $N$ in a
cluster grows with the size $r$ as
\be
\label{carp} 
N(r) \propto r^{1.5}\,\,.
\ee

Carpenter (1938) regarded this relation as a cosmic
restriction so that a cluster of a given extent
may have no more than a limited number of members. 
This relation extends
from
pairs of galaxies to large systems of hundreds of
members. This showed
for him that small groups and large clusters do not
essentially differ:
{\small \it  the objects commonly recognized as physical
 clusterings are merely the extremes of a nonuniform
though not
 random distribution which is limited by density \ldots}

A next step was made by Kiang (1967) from the Dunsink
Observatory in Ireland. After
analyzing the distribution of Abell's clusters together
with computer
generated artificial distributions, he concluded that:
{\small \it the model of simple clustering by uniform
clusters fails to represent
the world of Abell's objects, in the same way as it has
failed in the world
of galaxies.}
Here  he refers to the model
introduced by Neyman \& Scott (1952), where galaxies
occur
in clusters which are uniformly distributed throughout
space.
Kiang put forward a
hypothesis that \emph{clustering of galaxies occurs on
all scales}
 -- there are no clear-cut hierarchic
levels. He wished to visualize the arrangement of galaxies
so that the various clusters interpenetrate each other,
which may guarantee
that the average density does not depend on the volume.
He came close to the modern view of  fractal clustering,
but did not make the crucial step to the fractal
concept, and regarded
the average cosmic density the same everywhere.
Later Kiang \& Saslaw (1969) found that the clustering
extends at least to
scales of 100 Mpc.

Karachentsev (1968) added an important new aspect to
Carpenter's result.
He estimated average characteristics of 143 systems from
binary galaxies to superclusters. He found evidence that
both luminous and total (virial) mass densities are
decreasing
with increasing size of a system. This showed for the
first time that
the mass--radius behaviour of the hidden mass is also a
power law, but
with the exponent different than for the luminous
matter.

 De Vaucouleurs made the decisive step in recognizing
the cosmological
significance of the clustering of galaxies. Based on
his previous work and
the studies of Carpenter, Kiang,
 and Karachentsev, he
 calculated from new data the density of matter inside
 galaxy clusters of different sizes.
The results of his thinking he published in 1970 in the
article ``The Case for
 a Hierarchical Cosmology''. Following his earlier paper
(de Vaucouleurs 1961),
 he suggested the existence of a
 universal density-size law in the galaxy universe:
\be
\label{e3}
\rho(r) = \rho_0(r_0)(r/r_0)^{-\gamma}
\ee
where $\rho(r)$ is the mass density in the sphere of
radius $r$,
$\rho_0$  and $r_0$ are the density and radius
at the lower cutoff of the structure,
and $\gamma \approx 1.7 $ is the power-law exponent derived
on the basis of the available galaxy data.

De Vaucouleurs (1970) summarized all at the time
known properties of the galaxy clustering concluding that:
{\small \it In the 1930s astronomers stated, and
cosmologists believed, that,
 except perhaps for a few clusters, galaxies were
randomly distributed
 throughout space; in the 1950s the same property was
assigned to cluster
 centres; now the hope is that, if superclusters are
here to stay (and
 apparently they are), at least they represent the last
scale of clustering
 we need to worry about\ldots}

He considered two extreme cases for the behaviour of
the density-radius relation as a test for the nature
of galaxy clustering. First, the density may decrease
smoothly and
monotonically, which would happen if there is no
preferred sizes of
galaxy clusters. In fact, this is what now is called
stochastic
fractal distribution. Second, the density curve may have a
series of peaks
at several preferred scales for which he suggested the
values of 10 kpc
(galaxies), 100 kpc (pairs and multiplets), 1 Mpc
(groups and clusters),
10 Mpc (superclusters) and 100 Mpc (third order
clusters).
In retrospect we may note that later the correlation
analysis of observations
pointed at the  case without preferred sizes.

\subsection {Early arguments for the homogeneity of the
galaxy distribution.}

It is intriguing that parallel to the piling data on
galaxy clustering,
there was a growing number of arguments favouring the
uniformity of
the galaxy distribution. In this view clusters were
considered as exceptional
objects in the sky. We summarize below the evidence for
homogeneity, which
were presented before the 1970s:

\begin{itemize}
\item[$\bullet$]{\emph{Hubble's galaxy counts.}}
\item[$\bullet$]{\emph{Fluctuations in the number of
galaxies due to variable
Galactic dust extinction.}}
\item[$\bullet$]{\emph{Sandage-Tammann-Hardy argument
from the local linear
Hubble law.}}
\item[$\bullet$]{\emph{Isotropic distribution of distant
objects}}
\end{itemize}

\subsubsection{ Hubble's counts of bright
galaxies.  }
Hubble (1926), from his bright galaxy counts, concluded
that these correspond
to the expectation from homogeneity up to $m = 16$.
This was widely regarded
as evidence for a homogeneous cosmological model.
Thus he used the number-magnitude relation for galaxies
as a cosmological test.

\smallskip

{\small
The fundamental equation of stellar statistics,
when applied to objects having a power law radial number
density
distribution
$n \propto r^{-\gamma}$
or a fractal number--radius relation $N \propto r^D$,
implies that the number of objects $N(m)$ having the
magnitude less than $m$
follows the relation
\be
\label{Nmgamma}
N(m) = \frac{D}{5}m + \mathrm{const}\, .
\ee
with the fractal dimension $D = 3-\gamma$.
This result does not depend on the luminosity function
of the objects.
Hence for a homogeneous distribution ($\gamma = 0$;
$D=3$) one obtains
the classical Seeliger law
\be
N(m) = 0.6m + \mathrm{const}\,.
\ee
}

\begin {figure}
\centerline {\psfig {file=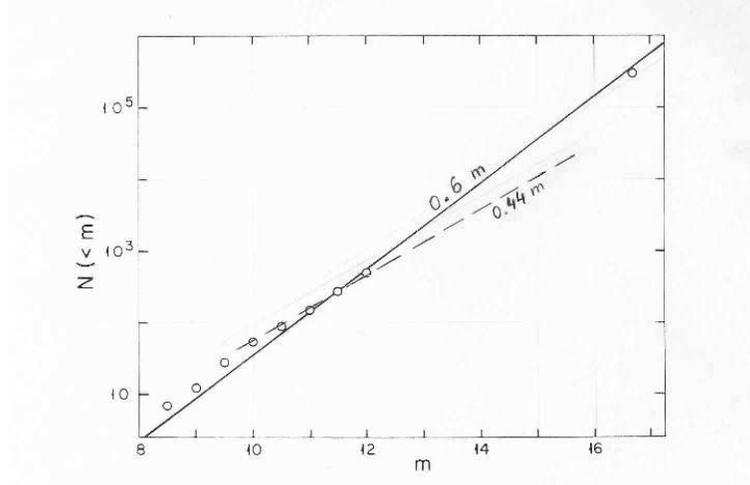,width=10cm}}
%\resizebox{\hsize}{!}{\includegraphics{fig7.ps}}
\caption{\footnotesize Hubble's (1926) early result on
the counts of
bright galaxies. The single point at 16 mag was based on
old data by
Fath. The straight thick line has the slope of 0.6,
which corresponds
to the homogeneous cosmological model. The dashed line
presents modern data from the LEDA database showing a
slope of 0.44
in the interval $10^m \div 14^m$,
corresponding
to the fractal dimension $D = 2.2$
(Teerikorpi 2004).}
\label {hubblebrightcounts}
\end {figure}

Hubble (1926) found that for photographic magnitudes
in the interval $8^m \div 12^m$ plus at the point $16.7$
the ``$0.6m$-law''
was valid, though he noticed a small systematic
deflection which
he ascribed to either observational errors or ``a
clustering of nebulae
in the vicinity of the galactic system'' (Fig.
\ref{hubblebrightcounts}).
In the interval $10^m \div 13^m$ this result was also
confirmed by Shapley
\& Ames (1932) in their catalogue, for the sum of the
counts in the northern
and southern skies.

%\begin {figure}
%\vspace {8cm}

For the bright galaxies ($m < 16$),  the major
shortcoming was the lack of data for
the magnitude interval $12^m \div 16^m$ (see Fig.
\ref{hubblebrightcounts}).
We will discuss modern data in this interval in
sect. 5.4.3.

Hubble was convinced that his data on bright galaxy
counts already
shows homogeneity. He considered that clusters of
galaxies contain only a small
fraction of all galaxies, and the true spatial
distribution is quite
homogeneous. Hubble's conclusion had a strong impact on
the theoretical cosmology,
as expressed by Einstein in 1933:
{\small \it Hubble's research has, furthermore, shown that these
objects are
distributed in space in a statistically uniform fashion,
by which
the schematic assumption of the theory of a uniform mean
density
receives experimental confirmation} (cited by Peebles
1980).
From this time on the picture of a uniform galaxy field
with
clusters as rare fluctuations became a paradigm of the
homogeneous galaxy universe.

Later Hubble's bright galaxy counts were used by
Sandage, Tammann \& Hardy (1972) to show the
incompatibility with de Vaucouleurs's
hierarchical model with $\gamma = 1.7$.
For this $\gamma$ Eq.\ref{Nmgamma} implies that
$N(m) = 0.26m + \mathrm{const}$, and this clearly
differs from
the value $0.6m$ from the counts by Hubble and Zwicky,
available at that time.
Hence they demonstrated that the hierarchical model
with the fractal dimension $D = 1.3$ cannot explain the
galaxy number counts.
Here we should emphasize that modern counts of bright
galaxies in the range
$10^m \div 14^m$ display a slope 0.44 which corresponds
to $D = 2.2$. This
will be discussed in sect. 5.4.3.

\subsubsection { Hubble's deep galaxy counts.  }
Hubble (1934, 1936) extended the number counts up to the
magnitude 21
in his massive galaxy count
program on 1184 photographs on random positions in the
sky.
Each plate covered 0.25 deg$^2$.
Now he did not confirm the $0.6m$ prediction at faint
magnitudes, but
instead the counts followed a $0.5m$ law.
He did not abandon the hypothesis of homogeneity, but
tried to explain this deflection as due to a redshift
effect on galaxy
magnitudes. He made a kind of K-correction and concluded
that
the $0.6m$ law may be obtained only in a non-expanding
universe. In fact,
to the end of his life (1953), Hubble regarded that the
cosmological redshift
may be caused by some other effect than expansion.

 Sandage (1995) has analyzed Hubble's counting programme
and found
three kinds of systematic errors which influenced
Hubble's calculations:
1) there were systematic errors in Hubble's photographic
magnitudes, making
them increasingly too bright for fainter objects; 2)
there was an error in
the K-term as applied by Hubble; 3) when calculating the
prediction of
spatial volumes in the Friedmann model corresponding to
different redshifts
Hubble used an incorrect kind of distance. In the light
of these problems,
one may now see that Hubble's old result on faint
galaxies cannot be used as
evidence neither for homogeneity nor non-expansion.

\subsubsection { Variable dust extinction.  }
One argument against the reality of superclusters
referred
to our ``dusty window'' to extragalactic space. Because
the early
evidence for superclustering of galaxies came from the
sky
distribution, a natural objection was that
one should take into account the variable light
extinction in different
directions of the sky, due to the cloudy dust
distribution in the Milky
Way.

Victor Ambartsumian
% (1908-1996)
developed the theory of a fluctuating
Galactic extinction and applied it to the counts of
extragalactic nebulae
(Ambartsumian 1940; 1951).
The paper by Neyman \& Scott (1952) was devoted to their
well-known
two level clustering model. With references to Charlier
(1922) and
Ambartsumian (1951) they emphasize that there are in
principle
two approaches to galaxy clustering: 1) galaxies are
really clustered in space;
2) the apparent clustering in the sky is caused by
variable extinction
due to interstellar dust clouds.
Which factor dominates for clusters and superclusters?
Later Holmberg (1975) and  Fesenko (1975) further
studied
the role of inhomogeneous dust in the apparent galaxy
clustering and concluded
that the observed clustering is essentially modified by
dust.

In the 1950s Zwicky (1955, 1957) proposed that also
intergalactic dust, concentrated
in clusters of galaxies, could cause extinction of
background clusters.
He studied the distribution of 921 clusters and found a
deficit of
clusters in regions around the Virgo, Coma and Ursa Major
clusters of galaxies. This he
interpreted as
obscuration by intergalactic dust.

Also Karachentsev \& Lipovetskii (1969) derived from counts of background
clusters a positive mean value for the light extinction in clusters
of galaxies. Their result was 0.2 magnitudes in the B band and they pointed
out that there might have been selection effects influencing Zwicky's own
counts: e.g., seen behind nearby clusters it is more difficult for
more distant clusters to fulfil the identification criteria.
We note that Mattila (1978) measured the diffuse light in the Coma cluster
and concluded that a part of it could be scattered light from intergalactic
dust inside the cluster. Later he measured together with Stickel et al.
(1998) the far-infrared emission of the dust in Coma. The emission was
detected and it was calculated that it could cause only about 0.2 magnitudes
or probably less of extinction.

Incidentally, Teerikorpi (2002) showed from the reddenings of quasars
seen through galaxy halos (optical spectra of those quasars contain
narrow absorption lines at much lower redshifts) that there is about
0.2 magnitudes of extinction per halo. Inside compact galaxy clusters
and around the cores of rich clusters such as Coma the galaxy halos may
almost overlap in projection and one might on this basis alone
expect an extinction of the order of 0.1 mag.

It is interesting to note that Zwicky \& Rudnicki (1963)
concluded,
{\small \it taking into account the effects of
interstellar and intergalactic
absorption ... that the results obtained confirm the
assumption that
the distribution of clusters of galaxies is uniform
within a space
whose indicative radius is of the order of $10^9$ pc}.
They inferred that
clusters have different sizes up to a maximum 40 Mpc and
there is no
superclustering.

However, the argument from extinction loosed its
power after massive measurements of galaxy redshifts led
to the
discovery of the very lumpy 3-d galaxy distribution. We
note that the study of
the galactic
extinction continues to be relevant for many
extragalactic subjects, including
the fluctuations in the cosmic background radiation.

\subsubsection { The classical ``linearity''
argument by Sandage,
Tammann \& Hardy.  }
De Vaucouleurs (1970) expected that mass density
fluctuations cause deviations
from the Hubble law and in particular in his
hierarchical model the Hubble expansion
rate should be reduced by gravitation inside the Local
Supercluster. His PhD student
Wertz (1971) developed a Newtonian expanding
hierarchical model and calculated
the predicted deflection from the linear Hubble law: the
Hubble ``constant'' would
increase with increasing distance within 20 Mpc from us.

By this time Sandage had collected redshifts for 82
first-ranked cluster galaxies,
which splendidly allowed one to test the prediction of
the hierarchical model.
This test was performed by Sandage, Tammann \& Hardy
(1972) who
confronted the observations with the calculations
by Wertz (1971) and Haggerty \& Wertz (1972) for the
predicted deflection from the
linear redshift--distance law. The result of the test
was striking:
the linear Hubble law with $H_0=$ constant was
established at all tested
scales, while the hierarchical model
predicted so strong a deflection that one should not see
any cosmological
expansion at distances closer than 20 Mpc. Later
observations only strengthened
the argument from linearity:
a smooth linear Hubble flow starts in the vicinity of
the Local Group, already
at 1.5 Mpc (Sandage \& Tammann 1975; Sandage 1986,
1987).

Actually this test had a deeper meaning than being just a probe of
the hierarchical
distribution of luminous
matter. It demonstrated a paradox: both empirical facts
were true, i.e.
the strongly inhomogeneous galaxy distribution in the
local universe and the
unperturbed linear Hubble law at the same scales.
Sandage, Tammann \& Hardy suggested
two possible solutions for this paradox: 1) the mass density
parameter
could be very small, $\Omega_0 \ll 1$, or 2) there
could be an invisible uniform
medium of high density. In both cases, the perturbations
of the Hubble law
would be tiny. In fact, from this post-classical
cosmological test a whole new
approach was developed, studying the properties of the
cold very local Hubble
flow (Chernin 2001; Baryshev, Chernin \& Teerikorpi
2001; Macci\`{o}, Governato \& Horellou 2004).

De Vaucouleurs (1972) did present evidence for a curved
Hubble law in the local galaxy universe (the Hubble ``constant''
inferred from brightest group members strongly increased
from small to large distances). Teerikorpi (1975a) studied the reality of
such a behaviour using spiral galaxies with known van
den Bergh's luminosity
classes $L_c$. These also showed a similar increase in
$H$, but with a systematic shift depending on $L_c$.
This gave the crucial hint that the phenomenon was
not real, but was caused by a selection bias influencing
distance measurements.
In Teerikorpi (1975b, 1982) the bias was satisfactorily
modelled both for
luminosity classes with Gaussian luminosity
functions and for the brightest
group member criterion.

\smallskip

{\small In Teerikorpi (1975a) the selection effect and
its influence on
the $V/R$ (``Hubble constant'') versus $V$ (radial
velocity) diagram
of luminosity classified was explained in a simple manner, which
serves as an
example how a strong bias may be caused by observational
limits.

Assume that there is a limiting magnitude $m_l$ for the
galaxy sample.
Then in the $M--logR$ ($R = $ distance in Mpc) diagram
only galaxies
below the line $M= m_l -25 5 \log R$ are available for
mapping the
kinematics of the local universe.

Now consider a class of galaxies with a true average
absolute magnitude $M_0$,
then the selection begins to affect distance
determinations at least at the
distance $R_0$ where $M_0 = m_l - 25 - \log r_0$. It is
easy to show that
if $R$ is the erroneous distance calculated from $M_0$
at the real distance
$R=V/H$, then the lower envelope of the points in the
$V/R$ vs. $V$
diagram is defined by $V/R = V/R_0$ (when $R > R_0$).
Note that the
slope depends on $R_0$, i.e. on the $M_0$ of the
luminosity class considered, when
the limiting magnitude is constant.
}

\subsubsection { Isotropy in a homogeneous
universe.  }
Classically, from isotropy around one observer, together
with the Copernican
cosmological principle (``all points are alike''), one
may infer the global
homogeneity of the cosmological fluid (Walker 1944; for
a simple geometric
argument, see Weinberg 1977, p. 24).
Before the 1970's there were
three major observational evidences for an isotropic
matter distribution
around us.

First, Hubble's deep counts
 of faint galaxies (up to $m \approx 21$) did not show
large differences in
different directions of the sky, after correction for
Galactic extinction.
It is true that Shapley (1934) noted a considerable
difference
in the numbers of bright galaxies (up to $m_{pg} = 13$)
for
the northern and southern galactic hemispheres. He also
found that at
$m \approx 18$ the galaxy numbers might vary across the
sky by a factor
of 2 on angular scales of $30^o$.
However, these deviations were viewed as local
fluctuations.

Second, the thousands of
faint radio sources in the early catalogues (4C, Parkes,
Molongolo, and
others) were found to be
uniformly distributed in the sky within statistical
uncertainty (e.g.
Holden 1966).

Third, already the first measurements by Conklin
\& Bracewell (1967) and Penzias, Schraml \& Wilson
(1969)
 of the cosmic background radiation
found a remarkable isotropy at the level $\Delta T / T <
10^{-3}$
on arc-minute scales.

These isotropies have generally been interpreted as
strong evidence
for homogeneity at scales larger than 1000 Mpc. Is it
possible to have
local isotropy within inhomogeneous distribution of
matter?
The answer is ``yes'' for self-similar statistically
isotropic fractal structures.
This question
is also connected with a more general formulation of the
Cosmological Principle,
which we will discuss in the last section.

\subsection {Results from angular galaxy catalogues}

\subsubsection { Main catalogues of
galaxies and clusters.  }

In the 1950's and 1960's the first deep wide area
catalogues of galaxies and
galaxy clusters appeared, based on inspection of the
monumental Palomar
Sky Atlas photographic survey. This was the golden age
of angular galaxy catalogues,
from which the distribution of galaxies was investigated
without distance
information from the redshift. This important 2-d epoch
in studies of galaxy
clustering came to its end in the 1970's when the data
analysis of all
available galaxy catalogues was completed.

The main catalogues were the Shapley--Ames (1932) Bright
Galaxy Catalogue,
Abell's (1958) catalogue of rich galaxy clusters, the
Reference Catalogue
by de Vaucouleurs \& de Vaucouleurs (1964), the
Catalogue of Galaxies
and Clusters of Galaxies by Zwicky et al. (1961-68), and
the Lick counts
in cells by Shane \& Wirtanen (1967).

\subsubsection { The angular correlation function
analysis.  }

The main mathematical instrument for analyzing the above
mentioned
catalogues was the angular correlation function (CF)
techniques, which is
described in detail by Peebles (1980).
The angular CF $w(\theta)$ is defined analogously to
$\xi (r)$ as the
excess probability relative to the Poisson expectation
to find an object
in the solid angle $\delta \Omega$ at the angular
distance $\theta$ from
randomly chosen objects in the sample:

\be
\delta P = n \delta \Omega [1 + w(\theta)].
\label{anglecorr}
\ee
Here $n$ is the mean surface density of objects in the
sky.

A practical estimate of $w(\theta)$ for a sample of $N$
galaxies
with known positions on the sky can be obtained by
counting objects
in rings of radius $\theta$ and width $\delta \theta$.
The normalization condition is

\be
N = n \int_{d \Omega_s}  [1+w(\theta)] d \Omega,
\label{anglenorm}
\ee
where $\Omega_s$ is the solid angle of the survey. An
important practical
point is that if the 2-D projection of an inhomogeneous
spatial distribution
is close to homogeneous, then this method cannot detect
this true inhomogeneity
but regards it as a structureless Poisson distribution.
We will see that this just happened with spatial
structures having fractal dimension $D \geq 2 $.

\subsubsection { The relation between angular and
spatial correlation
 functions.  }
The primary aim of the analysis of angular catalogues was
to estimate the
spatial correlation function $\xi (r)$ from the directly
measured angular
distribution of galaxies. The relation between $w
(\theta)$ and $\xi (\theta)$
was first derived by Limber (1953), already in the
context of galaxies (a useful
review is by Fall 1979).
For small angles $\theta$, \emph{Limber's equation} is:

\be
w (\theta) = \int_0^{\infty} dx x^4 \phi ^2\frac{ \int_{-
\infty}^{\infty} dy
\xi [(x^2 \theta^2 + y ^2)^{1/2}]}{[\int_0^{\infty} dx
x^2 \phi (x)]^2}
\label{limber}
\ee
Here $\phi (x)$ is the sample selection function,
defined as the fraction
of galaxies per unit volume of space observable at a
distance $x$ from
Earth.

Eq.\ref{limber} may be inverted analytically in order to
obtain $\xi (r)$. However,
a negative side is that this procedure requires
differentiation of observed data
which always contain some noise amplifying errors (Fall
\& Tremaine 1977).

An important particular case, which is useful in
practice, is a power law
solution of Eq.\ref{limber}. For the spatial correlation
function in the
form:

\be
\xi (r) =  B r^{-\gamma} = B r^{-(3-D)}
\label{ksipower}
\ee
the corresponding angular correlation function is simply

\be
 w ( \theta ) = A \theta^{1-\gamma} = A \theta^{D-2},
\label{wpower}
\ee
where the constant $A$ comes from observations and the
constant $B$ depends
on $\gamma$ and the selection function $\phi$ (Fall
1979).

\emph{Caution:}
The power law solution (eq.\ref{wpower}) exists only if
\be
\gamma > 1, \; \; \; \; D < 2
\label{gammalim}
\ee
As we already discussed in sec.2.4.5 according to
theorem on fractals projection, 
this means that the method does not work for fractal
structures
with the fractal dimension $D \geq 2$. However, below we
will see that
the real galaxy distribution is characterized
by the value of $D$ which is just
within this critical range!

\subsubsection { Hierarchical $D = 1.2$ models for
angular galaxy catalogues.  }
In their pioneering work Totsuji \& Kihara (1969)
derived from the
angular data of the Lick galaxy counts a power law
angular correlation function
\be
 w( \theta ) = A \theta^{-0.8},
\label{wpowerTK}
\ee
from which they finally obtained
\be
\xi_{\mathrm{TK}} (r)
= \left( \frac{4.7 h_{100}^{-1} \mathrm{Mpc}}{r}
\right)^{1.8}
\label{SWpowerlaw}
\ee

In 1973 Peebles started an extensive programme analysing
all angular galaxy and
cluster catalogues, and asserted that all catalogues are
characterized by almost the
same power law with $\gamma = 1.77$
and $r_0=5h^{-1}_{100}$ Mpc in scale intervals
$0.1 \div 10$ Mpc.
A review of these results is given in Peebles (1980; 2001).

In fact, this was a rediscovery of de Vaucouleurs's
cosmological
density--radius relation at small scales ($0.1 \div 10$
Mpc).
Totsuji \& Kihara (1969) and Peebles (1974a) had for the
first time found that the
galaxy correlation function is a continuous power law
with no peaks, as would
be expected from preferred scales (case 2 in
de Vaucouleurs 1970).
Thus it naturally reflects the self-similarity of
fractal structures.
The exponent $\gamma = 1.8$ corresponds to the fractal
dimension $D = 1.2$.

The new emerging picture of continuous hierarchy inspired
the construction
of protofractal models for the spatial galaxy
distribution.
Peebles (1974b) tried to find a general luminosity
function of galaxies
consistent with the power law hierarchy.
Soneira \& Peebles (1977, 1978) constructed a static
hierarchical model of the
galaxy universe. They used regular hierarchical  model of Fournier's type
with 12 levels of galaxy pairs, so that $k_N = 2$ and
$k_r = 1.76 $ and according to eq.\ref{dimknkr} the fractal
dimension $D=1.23$ or correlation exponent $\gamma = 1.77$.
The positions of galaxies are
assigned according to the pair hierarchy,
the luminosity function and apparent
limiting magnitude were
used to create a magnitude limited sample, and finally
the galaxies were
projected on the sky of an observer.
Soneira \& Peebles (1978) compared this binary hierarchy
with Lick galaxy counts and concluded that even such
a simple model reproduced well the angular correlation
function of the observed galaxy distribution.
Soneira \& Peebles (1977) for
Zwicky et al.(1961-68)
\emph{Catalogue of Galaxies and Clusters of
Galaxies},
inferred that there is no
evidence for a spatially
uniform population of field galaxies. This confirmed the
general tendency
for galaxies to appear only in clusters.

\subsubsection { Tallinn 1977 conference  }

A landmark event in the study of the
galaxy distribution was the IAU Symposium N79
\emph{The large scale structure of the Universe}
held in Tallinn, Estonia, September 1977. Leading
astronomers presented observational and theoretical
works on galaxy clustering and it was the first
wide discussion of all existing arguments for
and against homogeneity at extragalactic scales.

Peebles (1978) presented a review of the angular
correlation function analysis applied for all available
angular galaxy catalogues. He emphasized that
{\small \it "a small systematic error in the
angular distribution
can be translated into a very
large error in the estimate
of the spatial clustering"} and that
{\small \it "redshift data
will allow us to avoid this problem"}.
Peebles argued that at distances $\leq 10$ Mpc
the galaxy clustering is described by power law,
but at larger scales the galaxy distribution
became homogeneous. He referred the first time to the book
by Mandelbrot (1977), where the idea of the fractal was
extended to extragalactic scales.

Joeveer \& Einasto (1978) presented an analysis of existing
redshift data and concluded that the galaxy Universe
has a cell structure with the mean diameter of voids
about $50\,h_{100}^{-1}$ Mpc. The presence of
large structures and holes
of various sizes was also demonstrated at the
symposium by de Vaucouleurs, Tully, Fisher, Abell,
Tifft, Gregory, Kalinkov and others.

In his concluding remarks Longair (1978) noted that
{\small \it "Everyone seemed to agree about the existence
of superclusters ... systems on scales $\sim 30 - 100$ Mpc."}
However, when he refers to the results of angular correlation
function analysis by Peebles' group who derived a
homogeneity scale at 10 Mpc, Longair asserted
{\small \it "I am still a firm believer in the basic
correctness of the covariance analysis"}. Though the
results presented by Kalinkov's team about the third-order
clustering point to the tendency of continuous galaxy
clustering. Longair emphasized that {\small \it
"One wonders whether their existence is consistent with
the isotropy of the distribution of extragalactic radio
sources and of the microwave background radiation"}.

\subsubsection { First evidence for the $D = 2$
distribution}

Though from the angular correlation function analysis
the value of the fractal dimension $D\approx  1.2$
($\gamma =1.8$)
was derived, there were other evidences  pointing at
the fractal dimension $D\approx 2$. Historically, it
is interesting that Lundmark (1927) made in effect the
first observational estimate of the fractal dimension
on the basis of Charlier's model, noting that
the second criterion is fulfilled, which corresponds to $D = 2$.

Baryshev (1981) discussed some observational and theoretical
arguments in favour of the fractal dimension $D\approx 2$.
Using such observational cosmological data as: 1) the galaxy number
counts $N(m)$ in a wide magnitude interval from $2^m$ up to
$24^m$; 2) the virial mass density---radius relation
$\rho_{vir} (R)$; 3) the peculiar
velocity dispersion---radius relation
$\sigma_v (R) $, he concluded that a hierarchical model
of galaxy distribution with $\gamma = 1$ (fractal dimension
$D=2$) is consistent with the observations.

In that paper a new theoretical argument on a special property
of fractals with the dimension $D=2$ was also presented. It comes from
Bondi's (1947) consideration of the global
gravitational redshift part of the cosmological redshift.
For a homogeneous matter distribution in the
case of $z << 1$
the gravitational cosmological redshift is:
\be
\label{z-grav-hom}
 z_{grav}=\frac{\delta  \phi (r)}{c^2}=
\frac{1}{2} \frac{GM(r)}{c^2r}=
\frac{1}{4} \Omega_0 \left( \frac{r}{r_H}\right)^2
\ee
were $\delta \phi (r) = \phi (r) - \phi (0)$ is the
gravitational potential difference between
the observer and the source, $r_H = c/H_0$ is the Hubble radius. 

It is very important
that from the causality principle it follows that
the source is in the
centre of the matter ball with radius equal to
the distance $r$ between the source and observer.
Note that Zeldovich \& Novikov (1984, p.97) and
Peacock (1999, problem 3.4) put the observer to the centre
of the ball and hence got the gravitational blueshift
instead of Bondi's gravitational redshift. However
such a choice of the reference frame violates the causality
in the process considered. Indeed, the event of emission
of a photon by the source  (which marks the centre of the ball)
must precede the event of detection of the photon by an observer.
The latter event marks the spherical edge where all potential
observers are situated.

Generalizing eq.\ref{z-grav-hom} to the case of
a fractal distribution where $M(r)\propto r^D$ one may
derive the following relation for the gravitational
part of the cosmological redshift within the fractal galaxy
distribution:
\be
\label{z-grav-fract}
z_{grav}=
\frac{4\pi G\rho_0 r_0^2 }{c^2D(D-1)}
\left( \frac{r}{r_0}\right)^{D-1}
\ee
where $\rho_0, r_0$ are the density and radius of the zero
level of the fractal structure (galaxies in our case).

For the fractal structure with $D=2$ the cosmological
gravitational redshift is a linear function of distance:
\be
\label{z-grav-fract-2}
z_{grav}(r)\;=\;
\frac{2\pi G\rho_0 r_0 }{c^2}\; r\; = \;
\frac{H_g}{c}\; r
\ee
where $H_g$ is the gravitational Hubble constant, which
may be expressed as
\be
\label{H-grav}
H_{g}\;=\;
2\pi \rho_0 r_0\; \frac{G}{c}
\ee
For a structure with fractal dimension $D=2$ the
constant $\beta =\rho_0r_0$ may be actually viewed as
a new  cosmological fundamental constant. If the value of
the constant $\beta = 1/2\pi $ g/cm$^2$
( e.g. $\rho_0 = 5.2\times 10^{-24} $
g/cm$^3$ and $r_0 = 10$kpc), then
$H_g=68.6$ (km/s)/Mpc. So the linear Hubble law within
the fractal structure with $D=2$ is possible, though it
requires a very large amount of fractal-like distributed dark matter.

\subsubsection {Why did angular catalogues lose the $D=2$ structure?}

An explanation of why it is difficult to study
a fractal structure with dimension $D\geq 2$
from angular distribution of galaxies
was given by Baryshev (1981).
If we model a part of the fractal as a spherical cluster of particles
inside the radius $R$, then one can derive the surface distribution
$F(\sigma )$
of the particles, projected on the sky,
using the Abel equation
\be
\label{F-s}
F(\sigma )= 2\int_{\sigma}^{R}\rho (r)
\frac{r\;dr}{\sqrt{r^2-\sigma^2} } \;
\ee
where $\sigma$ is the projected distance from the centre of the sphere.

For a power-law representation of a spherically symmetric fractal
structure
$\rho (r)=\rho_0(r_0/r)^{3-D} \propto r^{-\gamma}$
it is possible to obtain analytical solutions in closed form for
fractal dimensions $D$  3, 2, and 1.
For a homogeneous ball ($D=3$, $\gamma = 0$)
\be
\label{F-s-3}
F(\sigma )= 2\rho_0 r_0\;
\frac{R}{r_0} \sqrt{1 - \frac{\sigma^2}{R^2}}  \; .
\ee
For a structure with $D=2$ ($\gamma = 1$) we get
\be
\label{F-s-2}
F(\sigma )= 2\rho_0 r_0\;
\left[
\ln \left(1+ \sqrt{1 - \frac{\sigma^2}{R^2}}\right)
+ \ln \frac{R}{\sigma} \right]
  \; .
\ee
Finally,  for $D=1$ ($\gamma = 2$) the surface density will be
\be
\label{F-s-1}
F(\sigma )= 2\rho_0 r_0\;
\frac{r_0}{\sigma } \arccos \frac{\sigma}{R}  \; .
\ee

Hence for the values of $\sigma << R$ the surface density behaves
approximately as
\be
\label{F-s-small3}
F(\sigma ) \sim  \mathrm{const}\,, \,\, \,\,\, 
\mathrm{for} \,\,\,\, D=3 \, ,
\ee
\be
\label{F-s-small2}
F(\sigma ) \sim  \ln \sigma \,, \,\,  \,\,\, 
\mathrm{for} \,\,\, D=2 \, ,
\ee
\be
\label{F-s-small1}
F(\sigma ) \sim \sigma^{-1} \,, \,\, \,\,\, 
\mathrm{for} \,\,\, D=1 \, .
\ee
Both for $D=2$ and $D=3$ the surface density varies slightly,
which means that the projected distribution appears on the sky
with a homogeneous surface density.

This derivation demonstrates that the angular correlation
function analysis becomes inefficient for structures
with the fractal dimension close to or larger than 2, because
the information on the 3-d structures with $D\geq 2$ is lost
as the projected on the sky 2-d distribution approaches homogeneity.
As we already emphasized 
this result is a consequence of the general theorem on fractal projections
(sect. 2.4.5) with its critical dimension $D_{pr} = 2$.

\section {Debate on fractality: the epoch of spatial maps}

During the 1980s several galaxy redshift catalogues (see Table 1)
became available for the 3-d analysis of spatial maps.
This brought into light
unexpected news about different behaviour of the $\xi$-correlation
function for different samples of galaxies and clusters of galaxies.
It was then realized within some research teams that the fractal
galaxy structure can naturally explain
the peculiarities of the measured $\xi$-function, and also that
the appropriate mathematical tool for the analysis of fractals is the
method of $\Gamma$-function.
Since then the fractal approach has given rise to new fruitful directions
for observational and theoretical studies.
 
\subsection {The fractal breakthrough in the 1980s.}

After the concept of the fractal entered the scene of the LSS of the
Universe, the cosmological community as if divided into two parts.
One continued to use the $\xi$-function method, which 
led to the conclusion about the absence of a galaxy "fair sample".
The second, smaller, group started to
apply the fractal approach based on
 the $\Gamma$-function method,
and they obtained concordant results from different 
galaxy samples.

\subsubsection {Davis \& Peebles $\xi$-function analysis
of the CfA sample of galaxies }

The paper by Davis \& Peebles (1983; hereafter DP83) was in many ways
classical. It presented the first systematic analysis of the CfA data
by the $\xi$ correlation function method.
The CfA catalogue was the result of the first large redshift survey
of the Harvard-Smithsonian Center for Astrophysics (CfA), which was
complete to $m_B = 14.5$ in the sky regions $(\delta > 0 \, , \,\,
b > 40^o)$ and $(\delta \geq -2.5^o \, , \,\, b < -30^o)$). It contained
2400 galaxies with redshifts.

DP83 extracted a volume-limited subsample with
$M_B < -18.5 + 5 \log h_{100}$ which contained 1230 galaxies in the
Northern zone and 273 galaxies in the Southern zone.

In DP83 used the concepts of redshift-space and real-space correlation
functions were utilized, 
developed by Peebles (1980). They also took 
into account the peculiar velocities, using the method, which we
considered in sect. 3.2.5. 
They found the rms line of sight
peculiar velocity distribution
\be
\sigma_v (r) = 340 \pm 40 (\frac{r}{\mathrm{Mpc}\,h_{100}^{-1}}
)^{0.13 \pm 0.04} \, \mathrm{km/s}
\label{v12}
\ee

From the 3-d rms peculiar velocity
the ``cosmic energy equation'' gives
(Peebles 1980; Eq.74.9):
\be
<v_{\mathrm{pec}}>^{1/2} \approx 850 \Omega_0^{1/2} \mathrm{kms}^{-1}
\label{cosmic_eneq}
\ee
which may be used to estimate the density parameter $\Omega_0$. For
CfA DP83 obtained $\Omega_0 \approx 0.2$ for the component of matter
clustered with the galaxy distribution on scales $r < 1h_{100}^{-1}$ Mpc.

DP83 suggested the $\xi$-function estimator which later came to be called
standard. The calculation of data--random pairs was introduced for
a correction of the edge effect and to reduce the shot noise on small
scales.

The main conclusion of DP83 was that the real-space two-point correlation
function after the projection has a power-law form $\xi \propto r^{-\gamma}$

\be
\xi_{\mathrm{DP}} (r) =
\left( \frac{5.4h_{100}^{-1} \, \mathrm{Mpc}}{r} \right)^{1.74} \, ,
\label{corDB83}
\ee
in the surprisingly wide interval of scales
\be
10\, h_{100}^{-1}\, \mathrm{kpc}\, <\, r \, < \, 10\,
 h_{100}^{-1}\, \mathrm{Mpc} \, .
\ee
Together with estimated errors the $\xi$-function parameters were
$r_0 = 5.4 \pm 0.3 \, h_{100}^{-1} \, \mathrm{Mpc}$ and
$\gamma = 1.74 \pm 0.04$.
It is important that for scales $r >10\, h_{100}^{-1}\, \mathrm{Mpc}$
the $\xi$-function drops, changes sign, and starts to oscillate
near the zero-level. From what we considered in sect.3.2.3., this is
exactly what is expected for the $\xi$-function estimator.

After this pioneering work, the values of the unit scale $r_0 \approx
5\, h_{100}^{-1}$ Mpc (defined as $\xi (r_0) = 1$) and the correlation
exponent $\gamma \approx 1.8$ have been  
generally considered as standard cosmological numbers
(see Peebles (2001) in the conference ``Historical Developments of
Modern Cosmology''). 

As we discussed in sec.3 the $\xi$-function method actually
gives a distorted value for the intrinsic power-law
exponent $\gamma_{true}$ of a fractal structure, hence
these results contain systematic errors.
However at that time it seemed that the first 3-d map gave results which are 
consistent with analysis of angular catalogues and that the galaxy
distribution becomes homogeneous  
on scales larger than $20\, h_{100}^{-1}$ Mpc 
.

\subsubsection {The puzzling behaviour of the $\xi$-function }

When the first redshift surveys were studied
by means of the $\xi$-function method, a new unexpected problem
appeared. The characteristic length $r_0$ was found to be dependent on
certain parameters of the samples, such as the depth of a survey,
the type and luminosity of galaxies and clusters, and the mean separation
between the objects in the sample. 

Contrary to the expectation that ``the spatial correlation function of
galaxies is quite small for separations greater than about $20 h_{100}^{-1}$
Mpc'', it was found by Bahcall \& Soneira 1983 and Klypin \& Kopylov 1983
that the characteristic length $r_0$ 
(and hence the amplitude of the $\xi$-function)
become quite large for clusters of galaxies.

Bahcall \& Soneira (1983) calculated the redshift-space $\xi$-function
for a complete sample of $N=104$ Abell clusters with the distance class
($\leq 4$), and obtained the following estimates of its parameters:
\be
\label{clust0}
r_0^{cl} \approx 25\,h_{100}^{-1}\mathrm{Mpc}\,,
\,\,\,\,\gamma ^{cl} \approx 1.8
\ee

Klypin \& Kopylov (1983) studied another sample of Abell clusters
with $z<0.08$ and $\| b \| \geq 30^o$. 
Their catalogue contained $N=158$ rich clusters of galaxies including
redshifts measured with the 6-meter telescope of the 
Special Astrophysical Observatory of USSR Academy of Sciences.
The redshift-space $\xi$-function for the sample had the following
parameters
\be
\label{clust1}
r_0^{cl} \approx 25\,h_{100}^{-1}\mathrm{Mpc}\,,
\,\,\,\,\gamma ^{cl} \approx 1.6
\ee
These results revealed a significant discripancy between the unit
scales $r_0$ for galaxies (5 Mpc) and for clusters (25 Mpc).

Moreover, when the $\xi$-function was calculated for superclusters
of galaxies (Bahcall \& Burgett 1986; Lebedev \& Lebedeva 1988),
even larger scale were found:
\be
\label{clust2}
r_0^{cl} \approx 60\,h_{100}^{-1}\mathrm{Mpc}\,,
\,\,\,\,\gamma ^{cl} \approx 1.8
\ee
According to these data the correlation length $r_0$ increases from
5 to 60 $h_{100}^{-1}$Mpc when one considers increasingly massive
objects in the universe.

An important property of this new effect was also found by
Einasto, Klypin \& Saar (1986) who studied the behaviour of $r_0$
within  galaxy and cluster
samples having increasing volumes. They used a cubic geometry
for  galaxy samples with the edge size $l=R_s$ (which is the depth
of a survey) and found an approximately linear relation beween
the unit scale and the depth of the sample:
\be
\label{clust3}
r_0 \propto R_s\,.
\ee
In the framework of Gaussian density fluctuations on a homogeneous
background this behaviour of $\xi$-functions is an enigmatic fact, and
to explain it several possibilities were discussed
(Kaiser 1984; Bardin et al.
1986; Davis et al. 1988; Bahcall 1988). 

The most popular one is Kaiser's idea
of \emph{biased galaxy formation}, based on a possible relation between
the correlation functions for galaxy clusters and for the underlying
mass density field. Here clusters are regarded as rare high density spots in  
the density field, so that one might expect
\be
\xi_{\mathrm{clusters}} = b \xi_{\mathrm{density}} (r) \, ,
\label{clusterbias}
\ee
where $b$ is the bias factor which is about 10 if 
$\xi_{\mathrm{density}} = \xi_{\mathrm{galaxies}}$. This means that
galaxies in clusters are formed from rare peaks above some global
threshold in the primordial density field.
However, the
validity of this explanation in the case of Gaussian density fields
was recently
critisized by Gabrielli, Sylos Labini \& Durrer (2000). They demonstrated
that the increasing sparseness of peaks over the threshold in Gaussian random
fields does not explain the observed increase of the amplitude of
the correlation function $\xi (r)$.

Other alternative possibilities, like local
inhomogeneities, corrections for Galactic extinction, and luminosity
segregation (Davis et al. 1988), look surprising and demand careful
future studies with much larger galaxy samples.

\subsubsection {Pietronero's solution of
the mystery of $r_0$ }

A radically new interpretation of the observed $\xi$-function
behaviour was found by Pietronero (1987) within the fractal
approach to galaxy distribution.
In this classical paper
he introduced the $\Gamma$-function method for the 3-d galaxy map
analysis and derived the relation between the $\Gamma$ and $\xi$
functions which we considered in sec.3.4. 

For a spherical (or cubic) galaxy sample with  the depth $R_s$
and a fixed luminosity of galaxies Pietronero (1987) obtained 
for the characteristic scale $r_0$, defined as
$\xi(r_0)=1$, the relation
\be
r_0 =(\frac{3-\gamma}{6})^{1/\gamma} R_s
\label{r_0-R_s-2}
\ee
Hence within a fractal model 
one expects a linear dependence of $r_0$ on $R_s$.
An important consequence of eq.\ref{r_0-R_s-2}
is that the increasing amplitude of $\xi$-function and the corresponding
increase of $r_0$ for samples with larger $R_s$ is not due to a 
larger correlation length, but is simply an artificial effect
caused by the definition of the reduced correlation function.
In Fig.3 of Pietronero (1987) it was clearly demonstrated
that the cause of the increasing amplitude of $\xi$-function
is the increasing depth of a sample within the fractal structure.

If the picture of the universal fractal galaxy distribution
is true, then
the method of $\Gamma$-function, appropriate for fractal structures,
will reveal a power-law behaviour for future still deeper galaxy
samples. The requirement of spherical geometry for a sample
is the most important restriction for deep galaxy surveys.

The first analysis of the CfA galaxy redshift catalogue 
by means of the $\Gamma$-function method was performed
by Coleman, Pietronero \& Sanders (1988). They found a power-law
$\Gamma$-function with $\gamma=1.5 \pm 0.2$, for VL samples with
$N_{gal}=226$ and 442. The fractal behaviour was detected on the
interval of scales from $1\,h_{100}^{-1}$ Mpc up to $20\,h_{100}^{-1}$ Mpc
without any characteristic scale,
contrary to the homogeneity
scale $r_0 = 5\,h_{100}^{-1}$ Mpc 
derived by the $\xi$-function method.

\subsubsection {Cellular fractal structure of the Universe }

Just a few months after Pietronero's (1987) paper, the fractal
interpretation of the "correlation length" versus depth was
applied to redshift data by Calzetti, Einasto, Giavalisco,
Ruffini, \& Saar (1987). They confirmed the linear relation
between $r_0$ and $R_s$, when the depth changed from  
$5\,h_{100}^{-1}$ Mpc up to $50\,h_{100}^{-1}$ Mpc.

Then in a series of papers of a team led by Ruffini
a cellular model of the Friedmann universe was developed, where within
cells with sizes of about 100 Mpc the distribution of galaxies has
the fractal dimension $D \approx 1.2$, and on larger scales the universe
becomes homogeneous (Ruffini, Song, \& Taraglio 1988; 
Calzetti, Giavalisco \& Ruffini 1988; 
Calzetti, Giavalisco \& Ruffini 1989) .
Their main conclusion was that de Vaucouleurs's density law may be
reconciled with the homogeneous Friedmann model if there is
a maximum scale of fractality.

The Ruffini at al. model is based on the assumption that there are
massive dark matter particles, called "inos", which obey Fermi
statistics and are responsible for the initial density fluctuations of
the cellular structure formation. The characteristic value of the
"ino" rest-mass-energy is $0.4 \div 10$ eV, and the corresponding
density parameter is $\Omega_{inos}=(0.4 \div 1)$. An expected
value of the size of the fractal cell is about 100 Mpc, which
is determined by the Jeans length at the epoch when "inos"
decoupled from matter. The angular scale of the corresponding
CMBR fluctuations is about 1 degree. 

However, they used the lower value of the fractal dimension $D \approx 1.2$ 
derived from the $\xi$-function analysis, which is actually a distorted
value of the true fractal dimension $D \approx 2$ obtained by means of
the appropriate $\Gamma$-function method. 
It would be interesting to reconsider their model for $D \approx 2$.

\subsubsection {The multifractal confusion
from the $D_2 = 1.2$ estimation using the $\xi$-function method }

Almost at the same time another group of astronomers started
to apply the fractal approach for a description of the galaxy distribution.
However, now we may see that their work
was affected by a  distorted estimate of
the correlation fractal dimension from the $\xi$-function
method.

Jones, Martinez, Saar \& Einasto (1988) and 
Martinez \& Jones (1990) noted that when they applied the methods
of box-counting and minimal spanning tree for determining the
Hausdorf dimension $D_H$ of the CfA redshift survey, they obtained
the value $D_H = 2.1 \pm 0.1$. However, from the $\xi$-function
for the same galaxy catalogue they concluded that the correlation 
dimension differs from this value, $D_2=1.2$, hence 
{\it "the Universe is not a simple fractal. It is a more complex
structure, a multifractal"} (Martinez \& Jones 1990).
An analogous conclusion was made by  Klypin et al. (1989) and
Balian \& Schaeffer (1989).

Here we have an example of how inappropriate methods of data 
analysis may lead to erroneous theoretical conclusions.
Indeed, as we discussed in sec.3.4 the estimation of the fractal
dimension from the $\xi$ correlation function as $D_2=3-\gamma_{\xi}$
at scales close to $r_0$ gives a distorted value for the true
codimension $\gamma$ because there 
$\gamma_{\xi}(r=r_0) \approx 2\gamma$.
Hence to calculate the true value of the correlation dimension
from the slope of the $\xi$-function near $r=r_0$ 
one should take into account this distorsion, so
$D_2=3-\gamma = 3-(\gamma_{\xi}(r=r_0)/2) = 2.1$ for 
the observed slope $\gamma_{\xi}(r=r_0)=1.8$.

Therefore one may conclude that actually $D_2 \approx D_H$, i.e. 
the correlation dimension is consistent with the Hausdorf dimension
for the CfA catalogue and there is no need for multifractality
based on a difference between these dimensions.
Modern results of $\Gamma$-function analysis for 
CfA, 2dF,SDSS and other 
galaxy redshift surveys confirm the value for
the correlation dimension $D_2 \approx 2$, and hence
eliminate the above confusion on multifractality (see sec.5.3).

\subsubsection {Balatonfured 1987 conference and
observational evidence for very large structures}

Parallel with developments of statistical methods for the analysis
of fractal galaxy distribution in the 1980s, new observational
evidence appeared supporting  the existence of galaxy structures
with sizes much larger than the characteristic homogeneity
scale $r_0$ derived from the $\xi$-function analysis.

The results of a decade of intensive research after 
the Tallinn'77 conference were discussed at the 130th IAU Symposium
"Large Scale Structures of the Universe", held in Balatonfured,
Hungary 1987. It happened that this symposium became the last one
on cosmology for Yakov Zeldovich (1914--1987) and for one of his
talented pupils Victorij Shvartsman (1945--1987) who had just started
to investigate the large scale structure of the Universe
at the Special Astrophysical Observatory (SAO)
of the USSR Academy of Sciences.   

At the Balatonfured'87 conference new observational data
on the reality of galaxy structures with sizes of about 100 Mpc
were presented. Huchra, Geller, de Lapparent \& Burg (1988)
discussed an extension of the CfA redshift survey, which for several
years was the main test bench for different statistical methods
of the galaxy distribution analysis. They concluded that empty 
regions (voids) and filaments are common in the observed galaxy
distribution, and the sizes of voids achieve $50\,h_{100}^{-1}$
Mpc, which is much larger than the characteristic $\xi$-function
scale $r_0 \approx 5\,h_{100}^{-1}$ Mpc.

Karachentsev \& Kopylov (1988) reported the results of a spectral
survey of 245 galaxies with $m_B \leq 17.5$
in a narrow strip which passed through the Coma cluster. They confirmed
a bubble-like type of structure within the Coma supercluster
and estimated the parameters of the $\xi$ correlation function:
$r_0=22\,h_{100}^{-1}$ Mpc and $\gamma = 1.5$. The average
size of 14 voids was estimated to be about $25\,h_{100}^{-1}$ Mpc.

The most prominent structures were discovered by studies
of rich galaxy clusters. Tully (1986, 1987) analyzed the
distribution of 47 Abell clusters within a region up to $z=0.1c$
and found a flat structure having a size of about 
$300\,h_{100}^{-1}$ Mpc. This is called the Pisces--Cetus
Supercluster Complex.

Similar results for an  even deeper survey of galaxy clusters
were obtained by
Kopylov, Kuznetsov, Fetisova \& Shvartsman (1988). They presented
the first result of the program "The Northern Cone of Metagalaxy"
which included measurements of redshifts up to $z=0.28$ for
58 rich compact clusters of galaxies inside the cone with
$b^{II}>60^o $. From these data they made a preliminary 
statement on the existence of inhomogeneities in the distribution
of galaxies on scales up to $500\,h_{100}^{-1}$ Mpc.

In the Summary of the conference made by Peebles (1988)
one may read: { \it There is considerable evidence of structure
on scales $ \geq $ 50 $ h^{-1} $ Mpc, but I think it is fair
to repeat the old questions: could this be an artifact of errors
in the catalogues? Could the eye be picking patterns out of
noise? If the answers were definitely "no" it would be very
damaging for scale invariant cold dark matter. We all will be
following the debate with great interest. } 

It took one more decade of hard observational work to disclose
the reality of such super-large structures in galaxy
distribution, but the debate is still going on how damaging
it actually is for the CDM models.

\subsection {Further steps in the debate}

After the new idea of fractality entered the studies
of large-scale galaxy distribution there was a period of
strong opposition from those who used conventional methods of analysis.
Fortunately in science the collision
of ideas is actually needed for a deeper understanding of the universe.
This was also the case with the fractal galaxy distribution,
which gave an alternative to homogeneous  matter
distribution in the Universe.

\subsubsection {Princeton ``Dialogues'96'': 
Davis's evidence for
homogeneity at scales larger $20\mph$ }

In 1996 an international astronomical meeting under the intriguing
title "Critical Dialogues in Cosmology" was held in Princeton.
Remarkably the first subject which opened the conference was
the dialogue between Marc Davis and Luciano Pietronero on the
homogeneity of the galaxy distribution.

Davis (1997) presented the position "that there is overwhelming
evidence for large scale homogeneity on scales in excess of
approximately $50 \mph $, with a fractal distribution of matter
on smaller scales". He emphasized that the observed correlation
function $\xi(r)$ is well characterized by a power law,
$\xi(r) \approx (r/r_0)^{-\gamma}$, with $r_0 \approx 5\mph$
and $\gamma = 1.8$. Hence the fractal dimension at scales
$r < r_0$ is $D=1.2$.

Davis' arguments for homogeneity were:
\begin{itemize}
\item {\it D1. Isotropy of the CMBR, X-ray and radio source counts.}
\item {\it D2. Observed counts of galaxies for magnitude range
$14 < m < 18$ have slope $0.6m$.}
\item {\it D3. The observed angular correlation function $w(\theta)$ 
is reliable for recovery of spatial correlation function
$\xi(r)$.}
\item {\it D4. Analysis of four VL samples from 
1.2 Jy IRAS redshift survey give for $\xi(r)$
standard values of $\gamma$ and $r_0$ when the volume limiting
radius is increased from $60\mph$ up to $120\mph$.}
\item {\it D5. "The end of greatness" seen 
from the LCRS redshift survey.}
\item {\it D6. Ly-$\alpha$ clouds detected in qso absorption spectra
appear to be very nearly uniformly distributed in space.}
\end{itemize}

After discussing his arguments Davis concluded:
"The measured two-point galaxy correlation function $\xi(r)$
is a power law over three decades of scale and approximates
fractal behavior from scales of $0.01\mph < r < 10\mph$, but
on scales larger than $\approx 20\mph$, the fractal structure 
terminates, the rms fluctuation amplitude falls below unity,
and the Universe approaches homogeneity, as necessary to make
sense of a FRW universe."

\subsubsection {Princeton ``Dialogues'96'':
Pietronero's arguments for
fractality  }

Pietronero presented the statistical method ($\Gamma$-function
analysis) which is relevant for study the fractal structures,
compared it with the $\xi$-function method, and demonstrated
the first results of application of both methods to the available
redshift catalogues (Pietronero, Montuori, Sylos Labini 1997).

Pietronero's main
arguments for the fractality of the galaxy distribution were:
\begin{itemize}
\item { \it P1. The projection effect of a spatial fractal structure
may lead to the observed isotropy on the sky for the
angular distribution of astrophysical sources.}
\item { \it P2. A small number effect in the counts of bright galaxies
may lead to observed $0.6m$-law even for fractal distribution.}
\item { \it P3. The method of $ \xi $ correlation function gives artificially
distorted values both for the fractal dimension $D$ and for 
the homogeneity scale $r_{hom}$. }
\item { \it P4. The method of $\Gamma$-function (conditional density)
is appropriate for estimation of the true value of the fractal
dimension $D$ and for detection of the crossover to homogeneity.}
\item { \it P5. The $\Gamma$-function analysis of available 3-d galaxy
catalogues, CfA, PP, IRAS, LEDA, LCRS, ESP, gives the value
$D=2.0 \pm 0.2$ for the fractal dimension at scales up to the
radius $R_{max}^{sph}$ of the largest sphere that can be
contained in the sample.}
\item { \it P6. The homogeneity scale is not yet reached in existing
galaxy catalogues and may be as large as $150 \mph$ (LEDA result)
and even as $1000 \mph$ (number counts of ESP redshift galaxy
survey).}
\end {itemize}

Both sides of the discussion agreed that the galaxy
distribution is a fractal structure at least within
scales $0.1 \div 10 \mph$. But they disagreed about the values
of the fractal dimension (Davis for $D\approx 1.2$ and
Pietronero for $D \approx 2$) and about the value of the homogeneity
scale (Davis for $R_{hom}\approx 20\mph$ and
Pietronero for $R_{hom} \geq 150\mph$).
Pietronero also emphasized that possible existence of 
uniformly distributed
dark matter may reconcile Friedmann homogeneous
model with observed visible fractal structure.

%Fortunately future all-sky observations of all galaxies
%up to sufficiently faint magnitude will allow to resolve
%this debate. The first step in this direction is the SLOAN
%redshift survey which will cover $\pi $ steradian on the sky
%and reach all galaxies up to about 19 magnitude.

\subsubsection {The problem of sky projection of fractals}

One of the strongest argument for the value of fractal dimension
$D=1.2$ and the homogeneity of galaxy
distribution in space at scales larger than $r_0=5 \mph $ 
was the claim that the analysis of galaxy angular catalogues
led to just such values for the parameters of the angular
correlation function.

Indeed,
starting with Totsuji \& Kihara (1969) in all angular galaxy
gatalogues the analysts found the universal behaviour of
the angular correlation function $\xi_{ang} (\theta)
\propto \theta ^{-\alpha}$ with $\alpha \approx 0.8$.
For scales $r < r_0$
 this would correspond to the fractal dimension $D$ in 3-d space
$D = 2 - \alpha \approx 1.2$. As this is less than 2, then
according to the fractal projection
theorem (sect.2.4.5) one can estimate the true
fractal dimension from the galaxy distribution projected on the sky.

Unfortunately, this logic has a flaw. Indeed, if the real
spatial galaxy distribution is a fractal structure with
$D \geq 2$ we would not be able to detect it from observations
of the angular distribution of galaxies. This is because according to
the fractal projection theorem the projected distribution imitates
a homogeneous surface density. Consequently, 3-d maps are necessarily
required to discover fractal structures with $D \geq 2$.

It is an intriguing fact that the fractal analysis of modern
extensive 3-d galaxy maps has revealed $D \approx 2.2$,
just putting the fractal dimension into that critical interval.

A decade after the first warning on the possible ``conspiracy'' of structures
with $D\geq 2$ by Baryshev (1981), detailed studies started to appear
on the complex problem of angular projections (Dogterom \$ Pietronero 1991;
Coleman \& Pietronero 1992; Durer et al. 1997; Montuori \& Sylos Labini
1997; Eckmann et al. 2003).   

{\it The method of angular $\Gamma$-function. }
The best demonstration of the consistency of the observed angular
and spatial fractal structures with $D \approx 2$ was given by
Montuori \& Sylos Labini (1997).  
They studied 3-d maps together with the corresponding angular
distributions from several redshift catalogues: CfA1, SSRS1,
Perseus-Pisces, APM bright galaxies, and Zwicky galaxies.
For an undistorted estimation of the correlation exponent in
angular data they used the conditional surface density
$\Gamma_{\mathrm{ang}}$
\be
\Gamma_{\mathrm{ang}}(\theta) = \frac{1}{S(\theta)} \frac{dN(\theta)}{d\theta}
           = \frac{B_{\mathrm{ang}} D}{2\pi} \theta^{-\alpha}
\label{gamma_ang}
\ee
where $S(\theta)d\theta$ is the solid angle element ($S(\theta) \approx
2 \pi \theta$ approximated for small angles $\theta \ll 1$),
$N(\theta) = B_{\mathrm{ang}} \theta^D$ is the number of galaxies
in the ``polar cap'' with the radius $\theta$, $D$ is the fractal
dimension of the 3-d structure, which with the condition $0 \leq D < 2$
coincides with the fractal dimension $D_{\mathrm{pr}}$ of the projected
structure according to the fractal projection theorem (see sect.2.4.5).
$\alpha$ is the angular correlation exponent related to the
fractal dimension $D$ as
\be
\alpha = 2 - D = \gamma - 1
\ee 
The last equality follows from the relation between the spatial
and angular $\Gamma$-functions. The first one is
the usual 3-d $\Gamma$-function
$\Gamma (r) \propto r^{-\gamma}$ with $\gamma = 3-D$, and the second one
is  the angular $\Gamma$-function $\Gamma_{\mathrm{ang}} \propto
\theta^{-\alpha}$ with $\alpha = 2 - D$.

The result of this angular and spatial $\Gamma$-function analysis
demonstrated very clearly that the fractal dimension of the 
observed 3-d structure is
\be
D = 1.9 \pm 0.1
\ee
which was derived from independent analyses of the angular
catalogues and 3-d maps. The angular correlation exponent
was $\alpha = 0.1 \pm 0.1$ and the spatial correlation
exponent was $\gamma = 1.1 \pm 0.1$ for all above mentioned
catalogues.

The most amazing fact is that the previously derived ``universal'' value
of the angular correlation exponent $\alpha_w = 0.8$
is an artificial effect caused by the method
of the angular $\xi$-correlation function $w(\theta )$ (sect. 4.4).
This reduced correlation function gives systematically distorted
values of the true correlation exponent due to the normalization
condition.

We see here again that the whole story of how to find the true
correlation exponent revolves around the difference between
 the power-law behaviour of the complete correlation function and
the corresponding non-power-law reduced correlation function.
This happens both for angular and spatial distributions.

\subsubsection {Modern research topics related to large
scale fractality}

The debate on the nature of the large-scale structure of
the visible matter in the Universe inspired many astronomers
and physicists to study different aspects of the fractality 
galaxy distribution. 
The spectrum of subjects is very wide and shows that
the principal questions, 
already raised by Einstein(1917, 1922)
and Selety(1922, 1923) on the properties
of hierarchical cosmological models,   
now are under careful investigation
and actually they generate new branches of cosmological
physics. 

In Table 3 we present a list of main reseach topics together
with corresponding references which demonstrate
the variety of cosmological aspects touched
by the fractality of the large scale structure of the Universe.
We shall discuss some results of the studies
in the sec.6 of the review.

\begin{table}
\label{topics}
\caption{
Main  research topics related to
large scale fractality.}
\vskip 0.3cm
\begin {tabular}{|c|c|}
%  \hline
%\multicolumn{3}{|c|}{\emph{The debate on large scale
%fractality}} \\
% \hline
% \hline
%\multicolumn{2}{|c|}
%{\emph{}} \\
 \hline
%& & \\
 Subject & References \\
%& & \\
%& & \\
\hline
 conditional density & Pietronero 1987; Coleman,
Pietronero 1992;
Gabrielli et al. 2004    \\
 new methods of data analysis & Bharadwaj et al.
1999; Best 2000; Martinez,
 Saar 2002;  \\
 2-point conditional column density & Baryshev,
Bukhmastova 2004; Vasiliev 2004 \\
%& & &\\
 \hline
 fractal dimension  &  Coleman et al. 1988;  Klypin et al. 1989;
 Lemson, Sanders 1991; \\
 of galaxy distribution
 & Jones et al. 1988;  Martinez, Jones 1990; Jones et al. 1992;\\
 scales of fractality & Sylos Labini, Montuori,
Pietronero 1998 \\
 mass-radius relation for all scales & Sidharth 2000;
Teerikorpi 2001; Rost 2004\\
\hline
 local radial galaxy distribution  & Sandage 1995;
 Teerikorpi et al. 1998; Teerikorpi 2004 \\
 number counts, normalization & Baryshev 1981;
Joyce, Sylos Labini 2001;
                                               Courtois
et al. 2004         \\
luminosity function, multifractals & Sylos Labini,
Pietronero 1996  \\
% &&\\
\hline
 dependence of correlation function & Einasto et al.
1986; Calzetti et al.
                                            1987; Davis
et al. 1988 \\
on depth, luminosity &  Norberg et al. 2001, 2002 \\
 type of object & Klypin, Kopylov 1983; Bahcall 1988;
Bahcall et al. 2003 \\
  peculiar velocities &  Zehavi et al. 2002, 2004;
                                            Hawkins  et
al. 2003 \\
 \hline
local fractal dimension & Tikhonov, Makarov, Kopylov
2001; Tikhonov, Makarov 2003 \\
 linearity and coldness of & Sandage et al.1972;
Sandage 1986, 1987;

Karachentsev et al.1996 \\
 the local Hubble flow   &  Ekholm et al. 2001;
Karachentsev et al.

2003a,b; Whiting 2002  \\
\hline
 protofractal (hierarchical) models & Wertz 1971;
Wesson 1975; Soneira, Peebles

1977, 1978 \\
 relativistic fractal models & Bonnor 1972; Ruffini
et al. 1988;
                            Ribeiro 1993; Gromov et al
2001 \\
cosmological tests, fractal universe &  Fang et al
1991;
          Baryshev et al. 1994; Joyce et al. 2000 \\
\hline
 cosmological gravitational redshift & de Sitter
1917; Bondi 1947;
                               Baryshev 1981, 1994 \\
\hline
  local tests of cosmological vacuum & Chernin 2001;
Baryshev et al. 2001; Axenides et al. 2002  \\
 and dark energy within fractals & 
 Chernin et al. 2004; Maccio et al. 2004 \\
 \hline
 statistical mechanics of
 & Perdang 1990; de Vega, Sanchez, Combes 1996, 1998  \\
 self-gravitating fractal gas, $D = 2$
 & Combes 1998; Huber \& Pfenniger 2001;  \\
\hline
 N-body simulation,  & Governato et al.
1997;
 Moore et al. 2001     \\
local structure & Klypin et al. 2003; Maccio et al. 2004 \\
 initial conditions and discreteness  & Bottaccio
et al. 2002;
Baertschiger et al 2002; 2004  \\
 stability, velocity, force & Gabrielli, Sylos Labini,
Joyce, Pietronero 2004 \\
\hline
 origin and evolution of & Haggerty 1971; Peebles
1974a; Alfven 1982;

Lerner 1986 \\
 large scale fractals & Ostriker et al.1981;
Pietronero et al.1986;

Schulman et al.1986  \\
          & Szalay, Schramm 1985; Maddox 1987; Luo,
Schramm 1992      \\
\hline
 cosmological principle & Mandelbrot 1975, 1977;
Pietronero, Sylos Labini 1995; \\
 and fractality, & Rudnicki 1995; Wu, Lahav, Rees
1999; Baryshev,Teerikorpi 2002 \\
 isotropy and homogeneity & Mandelbrot 1989; Sylos
Labini 1994  \\
\hline
\end {tabular}
\end{table}

\subsection {Recent results from the $\xi$ and $\Gamma$ 
functions analyses}

As we discussed above the crucial parameters 
for correlation analysis of a galaxy distribution are:
1) the average separation distance between
nearest neighbour galaxies $R_{sep}$,
2) the radius of the maximum sphere completely 
contained in a sample $R_{max}^{sph}$,
3) the absolute magnitude interval $\triangle M_i$
of selected galaxies,
and 3) the number of galaxies in a volume limited sample $N_{gal}$.
To extract  reliable information on the correlation exponent
and a homogeneity scale one shoud be aware of the restrictions
of the method used for the estimation.
We shall see that within the common interval of applicability
both $\xi$ and $\Gamma$ functions analyses give compatable results.

\subsubsection { The redshift space $\xi$-  and $\Gamma$- functions }

{ \it  $\xi$- and $\Gamma$-function analysis of the 2dF data. }
The final release of 2dF galaxy redshift survey (Colless et al. 2001; 2003)
opens a new possibility for performing different kinds of statistical
analysis of large samples of galaxies.

The 2dF galaxy redshift survey 
contains about 220 000
galaxies in two (NGP and SGP) narrow slices of about 
$90^o \times 15^o$ (SGP) and $75^o \times 10^o$ (NGP) 
complete up to $b_j = 19.5$, with the effective reshift 
$z_s \approx 0.15$,
and the effective absolute magnitude
$M_s - 5\,log\,h_{100} \approx -20.0$, corresponding to
the luminosity $L_s \approx 1.4 L^*$ (Norberg et al. 2002). 

These data were analyzed using the
$\xi$-function method of the reduced correlation function by
Hawkins et al. (2003).
The redshift-space correlation function was  
approximated by two different power-law forms (their figs.5,6,7),
first, as
\be
\xi_{z} (s) =
\left( \frac{13\,h_{100}^{-1} \mathrm{Mpc}}{s} 
\right)^{0.75} \, ,
\label{cor2dF1}
\ee
for the interval of scales $0.1 < s < 3 \,h_{100}^{-1}$ Mpc,
and second, as
\be
\xi_{z} (s) =
\left( \frac{6.82\,h_{100}^{-1} \mathrm{Mpc}}{s} 
\right)^{1.57}
\label{cor2dF2}
\ee
at scales $3 < s < 20 \,h_{100}^{-1}$ Mpc. For larger scales
$30 \div 60 \,h_{100}^{-1}$ Mpc,
$\xi_{z}(s)$ becomes negative. 

Such a behaviour of the $\xi$ correlation function is consistent
with the eq.\ref{ksi_power}. It means that
 2dF galaxy distribution within the interval of scales
$0.1 < s < 3 \,h_{100}^{-1}$ Mpc may be considered as a fractal
structure with the fractal dimension $D=3-\gamma=2.25$.
For scales $s>3$ Mpc the $\xi$-function continuously changing
its slope and at scale $r_0 \approx 5h_{100}^{-1}$Mpc 
the exponent becomes $\gamma_{r_o}=2\gamma = 1.5$ in perfect
accordance with eq.\ref{gamma_ksi_r0}.
Hawkins et al. (2003, fig.7) found that 2dF sample has 
$\xi_z(s)$ which is similar to Las Campanas and SDSS
large slice-like surveys. 

The $\Gamma$-function method was applied to the 2dF VL samples
by Vasiliev (2004) and Vasiliev et al.(2005),
where they obtained that the conditional density for the 2dF
data have power-law with the fractal dimension $D=2.2 \pm 0.2$
at scales $0.5 < s < 40 \,h_{100}^{-1}$ Mpc.

The value  $D=2.2 \pm 0.2$
of the fractal dimension is consistent with results
obtained by Sylos Labini, Montuori \& Pietronero (1998) for all
 at the end of the 1990s avalable
galaxy redshift catalogues: CfA, Perseus-Pisces,
SSRS, IRAS, APM-Stromlo, LEDA, Las Campanas, ESP.
They used $\Gamma$-function analysis so the probed scales were limited
by the radius of maximum sphere $r_{sph}^{max}$
completely embedded in the geometry of a catalogue, i.e. about
$20 \mph$ for the existed slice-like surveys 
and $100\mph $ for LEDA sample.

{ \it SDSS: results from the $\xi$ and $\Gamma^*$ analyses. }
The Sloan galaxy redshift survey with its million 
galaxy redshifts
and the wide sky coverage of about $\pi$ steradians is 
the ideal
catalogue to settle definitely the on-going fractal 
debate.
Up to now, however, only narrow ($2.5^o \div 5^o$) slices -- 
like in
the 2dFGRS -- have been completed and published as EDR, 
DR1, and DR2
catalogues (see the web site of SDSS).

Zehavi et al.(2002) performed
the $\xi$-correlation function analysis of a sample of 29300 
SDSS galaxies
with radial velocities $5700 < cz < 39000$km/s and 
absolute magnitude
interval of $-22 + 5\log h_{100} < M_r < -19 + 5\log 
h_{100}$. 
The redshift space $\xi$ correlation function of this SDSS 
sample definitely has a non-power law (see their fig.5).
The authors took the scale interval 
$2 < s < 8 \,h_{100}^{-1}$ Mpc where they approximated
the $\xi$-function by the power-law
$\xi_{z} (s) = ( s/8.0h_{100}^{-1}\mathrm{Mpc})^{-1.2}$.
However
it is clear from fig.5 in Zehavi et al.(2002) that
the $\xi_z(s)$ has three characteristic intervals of scales:
1) for the interval $0.1 <s < 0.5\,h_{100}^{-1}$Mpc
the exponent $\gamma \approx 1.8$, 
2) for the interval $0.5 <s < 5\,h_{100}^{-1}$Mpc
the exponent $\gamma \approx 1$, 
and 3) for the interval $5 <s < 30\,h_{100}^{-1}$Mpc
the exponent $\gamma \approx 1.8$, 
As we discussed in sec.3.4 such behaviour of the $\xi$
correlation function is just as expected 
for the fractal structure if one takes into
account the characteristic scales 
$R_{sep}$, $r_0$, $R_{max}^{sph}$.

The first $\Gamma^*$ analysis of the SDSS Luminous Red Galaxy
sample was recently presented by Hogg et al. (2004) and also
discussed by Joyce et al. (2005). The LRG sample so deep
(average z about 0.3) that the radius of the maximum sphere 
reached the value $R_{max}^{sph}\approx 100 \mph$.
Hogg et al. (2004) for a sample with $N_{gal} = 3658$
found that the $\Gamma^*(r)$ has power-law
corresponding to the fractal dimension $D \approx 2$ for
the interval of scales $1 \div 25\,\mph$. For scales
$25 \div 70\, \mph$ there is a
deflection from the power law  and at scales
$70 \div 100\, \mph$ the $\Gamma^*$ achieves the constant
value. This was interpreted as a detection of the homogeneity scale 
$R_{hom}\approx 70\,\mph$ for the LRG SDSS galaxies.

It should be noted that in a slice-like survey, such as the above considred
sample of LRG, an artificial homogenization is possible
starting from scales of about $0.25 \,R_{max}^{sph}$, when the independent
spheres in transversal direction cannot be completely embedded in
the sample volume. Hence the above finding of a homogeneity 
should be in future reconsidered for larger spherical volumes.

\subsubsection {The problem of the peculiar
velocity field }

Results from the correlation analysis of 2dF and SDSS 
data show that the galaxy distribution at least within
interval of scales $0.1 \div 20 \,\mph$ is compatible with
a fractal distribution having the fractal dimension
$D \approx 2.2$.

As we discussed in sec.3 the restoration of the real-space
$\xi$-correlation function from the observed redshift-space
$\xi_z(s)$ involves a procedure of projection
which is defined only for structures with fractal dimension
$D < 2$ (i.e.$\gamma > 1$).
However the observed value of the fractal dimension
e.g. for 2dF galaxies was $D=2.25$ ($\gamma_s = 0.75$),
which  violates the necessary condition for the
restoration of the real-space correlation.
Therefore the results of papers based on the procedure
of projection should be reconsidered by using appropriate
methods of restoration.

Ignoring the problem of projection and using the standard
procedure of the restoration of the real-space $\xi$-function 
for 2dF data ( Hawkins et al. 2003) and for
SDSS data (Zehavi et al. 2002)
it was found that
the deprojected spatial correlation $\xi$-function has the 
canonical slope $\gamma \approx 1.7$ and the unit scale
$r_0 \approx (5 \div 6)\,\mph$
for the distance interval $0.1 < r < 15\,h_{100}^{-1}$ Mpc.
The derived  characteristic pairwise 
velocity dispersion
$\sigma_v = 500 \div 600$ km/s which is slowly decreasing 
with increasing
distance in the interval $0.1 \div 10\, h_{100}^{-1}$Mpc. 

Though the effect of peculiar velocities
leads to certain distorsion of the correlation function,
which should be studied by appropriate methods,
there is observational
evidence that the value of the distorsion
at scales $1 \div 10$ Mpc should be small because of the small 
observed velocity
dispersion $\sigma_v < 100$ km/s in the local
Hubble flow (Sandage 1999; Ekholm et al. 2001;
Karachentsev et al. 2003; Whiting 2003). This result is also
in contradiction with the much higher value of the velocity
dispersion derived from the $\xi$-function analysis.

Comparison of these results with N-body simulations in  
the $\Lambda$CDM Hubble
Volume is restricted by a priori unknown biasing 
factor between the
simulated cold dark matter
density (and velocity) field  and the real baryonic
distribution of luminous galaxies. 
The arbitrary bias as a function of  scale, 
$b(r)$, is still not predicted by the 
theory of the
large scale structure formation -- so there is freedom 
to choose this function
so that simulations can fit any observations.

\subsubsection { The problem of the dependence
of $r_0$ on $R_s$, $L$, $\bar{d}$ and galaxy type }

First of all as we already discussed in sec.3
the unite scale $r_0$  is not a proper
characteristic of galaxy clustering but contains
artificial distorsions due to individual properties
of the sample geometry and the total number of galaxies.
Hence a study of the relation between $r_0$ and other
physical parameters of a galaxy sample is not a correct
approach to galaxy clustering.

Secondly, the method of the projected $\xi$-correlation
function used by many groups
(Norberg et al. 2001, 2002; Hawkins 2003; Madgwick et al. 2003; 
Zehavi et al.2004a;b)
for derivation of the value of the real-space 
"correlation length"  $r_0$ utilized the procedure
which eliminates the fractal structures with $D \geq 2$. 

Therefore the observed relations between $r_0$ and 
$R_s$, $L$, $\bar{d}$, and galaxy type 
(Norberg et al. 2001, 2002; Hawkins 2003; Madgwick et al. 2003; 
Zehavi et al.2004a;b)
contain a mixture of artificial and real effects which
are difficult to separate.

An appropriate method of study of the correlation properties
of the fractal galaxy distribution is the dependence of the
fractal dimension on luminosity or galaxy type, which is
expected for multifractal structures (Pietronero 1987;
Sylos Labini \& Pietronero 1996; Gabrielli et al. 2004).
The main problems for such studies are the  small solid
angle in the slice-like galaxy catalogues and the small
number of galaxies in narrow absolute magnitude intervals. 
It is important to consider intervals of distances defined by
the characteristic scales of samples (sec.3.4.3).

\subsubsection { Power spectrum and intersection of fractals}

In sec.3.7 we considered the method of power spectrum
analysis of galaxy distribution.
As examples of its application to real data
we consider CfA and SDSS galaxy samples.

{ \it CfA redshift survey. }
Power spectrum (PS) of the CfA redshift survey was discussed
by Park et al.(1994). They extracted four volume limited samples
with about one thousand galaxies per sample and with the depths
60, 78, 101 and 130 $\mph$.

The PS is well described by two power laws:
1) at scales $5 \div 30\mph$ the spectrum is $P(k)\propto k^{-2.1}$,
2) at scales $30 \div 120\mph$ the spectrum is $P(k)\propto k^{-1.1}$.
Because the radius of the maximum sphere is about $30\mph$, this means
that the observed behaviour of the P(k) is consistent with the fractal
structure with fractal dimension $D=2.1$ up to scales $r=120\mph$.
At scales $r>R_{max}^{sph}$  the survey effectively becomes
2-dimensional and in accordance with the theorem of intersection
of fractal structure (sec.2.4.5)
the expected fractal dimension of the intersection should be
$D_{int}=D-1=1.1$, which is just observed.

{ \it SDSS galaxy redshift survey. }
The three dimensional power spectrum of several VL samples of the
early data release SDSS galaxy redshift survey was analyzed
by Tegmark et al.(2004). At their fig.22 decorrelated real-space
galaxy-galaxy power spectrum is presented, and again two
power-law presentation of the PS is possible:
1) at scales $10 \div 60\mph$ the spectrum is $P(k)\propto k^{-2}$,
2) at scales $60 \div 200\mph$ the spectrum is $P(k)\propto k^{-1}$.
Again as in the CfA case, this is consistent with the fractal
structure having the fractal dimension $D \approx 2$ at all
considered scales. The same behaviour of the power spectrum $P(k)$
was found also for the 2dFGRS sample (Tegmark et al. 2003). 

\subsection {Other results of the fractal 
approach}

\subsubsection {The two-point conditional
column density }

In section 3.6 we considered a new method of the fractal
approach which is based on the calculation of the
probability density to find a particle along the line
of sight under the condition that the line is ended by
two points of the structure. This method allows one to
extend the fractal analysis up to the depth $R_s$ of
a survey with the slice-like geometry.

The analysis of LEDA and SDSS galaxies (Baryshev \& Bukhmastova 2004)
and 2dF VL samples (Vasiliev 2004; Vasiliev et al.2005) with
the two-point conditional column density
method, essentially extends the interval of probed scales
from $20\mph$ (reached by the $\Gamma$-function method)
up to about $100\mph$. The result of the analysis is that
the fractal dimension $D\approx 2.2$ for the whole interval
of probed scales.

\subsubsection {The local number counts of all-sky bright LEDA galaxies}

The local galaxy number counts as a function of apparent magnitude have
two important applications in cosmology. Especially in the epoch of
angular galaxy catalogues, the counts were regarded as a strong argument
for a homogeneous galaxy distribution. Also, the local counts serve as
a reference value for deeper galaxy counts.
Here we give modern results on the the counts 
of bright galaxies on the basis
of the all-sky LEDA data base.

The Lyon extragalactic data base (LEDA) created by Georges Paturel in 1983
is a continuation of de Vaucouleurs's Reference Catalogue and its later
editions.
In fact,\emph{The Third Reference Catalogue} was already based on the LEDA
data. The LEDA extragalactic database offers currently a catalogue of
homogeneous
parameters of galaxies for the largest available whole sky sample.
Among its over million galaxies there are about 
50 000 galaxies with a measured
B magnitude brighter than 16 mag.

Made from the amalgamation of all available catalogues and
continually being completed with the flow of new data, the
completeness of the LEDA sample has been studied over the
years inspecting the counts (Paturel et al. 1994; Paturel et al.1997;
Gabrielli et al. 2004).
Simultaneously with completeness,the counts give information about
the slope of the bright end of the galaxy counts, hence on the spatial
distribution law.

Recently, Teerikorpi (2004), in connection with a study of the influence of
the Eddington bias on galaxy counts, investigated the LEDA galaxy counts in
the B magnitude range $10 \div 16$.
All such galaxies from LEDA were taken having its total B magnitude and
its $\sigma$ given,
and having galactic latitudes $\mid b\mid > 25$ deg.
The analysis of the counts indicated a slope of 0.44 in the B range
$10 \div 14$, which corresponds to the fractal dimension
$D=2.2$ up to scales about $100\mph$.

Rather similar results were obtained by Courtois et al. (2004a),
who calculated the
regression line up to $B =16$ for a somewhat different LEDA sample,
and derived the
slope of the counts of about 0.5.

\subsubsection {Radial number counts from the KLUN sample}

The usual studies of the large-scale structure utilize redshift as
distance indicator.
However, it is possible to use other distance measures, such as the
Tully-Fisher
and Faber-Jackson relations. Teerikorpi et al. (1998) derived the radial
galaxy spatial distribution around our Galaxy, 
utilizing over 5000 Tully-Fisher
distance moduli from the KLUN program.  First results give evidence for
a decrease
in the average density consistent with 
the fractal dimension $D= 2.2 \pm 0.2$
in the distance range $10 \div  100 \,h_{100}^{-1}$Mpc.  
This may be regarded as a new,
independent argument in favour of the results obtained by conditional
density methods,
based on redshift distances, which also lead to the fractal dimension
$D\approx 2$ at the same scale interval(Sylos-Labini et al. 1998).

The method of distance moduli, differs from usual galaxy counts in the
sense that the TF
distance moduli probe with a better spatial resolution the distribution
of galaxies.
A magnitude measurement provides a very poor distance estimate, while a TF
distance modulus has an error of about 0.5 mag. Furthermore, the method
developed by Teerikorpi et al.
(1998) takes into account the incompleteness of the sample, 
when it is applied.
It will be interesting to apply the method to the 
larger KLUN+ sample with its
20000 galaxies in the  future.

{ \it Conclusions from modern local number counts. }
The above results imply several important conclusions:

\begin{itemize}
\item[$\bullet$]{\emph{
there is no  $0.6m$-law in the bright galaxy counts, hence
no homogeneity up to $100h_{100}^{-1}$ Mpc};}
\item[$\bullet$]{\emph{
the observed $0.44m$-law of bright LEDA galaxies
and radial counts of KLUN galaxies are
consistenЕt with a fractal structure having
 $D=2.2$};}
\item[$\bullet$]{\emph{it is necessary to
revise the normalization of deeper
 galaxy counts}}
\end{itemize}

First, there is no slope of 0.6 for galaxies in the B magnitude
interval brighter than 14,
i.e. there is no uniformity in the galaxy distribution up to 5000 km/s
($100h_{100}^{-1}$ Mpc). 

Second, the observed number counts
$N(m)= 0.44m +const$ in LEDA, the radial distribution 
$N(r) \propto r^{2.2}$ for KLUN sample, and conditional
density for main redshift catalogues $\Gamma(r)\propto r^{-1}$
are consistent with a fractal galaxy 
structure having $D = 2.2 \pm 0.2$.

Third, the absence
of homogeneity on small scales influences the estimation of average number
and luminosity densities (Joyce \& Sylos Labini 2001).
In case of homogeneity, these densities are constant, while for a fractal
structure
they depend on the radius of the volume in which they are calculated. 
This means that
a usual normalization of deep galaxy counts based on the local
 homogeneity should
be revised to take into account the local radial inhomogeneity.

\section {Why fractality is important for cosmology}

The discovery of the fractal structure in  
3-d redshift galaxy surveys
opens new perspectives for understanding the origin
and evolution of the large-scale structure of the Universe.
One of the most important unsolved questions of theoretical
cosmology is how to describe non-analytical fractal sources
of the cosmological gravity field. Also much more studies 
should be done on the fractal  velocity field and its
evolution. This new situation in cosmology requires a careful
reanalysis of the logic and structure of modern world
models.

\subsection {Basic elements of cosmological models}

Modern cosmology is based on the following ``corner-stones'':
\begin{itemize}
 \item[$\bullet$]{\emph{cosmological principles,}}
 \item[$\bullet$]{\emph{fundamental physical theories, }}
 \item[$\bullet$]{\emph{cosmological observational data.}}
\end{itemize}
In all these parts the fractal approach plays an essential role.
Einstein's Cosmological Principle is extended up to 
Mandelbrot's CP. Fundamental physical theories include
the new fractal mathematics. Cosmological observations uncover
the scale-invariant properties of galaxy distribution.

\subsubsection {Three major empirical laws in cosmology}

The 20th century witnessed three major steps on the
ladder of key discoveries, which unveiled three cosmological
empirical laws:
\begin{itemize}
 \item[$\bullet$]{\emph{the cosmological redshift--distance law}}
 \item[$\bullet$]{\emph{the thermal law of cosmic 
microwave background radiation}}
 \item[$\bullet$]{\emph{the fractal law of galaxy distribution}}
\end{itemize}
Advances in astronomical instrumentation and spectroscopy
were necessary for the discovery of the galaxy universe and then
the Hubble law of redshifts in 1929. 
The development of
radio astronomical devices led to the discovery of the thermal ocean
of 3K photons in 1965.
Finally, as we have discussed  in this review,  gathering of
thousands of galaxy spectra with dedicated telescopes 
and applying appropriate methods of analysis, revealed
the fractality of the large-scale  galaxy
distribution in the last two decades.

\subsubsection {The theoretical basis of modern cosmology}

All four fundamental physical interactions -- the strong, the weak,
the electromagnetic, and the gravitational -- are used
for modelling and understanding
cosmological phenomena and for predicting
observed astrophysical effects.

But a special role in cosmology, among the fundamental 
interactions, is played by
gravitation, the true astronomical force. It appears as the dominating
force in the universe, starting from planetary and stellar scales
up to cosmological distances. Its is natural that the theory of gravitation
is at the heart of cosmology. It determines the large scale evolution of
matter in the universe.

Moreover, a cosmological model itself is the particular solution of 
the gravity field equations and this is why the gravity theory
lies in the foundation of the whole building of modern
cosmology (see discussion in Feynman et al. 1995;
Turner 2002a,b; Peebles 2002, 2003).
This also explaines why the astrophysical tests 
of gravity theories should be considered as crucial
cosmological observations (Baryshev 2003).

\subsubsection {The standard cosmological model}

At the start of the 21st century the standard cosmological model
continues to be the Friedmann (1922, 1924) model of expanding space.
It is  based on Einstein's
general relativity (GR)
and the cosmological principle of homogeneity.  The hot big bang
picture also includes the process of growth of local
inhomogeneities, caused
by the gravity of non-baryonic cold dark matter (CDM). The global
dynamics of the universe is at the present epoch
determined by the antigravity of dark energy.
The ``ordinary'' matter (stars, gas, dust), with which cosmology was
concerned most of the last century, now is thought to make mere
0.5 percent of the mass of the universe and therefore cannot influence
its expansion.

{ \it Equations for gravity field. }
Einstein's  equations of
general relativity gives the relation between the geometry of
space--time and the energy--momentum contents of matter
(we use the notations from Landau \& Lifshitz(1971) ):
\begin{eqnarray}
 \Re_k^i - \frac{1}{2}g_k^i\,\Re = \frac{8\pi G}{c^4}T_{k}^{i}\,,
\label{einstein}
\end{eqnarray}
in which  $\Re_k^i$ is the Ricci tensor, $g_{ik}$ is the metric tensor,
and
\begin{eqnarray}
 T_{k}^{i} = diag(\varepsilon,-p,-p,-p) \label{EMT}\,.
\end{eqnarray}
is the total energy-momentum tensor (EMT) of the cosmological fluid in
comoving coordinates. EMT includes two components: 1) ordinary matter
with positive pressure, and 2) exotic substance, called dark energy or
quintessence, with negative pressure.

{ \it Einstein's Cosmological Principle. }
The cosmological principle of uniformity implies that the density and pressure
of the cosmic fluid are functions of cosmic time only:
 \be \label{rhohom}
 \varrho(\vec{r},t) = \varrho(t)\,,
\ee \be \label{preshom}
 p(\vec{r},t)=p(t)\,.
\ee
 The total energy density $\varepsilon = \varrho c^2$
and pressure $p$ are given by sums of the above mentioned components:
\begin{eqnarray}
\label{eps2}
  \varepsilon =\varepsilon_{m} + \varepsilon_{de}\,,\qquad
  p = p_{m} + p_{de}\,.
\end{eqnarray}
Here index ``m'' relates to different kinds of ordinary  matter (dark
and luminous) with positive
pressure and with an equation of state
\begin{eqnarray}
 p_m = \beta\,\varepsilon_m,\qquad 0 \leq \beta \leq 1\,,
\label{pm}
\end{eqnarray}
e.g. $\varepsilon_m =\varepsilon_{CDM} + \varepsilon_{\mathrm{baryons}}
+ \varepsilon_{\mathrm{rad}} + \varepsilon_{\nu} $.
Index ``de'' (dark energy/quintessence) means the more exotic 
substance with negative pressure and with
an equation of state
\begin{eqnarray}
 p_{de} = w\,\varepsilon_{de},\qquad
 -1 \leq w \leq 0\,.
\label{pq}
\end{eqnarray}
As a particular case ($w=-1$) this includes Einstein's cosmological
constant $\Lambda$, which may be interpreted as cosmological vacuum.
Quintessence as a new cosmic perfect fluid with the equation of state
(\ref{pq}) was proposed by Caldwell, Dave \& Steinhardt (1998)
and Zlatev, Wang \& Steinhardt (1998) as a solution of the cosmological
coincidence problem (Peebles \& Ratra 2003).
Recently, values $w < -1$ have also been considered in the equation of
state for the so-called ``phantom energy'' (Caldwell, Kamionkowski \&
Weinberg 2003).

{ \it Friedmann equations. }
The Friedmann model is an exact solution of the Einstein's 
equations (eq.\ref{einstein}). Under the assumption of uniformity
(eqs.\ref{rhohom},\ref{preshom})
the Robertson-Walker
line element for homogeneous Riemannian spaces may be written
in the form:
\be
\label{rw1}
ds^{2} = c^{2}dt^{2} - S(t)^{2} d\chi^{2} - S(t)^{2} I_k(\chi)^{2}
d \omega^2
\ee
where $d \omega^2=d\theta^{2}+\sin^{2}\theta d\phi^{2}$,
$\:\chi,\theta,\phi$ are the "spherical"
comoving space coordinates, $t$ is the synchronous time coordinate,
$\:I_k(\chi) = \sin(\chi),\chi,\sinh(\chi)$
corresponding to curvature constant
values $\:k=+1,0,-1$ respectively and $\:S(t)$ is the scale factor.

Einstein's equations (eq.\ref{einstein})
in the case of homogeneity are directly
reduced to  the  Friedmann's equation, which may be presented in
the following form:
\be
\label{freq1}
\frac{d^{2}S}{dt^{2}}=
- \frac{4 \pi G}{3} S \left( \varrho+ \frac{3p}{c^{2}}\right)
\ee
Te Bianchi identity implies the continuity equation
\be
\label{cont}
\dot{\varrho} = -3\left(\varrho + \frac{p}{c^2}\right)
\frac{\dot{S}}{S}
\ee
which must be added to eq.\ref{freq1}.
Because the Lagrangian comoving coordinate $\chi$
does not depend on time,
one may rewrite eq.\ref{freq1}, using the definition of the
proper metric distance $r=S(t)\chi $,
as another form of the exact Friedmann equation:
\be
\label{freq2}
\frac{d^{2}r}{dt^{2}}= -\frac{GM_g(r)}{r^2}
\ee
where the gravitating mass $M_g(r)$ is given by
\be
\label{mgtot}
M_g = M_m + M_r + M_v
\ee
and contributions from matter, radiation and vacuum are
\be
\label{mgm}
M_m(r) =
\frac{4\pi}{3} \left( \varrho_m + \frac{3p_m}{c^{2}}\right)r^3
\ee
\be
\label{mgr}
M_r(r) = \frac{4 \pi }{3} 2 \varrho_r r^3
\ee
\be
\label{mgv}
M_v(r) = - \frac{4 \pi}{3} 2\varrho_v r^3
\ee
Solving the Friedmann's equation (Eq.\ref{freq2})
one finds the dependence on time for the metric distance
$\:r(t)$ or the scale factor $S(t)$. A classification
of two-fluid matter-dark energy Friedmann models is given
by Gromov et al. (2004).

{ \it Cosmological parameters. }
The FLRW model has two main parameters.
The Hubble parameter $ H = \dot{S}/S$ and the
deceleration parameter $ q = - \ddot{S}S/\dot{S}^2 $
which for the present time
$\:t_{0}$ are $\:H(t_{0})=H_{0}$ and $\:q(t_{0})=q_{0}$ respectively.

One also frequently uses the density parameter
$\Omega =\varrho/\varrho_{cr}$
where the critical density is
\be
\label{rhocr}
\varrho_{cr} = \frac{3 H^{2}}{8\pi G}.
\ee
Eq.\ref{freq2} may be written also in the form:
\be
\label{dec}
q = \frac{1}{2}\Omega \left(1 + \frac{3p}{\varrho c^2} \right)
\ee
where $\Omega, p ,\varrho$ are the total quantities, and
\be
\label{Omega}
\Omega=1+\Omega_k
\ee
with $\Omega_k = kc^2/S^2H^2$.

The old standard model has the following parameters
\be
\label{old}
\Omega_0 = \Omega_{(m)0} = 1, \;\; \Omega_v =0, \;\; q_0=0.5
\ee
The new standard model which is currently accepted has
\be
\label{new}
\Omega_0 \approx 1 , \;\;\Omega_m \approx 0.3,\;\;
\Omega_v \approx 0.7, \;\; q_0 \approx -0.6\,.
\ee

This means that in the Friedmann model 
the expansion of the present universe is
accelerated and that the dominant force in the universe is
cosmological antigravity of the vacuum.

\subsubsection { Fractal sources for gravity
field   }

The results of proper statistical analysis of spatial galaxy
distribution have revealed that the galaxy distribution is
essantially inhomogeneous at least up to scales about 100 Mpc.
It means that in modern cosmology one should consider more
general models than homogeneous Friedmann ones.

{ \it Mandelbrot's Cosmological Principle. }
The global mass-radius relation $M(r)$ is 
the main characteristics of the fractal matter distribution
which determines the fractal density field $\rho(r)$
as the sources of gravitational field.

In the next section we shall consider Mandelbrot's cosmological
principle , which is a generalization of the
Einstein's CP for the case of 
inhomogeneous cosmological models with isotropic fractal
structures , and may be written as
 \be \label{rho-mcp}
 \varrho(\vec{r},t) = \varrho(r,\,t)\,,
\ee \be \label{p-mcp}
 p(\vec{r},t)=p(r,\,t)\,.
\ee
It is important that here the variable $r$ is the radius of a ball
around each point of a structure (see sec.2.3).

Fractal cosmological models are based on 
solutions of gravity field equations with the sources
described by the fractal density law. 

{ \it Lema\^{i}tre-Tolman-Bondi model in general relativity. }
Lema\^{i}tre-Tolman-Bondi (LTB) models are exact solutions of
Einstein's
equations for 1) spherical symmetry, 2) pressureless
matter (dust) and 3) motion with no particle layers intersecting. 
Originally studied by Lema\^{i}tre, Tolman and Bondi,
these models are the simplest generalization of the
Friedmann-Robertson-Walker (FRW) models with a non-zero density gradient
(Bondi 1947).

The LTB model has been used for understanding the kinematics and dynamics
of galaxies around \emph{individual} mass concentrations. For
example, Teerikorpi et al. (1992),  
and Ekholm et al. (1999) could put in evidence the
expected behaviour in the Virgo supercluster: 1) Hubble law at
large distances, 2) retardation at smaller distances, 3) zero-velocity
surface, and 4) collapsing galaxies at still smaller distances.

Bonner (1972) was the first to apply the LTB model to the hierarchical
cosmology. He used de Vaucouleurs'
density law $\rho \sim r^{-\gamma}$ with $\gamma =1.7$. Ribeiro
(1992, 1993) has developed a numerical method for the solution
of the problem. 
A new approach to the
solution of LTB equations for a fractal galaxy distribution
with large scale asymptotic FRW behaviour was presented by
Gromov et al. (2001).

However, the application of LTB models to a fractal distribution
leads to a conceptual problem, because the original
LTB formulation contained a central
point of the universe, around which the density distribution is isotropic.
In a fractal distribution there is no
unique centre, but every object of the structure may be treated as a local
centre which accommodates the LTB centre.  Every structure point is surrounded
by a spherically symmetric (in average) matter distribution.

In this sense, the application of the LTB model to fractals means that there
is an infinity of LTB exemplars with centres on every structure point.  Their
initial conditions are slightly different, because for any fixed scale the
average density is approximately constant.  For different scales the density
is a power law. This excludes geocentrism and makes
possible the use of LTB models as an exact general relativistic
cosmological model where expansion of space becomes scale
dependent.
A similar problem of an apparent centre exists even
for Friedmann models where the rate of space expansion within distance 
$r$ from a fixed galaxy is determined by the total mass of the sphere around
this galaxy. For LTB
fractal models the space expansion at distance $r$ from a fixed point of the 
fractal structure is also determined by the
average mass of the sphere around this point.

\subsection {The origin and evolution of large scale
 fractals: challenges for theoretical models}

The observed fractality of large-scale  galaxy distribution opens
new aspects in the process of the structure formation.

\subsubsection {Hubble law within fractal galaxy distribution}

Two fundamental empirical laws which have been established from
extragalactic data are in apparent conflict with the SCM.
First, there is the power law density-radius relation
(de Vaucouleurs law) which corresponds to fractal struture with
fractal dimension $D\approx 2$ up to the scales of about 100 Mpc.
Second, Cepheids,
TF-distance indicator and Type Ia supernovae
confirm the linearity of Hubble's redshift-distance law within the
distance scales $1 \div 100 \mph$, just where the fractality exists.

As we noted in sec.4.3 Sandage, Tammann \& Hardy (1972) first
recognized the contradiction between the observed linearity
of the Hubble law and a possible hierarchical galaxy distribution.
They used the existence of the very local Hubble law as an argument
against the de Vaucouleurs hierarchical model, which was based
on calculations of the redshift-magnitude relation performed by
Haggerty \& Wertz (1972) for hierarchical cosmologies.

\begin {figure}
\centerline {\psfig {file=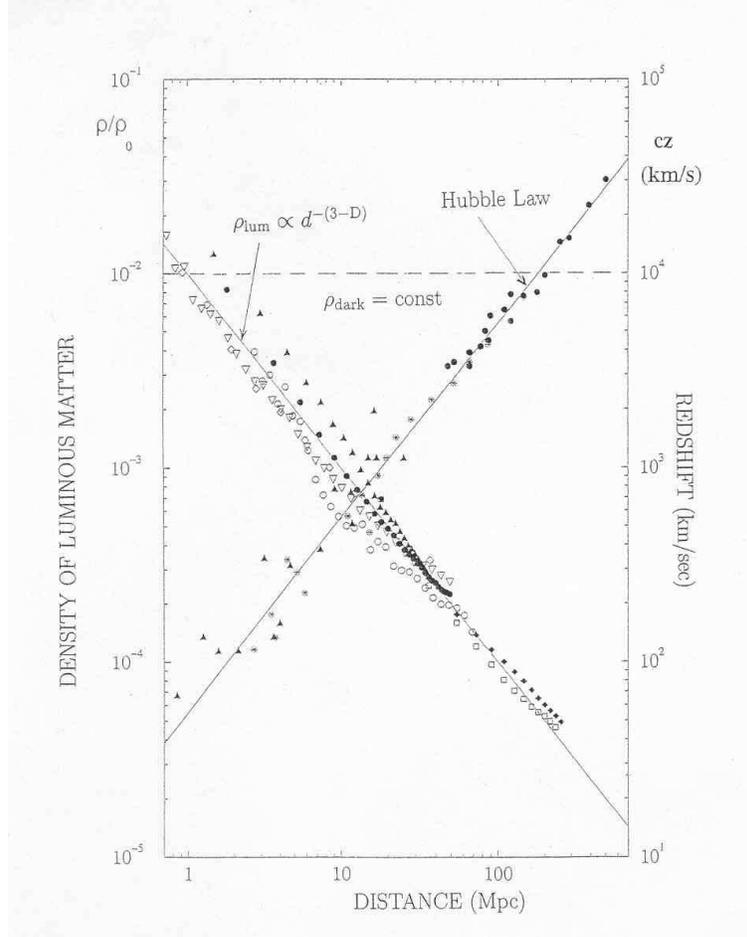,width=10cm}}
%\resizebox{\hsize}{!}{\includegraphics{fig7.ps}}
\caption{\footnotesize  The Hubble law within
the fractal galaxy distribution.
From Baryshev et al. (1998).
}
\label {hubbledevauparadox}
\end {figure}

A strong deflection from linearity expected within the fractal 
inhomogeneity cell was confirmed by Fang et al. (1991). 
The fractal aspect of this problem was discussed by
Baryshev (1992, 1994) and
Baryshev et al. (1998), who emphasized that the observed linear
redshift-distance relation inside the power-law fractal density
distribution creates the problem which they 
called the Hubble-deVaucouleur's paradox, i.e. the
observed coexistence of both laws on the same scales (see Fig.8). 
There are several possible solutions of the problem.

{ \it Asymptotically homogeneous LTB models. }
Using the
LTB model Gromov et al. (2001) found the necessary conditions
for the linear Hubble law existing within the fractal structure
with fractal dimension $D=2$. The larger the scale of homogeneity
$R_{hom}$ the smaller should be the density parameter $\Omega_m$,
e.g. for $R_{hom}=100$ Mpc the linear Hubble law exists at distances
$r>1$ Mpc if the density parameter $\Omega_m<0.01$.

{ \it Open Friedmann model. }
Joyce et al. (2000) suggested an open Friedmann model where
the decreasing density of the fractal matter component at distance
$R_{hom}$ becomes less than the density of the cosmic background
radiation, which plays the role of the leading homogeneous component
of the FRW universe. Hence the density parameter at the present
epoch is $\Omega_0=\Omega_{rad}$ and space expands
within the fractal structure. They also 
considered qualitatively the
modifications to the physics of anisotropy of the CBR,
nucleosynthesis and structure formation.

{ \it Influence of dark energy. }
Chernin et al. (2000),
Chernin (2001), Baryshev et al. (2001) suggested
that the cold local Hubble flow is a signature of the
dominance of the cosmological vacuum or dark energy.
Recent cosmological  N-body  simulations by Maccio et al.(2004)
confirmed that inclusion of cosmic vacuum in the calculations
leads to significantly lower velocity dispersion in Local
Volume--like regions than what happens without vacuum.

\subsubsection { Problem of the origin of the fractal structure}

Since the 1970s  Haggerty (1971), who worked
in the University of Texas' Center for Statistical
Physics, has developed 
a Newtonian model for the creation
of hierarchical self-gravitating structures.
This was based on  Prigogine's and Severne's study
of a non-Markoffian kinetic theory of binary gravitational
interactions with irreversible growth of correlational energy.

A scenario of explosive origing of large-scale structures was
suggested by Ostriker \& Cowie (1981) where large cavities and
the superclusters of galaxies at scales up to 100 Mpc.
This picture of galaxy formation is close to the way star formation
is viewed. Within this approach Schulman \& Seiden (1986) applied
the percolation theory and numerical simulations for the derivation
of the correlation exponent of the resulting structures.
They got $\gamma \approx 1$ which corresponds to the fractal
dimension $D \approx 2$. To obtain the commonly accepted value
$\gamma = 1.8$ they tried to find a mechanism for changing
the initial slope. 

Models for origins of the fractal large-scale structures
were considered by Szalay \& Schramm (1985), Pietronero \& Kupers (1986),
Maddox (1987), Luo \& Schramm (1992).

Statistical mehanics and thermodynamics of self-gravitating gas 
was considered by Perdang (1990), de Vega et al. (1996, 1998),
Combes (1998).
The intriguing result is that
Newtonian self-gravitating
N-body systems have a quasi-equilibrium fractal state 
with a dimension of $\approx 2$, though several problems are
still open and need more studies.

N-body simulations in order to study 
the gravitational growth and possible
origin of fractal structures from initial small density
fluctuations were made by Bottaccio et al. (2002),
Baertschiger et al. (2002, 2004). They discovered that
the discreteness has a
strong influence  on the results of 
standard cosmological simulations and noted that 
the Hubble time scale is too short for developing
observed fractal structures within the whole range of scales.

Large density fluctuations even at high redshifts lead to a
conflict with small temperature fluctuation of the CMBR and
to the problem of formation time for largest structures.
A possible solution is that the standard scenario of
gravitational growth of large-scale structure from small
initial density seeds should be revised. Also the observed
anisotropy of CMBR can be essentially distorted by the
interving matter, so the fractal initial conditions 
may not relate to the observed $\Delta T/T$ (Schwarz et al. 2004).

\subsection {The Cosmological Principle }

There is much confusion in the literature on how 
Mandelbrot's Cosmological Principle of Fractality
and its relation to the Einstein's Cosmological
Principle of Homogeneity should be understood. 
Below we compare these principles
and show that the principle of fractality is a natural
generalization of the principle of homogeneity.

\subsubsection {Einstein's Cosmological Principle}

Einstein (1917)  applied his general relativity
to the cosmological problem, i.e.\ for constructing a model of the
universe as a whole.
Not yet knowing about galaxies, he imagined a world filled with stars
and argued that the stars have a natural spatial distribution,
which is uniform: matter concentrations around any
preferred centre should with time evaporate and disperse uniformly
all over the universe.
Later, in correspondence with Selety, Einstein rejected
the hierarchical distribution of stars (see sec.4.1).

Einstein  postulated
a  uniform matter distribution, and put relativistic gravity
into cosmology. The result was a world with uniform geometry.
The Copernican cosmological principle has many faces, as
has been interestingly discussed by Rudnicki (1995) in his book.
Its one formulation ``all places in
the Universe are alike'' is naturally fulfilled in a homogeneous
world.  Besides the absence of a centre, another plus-side of
uniformity was a simplification of Einstein's equations, which
permitted him to derive the static spherical world model.
Finally,
Friedmann (1922)  liberated the universe from this stiff state, allowing the
uniformly distributed matter and space to expand.

The name Einstein's Cosmological Principle
 for the hypothesis of the homogeneity of large scale
matter distribution in the universe was coined by
 Edward Milne who analyzed the foundations of
 cosmology in the 1930s.
Milne formulated also an observer-oriented version of
 the Cosmological Principle as:
 \emph{the whole world-picture as seen by one observer (attached to
a fundamental particle or galaxy) is similar to the world-picture
seen by any other observer}.

In these early years of modern cosmology there
 was no direct observational evidence for the uniformity of the universe
 and it was theoretical reasoning which guided
the cosmologist.

{ \it Modern view on the Cosmological Principle. }
Every cosmological model has its beginnings in
cosmological principles, special kind of hypotheses which are
regarded as valid
for the whole, perhaps infinite universe, even though observations can be
made only from a finite part of it. Some principles touch questions of
epistemology and are tacitly supposed to be true, for example that
the knowledge about the whole universe is accessible for us, and
that the laws of physics are the same everywhere and produce
similar things.

In the current cosmological thinking the contents of the Cosmological
Principle are rather deep and complicated. We point out
three distinct statements which are usually regarded as a starting point
for constructing cosmological models.

\begin{itemize}
\item[CP1]{\it Physical laws the same in all space and time}
\item[CP2]{\it Fundamental physical constants are true constants}
\item[CP3]{\it Particular physical properties, including measuring
               standards, are the same in all space and time}  
\end{itemize}

When cosmology advances, these statements may require adjustments or
even radical changes in the light of new observations and theoretical
ideas. For example, multi-dimensional theories predict variations of
physical constants. However, all modern observational tests strongly
restrict any such variations (Uzan 2002). The requirement of the same
physical laws (CP1) also includes the possibility of discovering
new laws of cosmological physics which appear only on very large
spatial, temporal and mass scales. 
An important principle "More is Different"
was introduced by Anderson (1972), who emphasized that
each new level of complexity of material systems introduces
new laws for their behaviour.

In modern cosmology the \emph{Cosmological Principle} is understood
in a narrow sense as a hypothesis on the large-scale distribution of matter
in the
universe. For example, the statement that matter has a homogeneous and
isotropic distribution is at the basis of the standard cosmological model.
Nowadays observations
show that
Mandelbrot's cosmological principle of fractality has become an important
alternative for the spatial distribution of luminous matter.

\subsubsection {Derivation of uniformity from local isotropy}

Walker (1944), a British
mathematician who worked closely with Milne,
 proved that uniformity follows from his
 hypothesis of ``Local Spherical Symmetry'' which supposes that
 isotropy exists locally about each point of a Riemannian manifold.

A simple reasoning leading to homogeneity when there is isotropy
around each point, may be found in
the \emph{First Three Minutes} by Weinberg (1977) . He
shows how one can go from any one point to another arbitrary place along
circle arcs on which the density remains the same. Hence the density
is the same on every point.
However, strictly speaking this conclusion is based on a hidden
mathematical assumption of regularity, i.e.\ the existence of a smooth
density around each point,  only then from the left-hand side
does the conclusion follow:
\emph{Local Isotropy~+~No Centre~+~Regularity~ $\Rightarrow$~ Uniformity}.
Here ``local isotropy plus no centre'' means that all points are equivalent
and around each point the density law does not depend on the direction
(though it might depend on the distance from this point).
The ``regular'' matter distribution is described by continuous, smooth
mathematical functions.

In the chapter ``Simplifying assumptions of cosmology''
in his \emph{Introduction to Cosmology} Narlikar (1993)
explicitly introduces the assumption of the \emph{smooth fluid
approximation}, which essentially means going over from a discrete
distribution of
particles to a continuum density distribution.  This is what we call
``regularity''. It means that one may use the concept of the mass density
at each point of space, like in a fluid.
It is thus the union of local isotropy,
 no centre, and smoothness which gives homogeneity.
In this case the Cosmological Principle reduces to the statement
of the uniformity of the matter distribution.

\subsubsection {Mandelbrot's cosmological principle}

 In his book
 \emph{Fractals: form, chance, and dimension},
Mandelbrot (1977) foresaw that galaxies are fractal-like
distributed and gave the first mathematical description of the
fractal properties of such a distribution.
He recalls how around 1965, his ambition was to implement the
law of decreasing density with
a model where there is ``no centre of the universe'' or
``the center is everywhere".

Mandelbrot views the fractal galaxy distribution
as a major conceptual step in the description of the cosmological
matter distribution. It is a kind of
synthesis of hierarchical structures (``thesis'') and
homogeneity (``antithesis''), essentially based on randomness.
   Indeed, there is a fundamental difference
between true random fractals and stiff hierarchical protofractals.
Into protofractals the hierarchy is injected ``ex-nihilo'',
by defining explicitly its levels.  But fractals
internally contain a scale invariance (self-similarity) and the
impression of a hierarchy follows as an unavoidable consequence.
A useful example is the L\'{e}vy dust which is created by a random
walk process in which the direction of each step is chosen isotropically
and the length of a step follows a certain probability distribution.

Fractality carries within itself
also a trace of uniformity. Within a
fixed radius, i.e.\ for a fixed scale, every observer counts the
 same number of elements, \emph{on average}.  But upon  changing
 the radius, a ``new uniformity'' is found
with a new mean number density. Furthermore, there is no centre
for random fractals -- this is another ``relic'' from homogeneity.

Thus Mandelbrot made the first step for genuine
 fractals in cosmology, generalizing Einstein's
 cosmological principle corresponding to $D = 3$ by
% the cosmological principle of
fractality, which allows a non-uniform
galaxy  distribution with $D < 3$.  His
``Conditional Cosmographic Principle'' states that
all observers see similar cosmic landscapes around them, but
only under the condition that they make observations from a structure
element (galaxy).
``Conditional'' in the spatial context
emphasizes that each observer occupies a material element of the
structure (c.f. ``conditional density'').

Mandelbrot's cosmological principle of fractality --
that the observers attached to the material structure
elements are equivalent -- is close to
what Milne presented. Thus the fractality
of the universe perfectly satisfies Milne's Cosmological Principle.
It also automatically makes what Karachentsev (1975)
has called ``the ecological correction to the Copernican principle'' --
the real observer can \emph{live} only on or close to a material
celestial body. This is usually called the Weak Anthropic Principle.
In this sense the Copernican cosmological crinciple is contained by
Mandelbrot's principle of fractality and hence the assertions about an
``unprincipled''  fractal universe 
(see e.g. Coles 1998; Wu et al. 1999) are not
true.

{ \it Does isotropy always imply uniformity? }
The proof of uniformity is based on the density being smooth around all points,
which
is valid for regular distributions, but not for fractals.
It is smoothness  which wipes out fractality and makes uniformity.
 Thus strictly speaking from local isotropy and
the principle of no centre one can infer a fractality of the
structure, of which homogeneity is only a special case with
$D=3$ (see Sylos Labini, 1994).

Of course, there is never an \emph{exact} local isotropy around
every observer, not even in a uniform world, and still less inside
a fractal distribution.  Instead one may speak of a statistical
isotropy, so that the sky observed from any galaxy ``looks much
the same''. In particular the counts of galaxies as a function of
magnitude are similar around each galaxy, because radial distributions
of galaxies are similar.
It is natural to conjecture that for distributions made of
discrete points, there is a generalization of the above chain of
reasoning:
\emph{Statistical Isotropy~ + ~No Centre~ $\Rightarrow$~  Fractality}.
Statistical isotropy and no centre, without the restrictive assumption
of smooth mathematics, would thus lead to an isotropic fractal.

As considered in sec.2.4.5, the most important reason
for an apparent isotropic celestial distribution of galaxies
is the projection of a fractal structure
with fractal dimension $D\geq 2$ . According to the theorem
on fractal projections the resulting distribution will have the
fractal dimension $D_{pr}=2$, which means homogeneity
on a 2-d plane or isotropy on the celestial sphere. 

Two other factors which lead to apparent isotropy on the sky
are lacunarity and the luminosity function.
It is now known that the patchiness on the sky depends
 not only on the
fractal dimension, but also on the lacunarity,
which is a measure of how frequent large voids are.
Numerical simulations have shown that fractals with a small
lacunarity can have
rather smooth projections on the sky.
The second factor which smooths out the patchiness,
 is the large differences in the
 luminosities of celestial bodies.  As a result two objects
 with equal apparent brightness actually may have widely
 different distances. The mixing of nearby and distant objects
 hides clusters and fills in holes. For example, this
decreases the celestial anisotropy for very distant radio sources.

\subsubsection {Towards Einstein--Mandelbrot concordance}

We do know of genuinely uniform components of the universe: the
observed  photon gas of
 the cosmic background radiation,
the ocean of possible low-mass neutrinos, and maybe  more importantly,
the suggested physical vacuum or dark energy.
As the average density of the fractal matter decreases with
increasing scale, there will eventually be a scale beyond which the
density of the uniform component is larger than the density
of the fractal component.
Hence one may regard, after all,  the universe as homogeneous
on such scales. However, this is not due to the galaxy distribution,
but because of the uniformity of the relativistic matter component!

As to the fractal galaxy distribution, there are two alternatives --
a finite or infinite range of fractality.
True, there are no scale limits to a
pure mathematical fractal.
The name `fractal universe' is often linked with an
 infinite fractal.  Such a universe would have
zero average density.
But real physical objects usually have lower and upper cutoffs between
which the fractal properties are observed.
Thus it is also expected that
the fractal galaxy distribution appears only within a finite interval
of scales. One  possibility
is that it becomes homogeneous on some
maximum scale $R_{\mathrm{hom}}$.
Thus one may, as Mandelbrot did, allow for the possibility
that the matter distribution may become uniform on
 large scales, while being fractal on smaller scales.
With any uniform matter component, such as the photon-gas or
the cosmological vacuum, the universe becomes homogeneous
on a sufficiently large scale.
Such a universe would have a non-zero average density.
Thus the intuitions of both Einstein and Mandelbrot appear to have grasped
fundamental features of the universe.

\section {Concluding remarks}

The historical milestones of the path of the study of the large-scale
structure of the universe
are outlined in Table 2. There one may see that during the period
up to the 1970s many pioneering observational results and theoretical
considerations appeared. The idea of a hierarchical matter distribution
originated in a mathematical form thanks to Fournier, Charlier and Selety.
It is intriguing to see that the problems discussed by Selety and Einstein
in the 1920s, presently define whole directions in large-scale structure
physics (see Table 3).

Observers discovered on the sky a very lumpy distribution of galaxies
and prepared angular catalogues of galaxies and clusters, from which
the first
superclusters were detected up to sizes of 100 Mpc. These gathered clouds
over homogeneous cosmological models and there was a sharp debate about
the significance of such inhomogeneities.
At the time when there was no
extensive distance information, it seemed that convincing arguments
for a homogeneous galaxy distribution were found.
First, the number counts of bright galaxies appeared
to agree with the $0.6m$ law of homogeneity. Second, fluctuations in
Milky Way's dust extinction was a factor producing apparent clustering.
Third, the linearity of the Hubble law at small scales was seen as
evidence for the galaxy homogeneity at similar small scales.

In the meanwhile, a conceptual breakthrough was the suggestion by
Mandelbrot that the galaxy clustering may be described as a stochastic
fractal. After the nature of spiral nebulae was settled,
the discussion around the fractal dimension and the maximum scale of
galaxy clustering forms the core of the new Great Debate.

During those early years the correlation function method, used in statistical
physics for the description of homogeneous systems with small density
fluctuations, was extensively applied to the angular distribution
of galaxies. In an important step, Totsuji \& Kihara (1969) discovered the 
power law dependence
of the angular correlation function, with the exponent $\gamma = 1.8$
and a small homogeneity scale $r_0 = 5 h_{100}^{-1}$ Mpc. For a
comprehensive description of the statistical results from the angular
galaxy catalogue epoch see the review by Fall (1979) and the text-book by
Peebles (1980).

Since then
the correlation function has been applied for many angular and spatial
galaxy catalogues, and in the words by Kerscher, Szapudi \& Szalay (2000),
``{ \it the
two-point correlation function became one of the most popular statistical
tools in astronomy and cosmology}''.
These analyses created the feeling of confidence that the galaxy distribution
is uniform on scales larger than $10 h_{100}^{-1}$.

However,
when the study of the galaxy clustering entered the new epoch of 3-d maps,
a series of remarkable discoveries followed, which were unexpected
in view of the 5 Mpc ``correlation length''.
The first findings of real spatial large scale structures up
to 50 Mpc were reported at the Tallinn 1977 conference. After that larger
and larger structures have been discovered up to our days, proving that
the earlier detections of superclusters were not rare chance events,
but actually probed essential properties of the galaxy clustering.

A decisive step was the introduction by Pietronero (1987) of the statistical
method which
is proper for the analysis of a fractal distribution of galaxies.
This fractal-inspired method of conditional density has revealed the
scale invariant galaxy clustering with the fractal dimension $D \approx 2$
up to scales where the method is applicable (Sylos Labini et al. 1998).

Also, it has become clear that the classical 2-point correlation function
suffers from two major drawbacks. First, due to the projection effect
the angular correlation function cannot
detect spatial structures with the fractal dimension $D \geq 2$. Second,
due to the normalization condition, both angular and spatial
correlation functions yield systematically distorted values 
for the homogeneity
scale ($\theta_0$, $r_0$) and the correlation exponent ($\gamma_{ang}$,
$\gamma$), as compared with the true values of the complete correlation
function. The importance of these effects is seen from the derivation of
the correlation dimension $D_2 = 1.2$ from the $\xi$-method, when the
$\Gamma$-method gives $D_2 \approx 2$, for the main 3-d catalogues.
The effects explain why the results from the $\xi$ and
$\Gamma$ analyses have led to different conclusions about the homogeneity
scale and the correlation exponent.
These conflicting results on the value of the fractal dimension were also
a cause for the controversy around the multifractal
analysis of the galaxy data.

New extensive data, $200\,000$ galaxies for the 2dF survey
(instead of 2000 for the CfA), together with the $\Gamma$ function analysis
show that undistorted values of the correlation exponent is $ \gamma \approx
0.8$ which corresponds to the fractal dimension $D = 2.2 \pm 0.2$.
The scale of homogeneity $R_{\mathrm{max}}$ has a lower limit corresponding
to the
maximal sphere completely embedded into the volume of the galaxy survey,
which is about $40 h^{-1}_{100}$ Mpc for 2dF and $70h^{-1}_{100}$ Mpc
for the first release SDSS. In order to obtain an estimate  of
$R_{\mathrm{hom}}$ on larger scales, one must enlarge the solid angle of
the survey.
Also, indirect evidence exist, such as the  $450h^{-1}_{100}$ Mpc Sloan Great
Wall and $300h^{-1}_{100}$ Mpc inhomogeneities in the 2dF QSO distribution,
that $R_{\mathrm{hom}}$ could reach such large values.
An application of the new method of the two-point conditional
column density to the 2dF and SDSS data confirmed that 
$R_{hom}>100h^{-1}_{100}$ Mpc.

In order to obtain a reliable mass--radius
relation $M(r)$ from observational data it is essential to use appropriate
statistical methods of analysis (like conditional density $\Gamma(r)$)
and control the distance scales $R_{sep}$ and $R_{max}^{sph}$ which
determine the region of applicability of the used methods.

Three large numbers of galaxy redshift surveys needed for
further progress in the large-scale sctructure analysis: 
large solid angle
on the sky, large number of galaxies and large depth/separation
ratio. 

A special study of dark matter distribution is requiered,
which now is possible by using gravitational lensing technics
( Mellier 1999; Wittman et al. 2000; Gray et al. 2002; Refregier 2003).
The first results of the weak lensing analysis have shown that the dark  
and luminous matter are spatially similarly distributed.

We summarize
the following main challenges that the modern galaxy clustering
research faces:

\begin{itemize}
\item[$\bullet$]{ \it due to projection on the sky the angular distribution
                           of galaxies loses the information on structures
                       with fractal dimension $D \geq 2$;}
\item[$\bullet$]{ \it due to the normalization condition the $\xi$ function
                     estimators yield incorrect values of
                     the homogeneity scale $R_{\mathrm{hom}}$
                      and the fractal dimension $D$;}
\item[$\bullet$]{ \it the crucial role in statistical analysis
                     of a galaxy sample plays the three quantities:
                    the average distance between galaxies $R_{sep}$,
                    the radius $R_{\mathrm{max}}^{\mathrm{sph}}$ of
                      completely embedded spheres in the sample geometry,
                   and the total number of galaxies $N_{gal}$ 
                    in VL subsamples;}
\item[$\bullet$]{ \it essential fluctuations in the derived value of $D$
                     are caused by the 
                    ``cosmic variance'' (only one realization of
                      the stochastic process)
                    and the ``sampling variance'' (finite
                    available volume);}
\item[$\bullet$]{ \it the method of projected redshift-space correlation
                     function cannot be applied to fractal structures with
                     $D \geq 2$, hence new methods for extracting
                     real-space correlation properties of the galaxy
                      distribution are required;}
\item[$\bullet$]{ \it the pairwise velocity dispersion $500-600$ km/s on
                     scales $1 - 10$ Mpc, 
                     obtained by the $\xi$-function method, is surprisingly
                     high in comparison with the velocity dispersion
                     $\sigma_v < 100$ km/s in the Local Volume with $r < 10$
                     Mpc;}  
\item[$\bullet$]{ \it the differences in clustering of  galaxies with different
                    types and luminosities may require multifractal
                    analysis;}
\item[$\bullet$]{ \it understanding the origin and evolution of the observed
                      fractal distribution of galaxies may require a
                      revision of the standard picture of gravitational
                      growth of structure, by considering the consequences
                      of primordial fractal density fluctuations.}
\end{itemize}

We may conclude that the fractality of galaxy clustering has
become a fundamental empirical phenomenon of observational cosmology,
which should be explained by theoretical models of the Universe.

\section {Acknowledgements}

This work has been supported by Academy of Finland (project ``Fundamental
problems of observational cosmology'') and by the foundation 
Turun Yliopistos\"{a}\"{a}ti\"{o}. We are grateful to Luciano Pietronero
and Francesco Sylos Labini for valuable information, encouragement, and
criticism. We also thank Fred Rost for very helpful comments.

\end {document}